\documentstyle[11pt,./myext,epsfig,rotating]{article} 
\epsfverbosetrue
\title{Study of the determination of the SUGRA parameters using the 
                    ATLAS detector \\
     in the case of $L$-violating $R$ parity breaking}
\author{
Adrian MIREA and Elem\'er NAGY \\ \\
{\it\small Centre de Physique des Particules de Marseille,} \\
{\it\small Facult\'e des Sciences de Luminy,} \\
{\it\small IN2P3-CNRS, F-91288 Marseille, CEDEX 09 France} \\
}
\date{ }

\begin{document}           

\maketitle                 

\vspace{1cm}

\begin{abstract}
Simulating $pp$ collisions at LHC energies in the framework
of the SUGRA model and the detection of the produced leptons
and jets by ATLAS we demonstrate that a clean signature of SUSY can be 
obtained over a large domain of the parameter space 
in the case of $L$-violating $R$ parity breaking ($\lambda$ couplings).
The obtained signal allows the reconstruction of the SUSY particles and thereby
the precise determination of the model parameters $m_0$, $m_{1/2}$, tan$\beta$
and sign$\mu$.
\end{abstract}
\vskip 0.5 cm

\section{Introduction}

The ATLAS Collaboration has carried out a detailed study to detect the
SUSY signature in the framework of one of the most popular model,
SUGRA~\cite{SUGRA_rev},\cite{SUGRA_pap}. 
It has been shown~\cite{SUGRAincl} that if SUSY exists
at the electro-weak scale, it should be discovered by ATLAS and a general
method has been given to determine approximately the mass scale of the
SUSY particles. In subsequent papers~\cite{SUGRA12}--\cite{SUGRA5}
it was shown in five representative points of the parameter space
that some of the SUSY particles can be reconstructed and using the obtained
characteristics (masses, branching ratios) the model parameters can be
precisely determined \cite{Froid_LHCC}. All these studies have been carried out
assuming that $R$ parity is conserved.

In this note we consider that $R$ parity is broken in such a way that
the lepton number $L$ is violated through $\lambda$-type couplings. 
The present experimental limits~\cite{Rlimits}
cannot completely exclude such a scenario. In this case one of the prominent
signatures of SUSY, the missing energy is considerably weakened because
the lightest SUSY particle (LSP) is allowed to decay. Due to this decay
the lepton and/or jet multiplicity increases considerably and some efficient
cuts (e.g. lepton veto against the $t-\bar t$ background) cannot be applied.
On the other hand,
the decay products of the LSP in some cases allow its direct reconstruction.
Therefore the event topology and the search strategies are different of the
case when $R$ is conserved. This has motivated us to revisit
the feasability to detect SUSY and to determine the parameters of the SUGRA
model using the ATLAS detector.

In section 2 we give a brief description of the phenomenology of the $R$ parity
violation and the event generator used. 
Section 3 deals with the ATLAS detector and with the fast simulation
of its response. In section 4 we present the domain of the parameter space where
a SUSY signal can be expected by ATLAS. In the subsequent three sections the
reconstruction of the SUSY particles and the determination of the model parameters
are described in the LHC points 1,3 and 5 which represent a  heavy, light and
medium SUSY mass scale. We summarize the obtained results
in the concluding section.

\vskip 1.2 cm


\section{Basic Phenomenology}
\vskip 0.5 cm

\subsection{$R$ parity violation}
\vskip 0.5 cm

   $R$-parity has been introduced~\cite{Fayet1} in order to avoid fast nucleon decay
and flavor changing neutral currents (FCNC). If the multiplicative quantum number
\begin{equation}
R = (-1)^{3B+L+2S}
\label{eq:Rdef}
\end{equation}
is conserved it guarantees automatically baryon number ($B$) and lepton number
($L$) conservation. $R$ is +1 for Standard Model (SM) particles and its value is
-1 for their superpartners. The most important experimental consequences of the
conservation of $R$ are that super partners should be produced in pairs and the
lightest superpartner (LSP) should be stable. The LSP interacts weakly, therefore
the prominent signature of SUSY in case of $R$ parity conservation is a considerable
amount of missing (transverse) energy ($E_T^{miss}$).

    Although no violation of $B$ or $L$ has been observed
yet, there is no firm theoretical argument which would require exact conservation
of them and that of the $R$ parity. In fact the following term in the superpotential

\begin{equation}
W_{\rlap/R} = \lambda_{ijk}L_iL_jE_k^c + \lambda_{ijk}'Q_iL_jD_k^c + 
              \lambda_{ijk}''U^c_iD^c_jD_k^c 
\label{eq:Lag}
\end{equation}
which violates explicitely $B$, $L$ and $R$ parity, cannot be ruled out experimentally.
Here $L$ and $E$ are isodublet and isosinglet lepton, $Q$ and $D$ are isodublet and
isosinglet quark superfields, the indices $i$, $j$ and $k$ run for the three lepton
and quark families. The suffix $c$ denotes charge conjugate.
The first two terms violate explicitely $L$ whilst the last
one violates $B$. Present limits on the proton lifetime suggests that either the
$L$ or the $B$ violating terms (i.e. the corresponding $ \lambda_{ijk}$ couplings)
should vanish for the first family. Other experimental limits e.g. on lepton number violation:
double $\beta$ decay, or on $N-\bar N$ oscillation, etc. indicate that the couplings 
in Equ.~(\ref{eq:Lag}) shouldn't be expected to exceed a few percent, and usually are 
much smaller than the gauge couplings.
Even so, if $R$ parity is violated the topology of the expected SUSY signal changes
substantially. Since the LSP is no more stable, the missing energy is
considerably reduced. On the other hand the decay products of the LSP increase
the average number of jets and/or leptons in an event. In general, the event topology
depends crucially on the size of the couplings. If, e.g. the couplings are
of the order of $\sim 10^{-2}$ or larger, the mass spectra, branching ratios, etc.
will be different in the two cases where $R$ is conserved or violated.
If however the couplings are smaller than the above value, the
dominant effect of the $R$ parity violation is that the LSP becomes unstable.
An estimation of the LSP lifetime as a function of the couplings show~\cite{lifetime}
that we can distinguish four subcases giving rise to different detection
strategies: \\
\indent \indent \indent $(i)$ $10^{-4} \leq \lambda \leq 10^{-2}$ \\
\indent \indent \indent $(ii)$ $10^{-6} \leq \lambda \leq 10^{-4}$ \\
\indent \indent \indent $(iii)$ $10^{-9} \leq \lambda \leq 10^{-6}$ \\
\indent \indent \indent $(iv)$  $\lambda \leq 10^{-9}$ \\
In case $(i)$ only the event topology changes w.r.t. the case of $R$ conservation.
In case $(ii)$ one can observe a displaced vertex at the LHC energies.
In case $(iii)$ the LSP decays outside of a typical LHC detector, however
it can be catched by special purpose detectors~\cite{LSPdecay}. Finally
the case $(iv)$ cannot be distinguished experimentally from the case of $R$ parity
conservation.

In this study we have deliberately chosen to study case $(i)$ and compare
the result with the case of $R$ parity conservation, because it represents a more 
difficult experimental situation than case $(ii)$ where a displaced vertex could 
disentangle the LSP from the rest of the event. Moreover we have assumed that 
$\lambda_{ijk}' = \lambda_{ijk}'' = 0$ and only
one of the $\lambda_{ijk}$ coupling is different from zero in Equ.~(\ref{eq:Lag}). 
Nonzero $\lambda_{ijk}'$ and
$\lambda_{ijk}''$ are subject of other reports inside ATLAS~\cite{Jesper}.
The hierarchical structure observed for the Yukawa couplings in the
SM motivates our hypothesis above. The Lagrangian corresponding to
the superpotential of Equ.~(\ref{eq:Lag}) can be written
in terms of particle fields for our case as:

\begin{equation}
L^{\lambda}_{\not L} = \frac{1}{2}\lambda_{ijk}
(\overline{\nu}^{c}_{L_{i}} e_{L_{j}} \tilde e^{*}_{R_{k}} - 
e_{L_{i}} \overline{\nu}^{c}_{L_{j}} \tilde e^{*}_{R_{k}} + 
\nu_{L_{i}} \tilde e_{L_{j}} \overline{e}_{R_{k}} -
e_{L_{i}} \tilde \nu_{L_{j}} \overline{e}_{R_{k}} +
\tilde \nu_{L_{i}} e_{L_{j}} \overline{e}_{R_{k}} -
\tilde e_{L_{i}} \nu_{L_{j}} \overline{e}_{R_{k}}) + HC
\label{eq:Lagl}
\end{equation}
where $\nu_{L}$ and $e_{L,R}$ are lepton fields, the tilde
denotes the field of the superpartner, the $c$ stands for charge conjugation,
$*$ for complex conjugation  and $i,j,k$ are the flavor indices.
As stated before, if the $\lambda$ couplings are smaller than $10^{-2}$, 
which is our case, the sparticle mass spectrum
practically doesn't change and the main consequence of the $R$ parity violation is
the decay of the LSP. This process is depicted in Fig.\ref{LSPdecay1}
where we assume that the LSP is the lightest neutralino ($\tilde \chi^0_1$).
The decay proceeds through an $R$ conserving and an $R$ violating vertex
and in the final state there are always three leptons out of them at least
two of {\it different}
flavours, one neutral and the other two of opposite charges, since the
LSP is supposed to be neutral. 

The prominent signature of this type of SUSY event is the spectacular
increase of "stable" leptons: electrons and muons in the final state.
The neutral leptons, neutrinos, give rise to some missing transverse
energy, but its magnitude is much less than that if $R$ parity is conserved.
The flavour of the lepton in the final state depends on the values
of the indices $i,j,k$. Since $\lambda_{ijk}$ is antisymmetric in $i$ and
$j$ there are only 9 independent couplings which we choose as:
$\lambda_{121},\lambda_{122},\lambda_{123},\lambda_{131},\lambda_{132},
\lambda_{133},\lambda_{231},\lambda_{232},\lambda_{233}$. The first two
families, 1 and 2 give rise always to "stable" leptons, electrons and
muons, (and the corresponding neutrinos) in the final state. If
an index 3 appears, the lepton is not stable if it is a $\tau$,
and its decay products are most of the time different from electrons 
or muons. Since $e$ in Equ.~(\ref{eq:Lag}) is an isosinglet, $e_3$
is a $\tau$. The number of the stable leptons is the most
prominent for $\lambda_{121}$ and $\lambda_{122}$. It is less spectacular,
if an index 3 appears at the second place, and even less if the 3 appears
in the third place. Finally, if two indices have value of 3 one has the least
number of stable leptons. On the other hand, the number of the
neutrinos, and with that the magnitude of the missing energy 
increases in the order of the above mentioned cases. 
The expected extra number of stable charged leptons
and neutrinos per each LSP decay for the different $\lambda$ couplings calculated
by the program \cite{isajetR}  are given in
Table~\ref{tb:BRs}. It is clear that the average number
of the leptons for different flavours gives a strong hint
on the coupling which is realized. E.g. the coupling $\lambda_{122}$
gives rise predominantly to muons whilst the coupling $\lambda_{123}$
results in equal number of electrons and muons, etc.

\vskip -0.5 cm
\begin{Fighere}
\centering
  \epsfig{file=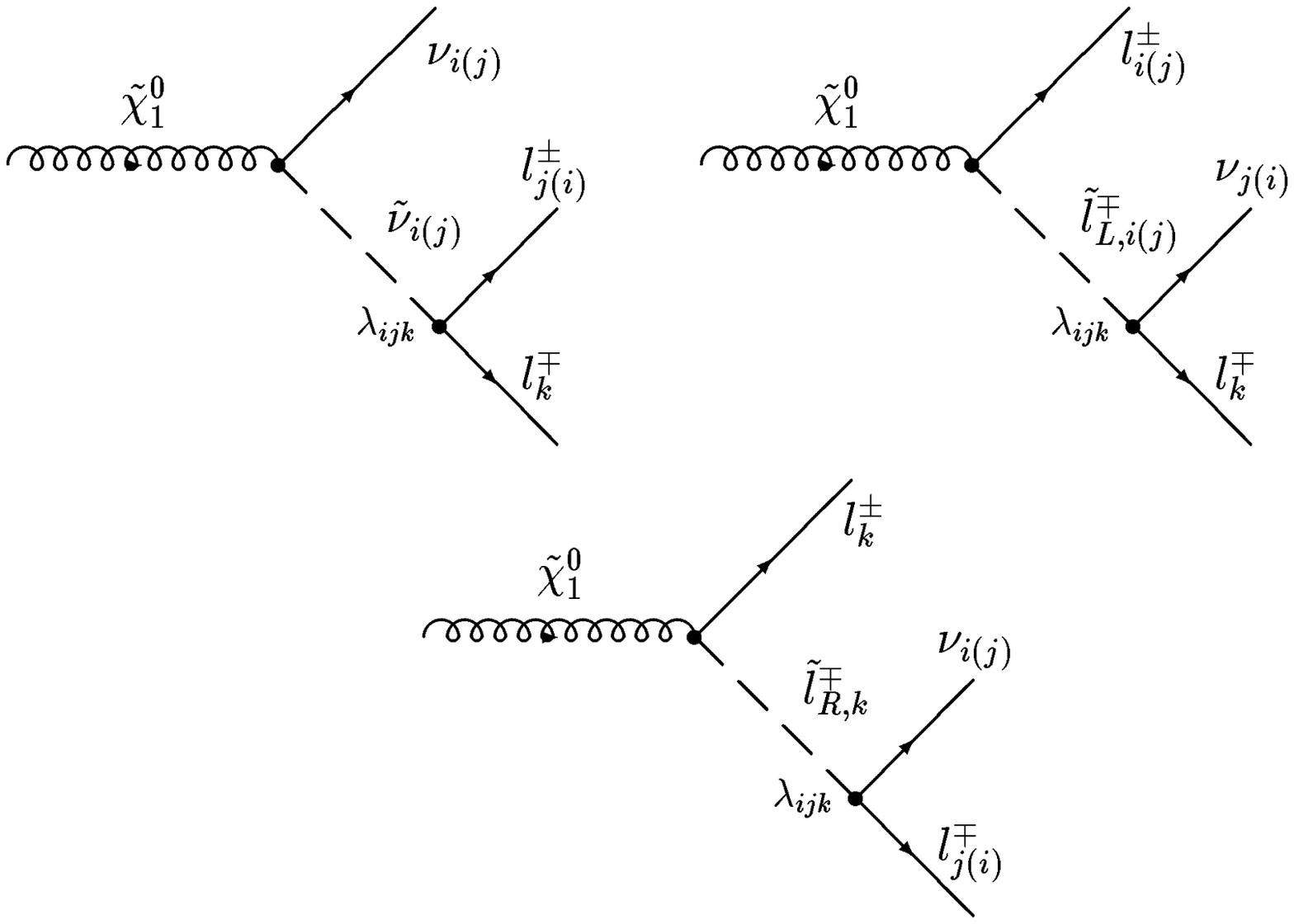,width=14cm  }
  \vskip -10.0 cm
\caption{\small The $\rlap/R$ decays of $\chi^{0}_{1}$ (assumed as $LSP$) through
$\lambda$ type couplings.}
\label{LSPdecay1}
\end{Fighere}
\vskip0.5cm

\vspace{-0.5cm}
\begin{Tabhere}
\centering
\caption{\small The number of produced stable leptons in the LSP decay
if it is the $\tilde \chi^0_1$. We have assumed $\sim$ 18\% branching
ratio for the semileptonic decay of the $\tau$.}
\vskip 0.5cm
\begin{tabular}{|c|c|c|c|c|} \hline 
$\lambda_{ijk}$ & Decay channel                                                   & $<N_e>$ & $<N_{\mu}>$ & $<N_{\nu}>$  \\ \hline 
 121            & $e^{\pm}\nu_{\mu}e^{\mp}$;$\nu_e\mu^{\pm}e^{\mp}$               &  1.5    &  0.5        &  1           \\ \hline
 122            & $e^{\pm}\nu_{\mu}\mu^{\mp}$;$\nu_e\mu^{\pm}\mu^{\mp}$           &  0.5    &  1.5        &  1           \\ \hline
 123            & $e^{\pm}\nu_{\mu}\tau^{\mp}$;$\nu_e\mu^{\pm}\tau^{\mp}$         &  0.68   &  0.68       &  2.36        \\ \hline
 131            & $e^{\pm}\nu_{\tau}e^{\mp}$;$\nu_e\tau^{\pm}e^{\mp}$             &  1.59   &  0.09      &  1.68        \\ \hline
 132            & $e^{\pm}\nu_{\tau}\mu^{\mp}$;$\nu_e\tau^{\pm}\mu^{\mp}$         &  0.59   &  1.09       &  1.68        \\ \hline
 133            & $e^{\pm}\nu_{\tau}\tau^{\mp}$;$\nu_e\tau^{\pm}\tau^{\mp}$       &  0.77   &  0.27       &  3.04        \\ \hline
 231            & $\mu^{\pm}\nu_{\tau}e^{\mp}$;$\nu_{\mu}\tau^{\pm}e^{\mp}$       &  1.09   &  0.59       &  1.68        \\ \hline
 232            & $\mu^{\pm}\nu_{\tau}\mu^{\mp}$;$\nu_{\mu}\tau^{\pm}\mu^{\mp}$   &  0.09   &  1.59       &  1.68        \\ \hline
 233            & $\mu^{\pm}\nu_{\tau}\tau^{\mp}$;$\nu_{\mu}\tau^{\pm}\tau^{\mp}$ &  0.27   &  0.77       &  3.04        \\ \hline
\end{tabular}
\label{tb:BRs}
\end{Tabhere}
\vskip 0.5cm
 
The presence or absence of $\tau$'s influence how easily the
SUSY particles can be reconstructed. On the one hand the signature
is weakened by the taus since they produce less stable charged
leptons, and these, partly originating from the $\tau$ decay, 
do not produce a sharp endpoint for the $\tilde\chi_1^0$ mass. On
the other hand, taus produce more missing transverse energy,
moreover the smaller number of stable charged leptons produced
decreases the combinatorial background in the mass reconstruction. 

\subsection{The SUGRA Model}

    The present study of $R$ parity violation is carried out
in the framework of the SUGRA model~\cite{SUGRA_rev},\cite{SUGRA_pap}.
Contrary to the minimal version of supersymetric models (MSSM)
which has a very large number of unknown parameters, the SUGRA model
is characterized only by 5 parameters which are the following: \\
\indent $(1)$ $m_0$, an universal scalar mass, \\
\indent $(2)$ $m_{1/2}$, an universal gaugino mass, \\
\indent $(3)$ $A_{0}$ a common trilinear interaction term, \\
\indent $(4)$ tan($\beta$), the ratio of the vacuum expectation values
of the two Higgs fields,\\
\indent $(5)$ the sign of $\mu$ of the Higgsino mass parameter.\\
The mass spectrum of the SYSY partners at the 
electro-weak scale as well as their decay branching ratios
are obtained from the above parameters
by solving the renormalization group equations (RGE). This is
performed in our case by the program of ISAJET~\cite{isajet}.

   The SUGRA model predicts a hierarchical structure of the masses
of the SUSY particles. 
The masses of the first two families of the squarks and of 
the sleptons are driven essentially by 
$m_{0}$ and $m_{1/2}$ through an approximate relation :

\begin{equation}
m^{2}_{\tilde f_{L,R}} = m^{2}_{0}+m^{2}_{f}+c(\tilde f_{L,R}) \cdot m^{2}_{1/2}+
D(\tilde f_{L,R})
\label{eq:squarkmasses}
\end{equation} 
where $c(\tilde f)$ are some numerical factors of order $5.5 \div 6.0$ for the 
squarks and of order
$0.15 \div 0.5$ for the sleptons, $D(\tilde f_{L,R})$ are the so-called D-terms (in general less
important). For the third family of the squarks and the sleptons 
there is a mixing due to the
corresponding Yukawa couplings, pushing down the mass of one of the sfermions 
and the other one in the opposite
direction. In some regions of the parameter space this could have the effect that the lighter
stau ($\tilde \tau_{1}$) could become the LSP but such scenarios are ruled out by cosmological
considerations. The gauginos $U(1),SU(2)$ and $SU(3)$ are driven mainly by $m_{1/2}$. 
Because the $U(1)$ and $SU(2)$ gauginos will mix with higgsinos to obtain the mass states 
($\tilde \chi^{0}_{i}$, $i=1,4$,
and $\tilde \chi^{\pm}_{j}$, $j=1,2$), part of that spectrum will depend on the $\mu$ parameter 
(in SUGRA $\mu$ is determined by the condition of electro-weak symmetry breaking and typically
$\mu \gg m_{1/2}$). 
We have the following approximate relations:

\vskip 1.0 cm
\begin{equation}
\centering
\begin{tabular}{c} 
$m_{\tilde g} \approx 2.4 \cdot m_{1/2}$  \\
$m_{\tilde \chi^{0}_{1}} \approx 0.4 \cdot m_{1/2}$  \\
$m_{\tilde \chi^{0}_{2}} \approx m_{\tilde \chi^{\pm}_{1}} \approx 0.8 \cdot m_{1/2}$  \\
$m_{\tilde \chi^{0}_{3}} \approx m_{\tilde \chi^{0}_{4}} \approx m_{\tilde \chi^{\pm}_{2}} \approx |\mu|$  \\
\end{tabular}
\label{eq:gauginomasses}
\end{equation}
\vskip 1.0 cm

The Higgs sector is composed in SUGRA by five mass states ($h^{0}$, 
$A^{0}$, $H^{0}$ and $H^{\pm}$).
In the case of the lightest one, $m_{h^0} \leq 130 \div 150$ GeV 
in about all SUGRA parameter space. 
The masses of the other Higgses are in general 
very heavy and depend on tan$( \beta )$.

The gaugino-higgsino mixing depends more strongly on the parameter
values, and this in turn determines the decay branching ratios.

\newpage

\section{The analysis chain}

\subsection{The event generators}

\indent \indent {\it \bf Simulation of the signal events}
\vskip 0.2cm

       Our basic program tool is ISAJET\footnote{We have used version 7.30}
\cite{isajet} which simulates 
for hadron colliders the production and decay of the supersymetric
particles, as well as the underlying event, i.e. the accompanying
partons and their hadronization.
    
    ISAJET does not include $R$ parity violation
in the SUGRA model. It has been introduced in detail in another
event generator, SUSYGEN \cite{SUSYGEN}, written for $e^+e^-$
collisions. 
Therefore, inspired by SUSYGEN,
a set of routines computing in detail all the $\rlap/R$ ($\lambda$-type) decays
of gauginos ($\tilde \chi^{0}_{i}$, $i=1,4$ and $\tilde \chi^{\pm}_{j}$, $j=1,2$) and
sleptons was written for ISAJET~\cite{isajetR}. This program is able to simulate production
of supersymetric particles in a hadronic collision with $R$ parity broken, if this latter
is accompanied by $L$ number violation.
To obtain correct predictions, the hypothesis of small values of the $\lambda$ couplings 
($\lambda \leq 10^{-2}$) must be used, which is our present case, 
in order to neglect any correction in the sparticle 
mass spectrum brought by RGE's. 
The dominant effect of the $R$ parity violation is the decay of the LSP.

   We have generated several million signal events. In Table~\ref{tb:sigstat}
one can see their repartition in the SUGRA parameter space and for the
type of the couplings. In the last column we quote the integrated luminosity
in terms of LHC months the generated events correspond to where we have
taken an LHC year equivalent to $10^4$ /pb (low lumonosity run with
appr. 1/3 of efficiency).

\vspace{-0.5cm}
\begin{Tabhere}
\centering
\caption{\small  The number of generated signal events }
\vskip 0.5cm
\begin{tabular}{|l|c|c|c|c|c|r|c|} \hline 
$m_0$ [GeV] & $m_{1/2}$ [GeV] &  $A_0$ & tan($\beta$) & sgn($\mu$) & coupling        & Events     & LHC month  \\ \hline 
 0 - 1500 & 0 - 1500      &  0     &  2           &  +1        & $\lambda_{123}$ &   256 000  &   ---      \\ \hline
 0 - 500  & 0 -  500      &  0     &  2           &  +1        & $\lambda_{123}$ &   300 000  &   ---      \\ \hline
 400      & 400           &  0     &  2           &  +1        & $\lambda_{122}$ &   400 000  & $\sim 36$  \\ \cline{6-8}
          &               &        &              &            & $\lambda_{123}$ &   400 000  & $\sim 36$  \\ \hline
 200      & 100           &  0     &  2           &  -1        & $\lambda_{123}$ & 1 000 000  & $\sim 1.5$ \\ \cline{6-8}
          &               &        &              &            & $\lambda_{122}$ & 1 000 000  & $\sim 1.5$ \\ \hline
 100      & 300           &  300   &  2.1         &  +1        & $\lambda_{123}$ &   450 000  & $\sim 36$  \\ \cline{6-8}
          &               &        &              &            & $\lambda_{122}$ &   450 000  & $\sim 36$  \\ \hline
\end{tabular}
\label{tb:sigstat}
\end{Tabhere}
\vskip 0.5cm
 
The samples in the first two raws were used to study inclusive reactions. The rest of the statistics is devoted
to the 1st, 3rd and 5th of the so called LHC points, where the determination of the SUGRA parameters
has been performed.

\vskip 0.2cm
{\it \bf Simulation of the background events}
\vskip 0.2cm

As mentioned, the signature of the signal events is the appearence of a large number
of high $p_t$ leptons accompanied by high $p_t$ jets and by a moderate amount of
missing transverse energy. Such event topology is also produced, albeit in a much
reduced level, by the decay of heavy SM particles and these events constitute the
main background. We have studied such processes using the ISAJET and PYTHIA~\cite{pythia}
event generators including initial and final state radiation. 
The background event statistics is listed in Table\ref{tb:bgstat}.

\vspace{-0.5cm}
\begin{Tabhere}
\centering
\caption{\small  The number of generated background events }
\vskip 0.5cm
\begin{tabular}{|l|c|} \hline 
 Reaction               & Number of events  \\ \hline 
 $t-\bar t$             & 1 200 000         \\ \hline
 $W-Z$, $W-W$ and $Z-Z$ &   600 000         \\ \hline
 $Z-b\bar b$            &   600 000         \\ \hline
 Drell-Yan              &   600 000         \\ \hline
\end{tabular}
\label{tb:bgstat}
\end{Tabhere}
\vskip 0.5cm

\subsection{Fast simulation of the ATLAS detector}

Due to the large number of events to be generated the detector response could not
have been simulated in detail, using e.g. GEANT~\cite{geant}. Instead a fast,
so called particle level Monte Carlo program, ATLFAST~\cite{atlfast} 
has been used\footnote{We have used the version 1.57}. 
In this program the ATLAS detector~\cite{ATLAS} is described by a simplified geometry 
and apart of the acceptance the detector response is parametrized. Below we repeat
the main features of these description. 

\vskip 0.2cm
{\it \bf Description of ATLAS by ATLFAST}
\vskip 0.2cm

The detector geometry is given in the variables of the pseudorapidity 
$\eta = - \mbox{ln tan}(\theta /2)$ and azymuthal angle $\phi$, where
$\theta$ is the solid angle of the particle produced in the interaction
point. The granularity in $\eta \times \phi$ is 0.1 $\times$ 0.1 for
$|\eta | < 3$ and 0.2 $\times$ 0.2 otherwise upto $|\eta | = 5$.
The produced particles except the muons and the neutrinos 
deposite their energies, smeared by a resolution function, 
and are integrated in individual $\eta$-$\phi$ cells.
The effect of the 2T solenoidal magnetic field on the deposited energy
of the charged particles is parametrized. The effect of cracks
in the calorimeter and in the trackers are taken into account by
the parametrized acceptance function. 

\vskip 0.2cm
{\it \bf Reconstruction algorithms}
\vskip 0.2cm

First clusters are created from the cell energies using a simple algorithm.
Next, isolated photons and electrons are reconstructed if 
the simulated particle falls in the acceptance region
(typically $|\eta | < 2.5$), its energy,
smeared by a parametrized resolution function,
matches that of a cluster, and in the case of electrons, the $\eta$ and $\phi$
values of the cluster and the electron are the same within the resolution. 
The energy of the reconstructed particles is that of the simulated one
smeared by a parametrized resolution, whereas their direction stays the same.

After having removed the clusters of the isolated photons and electrons
the reconstruction of the parton jets is carried out
by a simple fixed cone algorithm. Isolation criteria of jets
or charged tracks can be defined as a function of the deposited energy in a 
$\Delta R = \sqrt{(\Delta\eta)^2+(\Delta\phi)^2}$ cone.
The reconstructed jet energy in general doesn't match with the
true energy of the corresponding parton.
Therefore a correction factor has to be applied in a later phase
of the analysis. This correction factor, depending on the
cone size and on the momentum
of the jet, is established by comparing the reconstructed jet transverse momentum
with that of the original parton:  $R_{calib}^{b jet}= p_{t}^{b parton}/p_{t}^{b jet}$. 
It has been demonstrated that applying such factors
one can correctly reconstruct the position of the mass peak
of the Higgs boson~\cite{atlfast}. This correction factor depends
on the jet type and also if the jet contains prompt leptons. 
In Fig.\ref{p5selbjets} the correction is shown for the 
three different LHC points analysed. The observed difference
in point 3 and the other two points is due to the facts that :\\
\indent $(i)$ the production mechanisms of the $b$'s are different. In point
1 and 5 they originate mainly from the decay $h^0\longrightarrow b\bar b$, 
whereas in point 3 the production of the $b$'s is less correlated;\\
\indent $(ii)$ in point 3 there are more leptons produced and therefore
the probability is higher that a lepton is inside the jet-cone.\\
One has to stress that the dispersion of the correction factor
can be rather large, reaching even 25\%, especially at low $p_t$ ($\le $ 100 GeV). 
Fig.\ref{p5res} shows the invariant mass of the $b \bar b$ pair from the $h^0$ mass 
at the LHC points 1 and 5 after having applied the correction.

\vskip -0.5 cm 
\begin{Fighere}
\centering
  \mbox{\epsfig{file=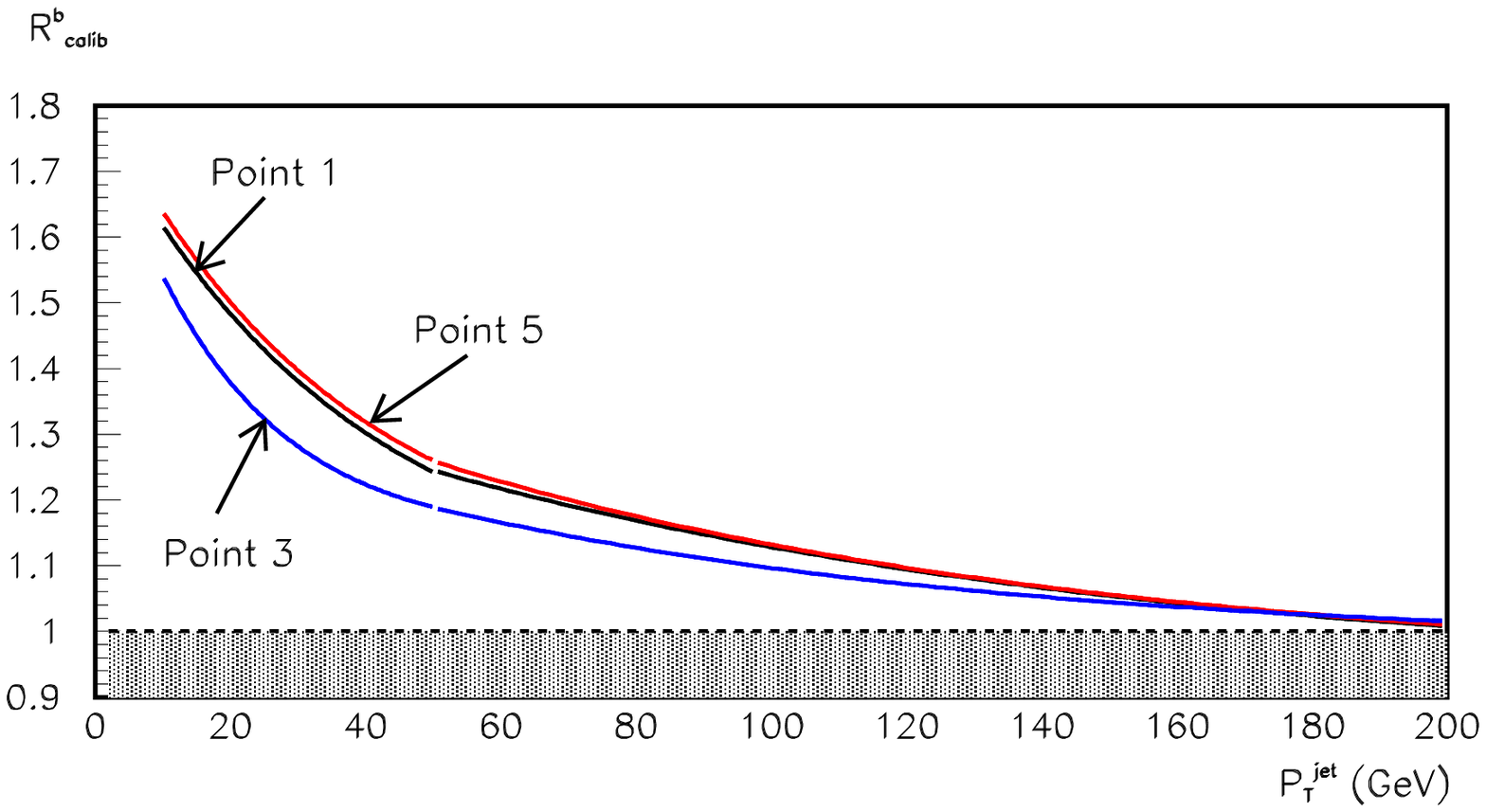,height=10cm,width=14cm  }}
\vskip -5.cm
\caption{\small  Calibration functions of b jets for SUGRA points 1, 3 and 5. }
\label{p5selbjets}
\end{Fighere}

\vskip -0.5cm
\begin{Fighere}
\centering
   \mbox{\epsfig{file=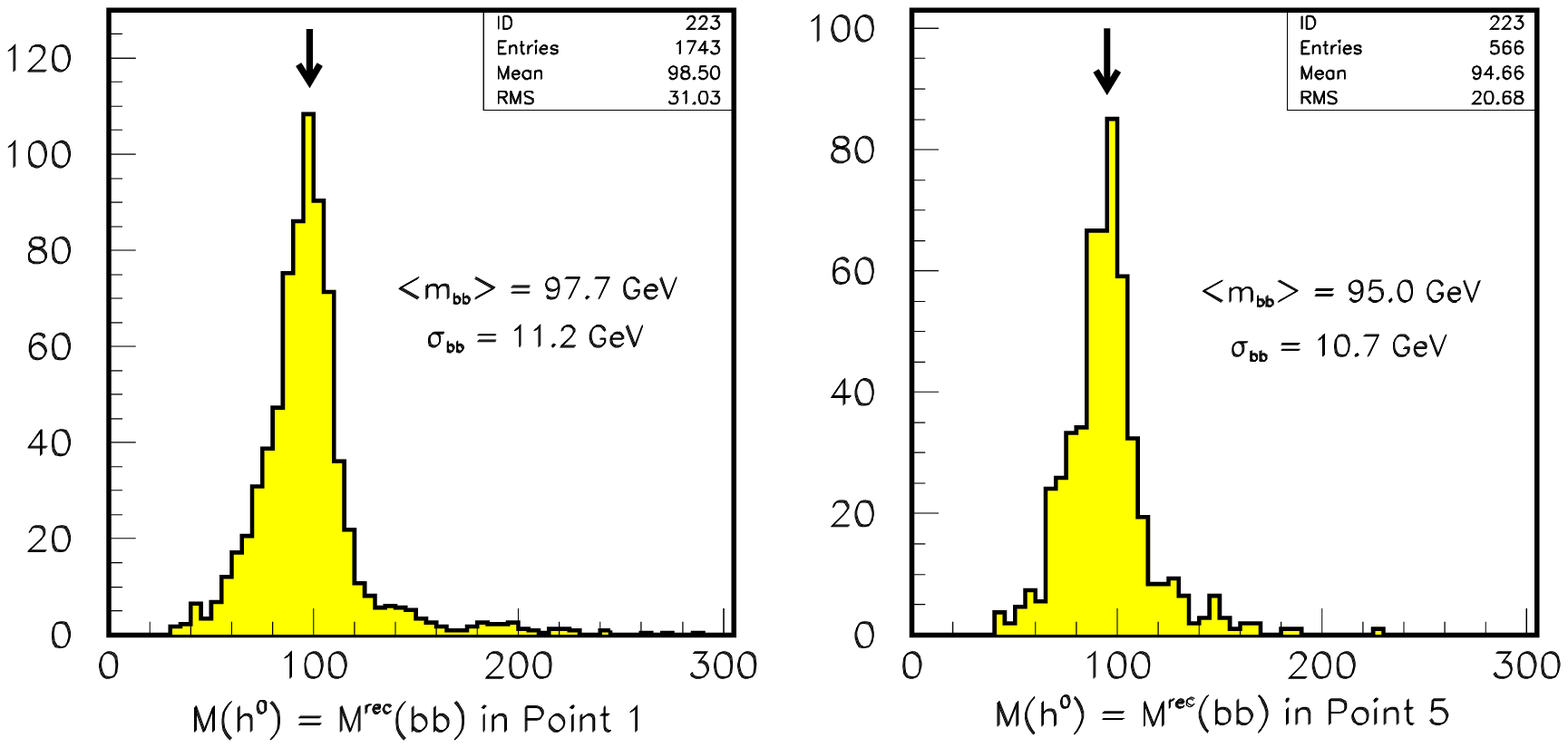,width=11cm  }}
\vskip -5.0cm
\caption{\small   
The effect of the resolution and calibration of b jets on the invariant mass distributions in
SUGRA points 1 (left) and 5 (right)  and for the $\rlap/R$ couplings $\lambda_{122}=10^{-3}$:
$M(h^{0}) \sim M(bb)$ - for reconstructed b jets originating from $h^0$'s.
The ISR, FSR and hadronization are switched on. The arrows point to the theoretical values of
$m_{h^0}$ in each SUGRA point.}
\label{p5res}
\end{Fighere}
\vskip 0.5 cm

Muons are reconstructed if they are in the acceptance region
(typically $|\eta | < 2.5$) and their energies and directions
are obtained from the simulated one by a parametrized smearing
function of ATLFAST.
In most cases these functions were determined or
cross checked by detailed simulation using GEANT. Finally,
the missing transverse energy is calculated.

\subsection{Detection and identification efficiencies}

The output of ATLFAST is written in a coloumn-wise ntuple (CWN)
for further analysis using PAW~\cite{paw}. This ntuple contains
all the reconstructed electron, muon and jet objects together
with the two transverse components of the missing energy.
In addition, we have stored among the originally 
simulated particles and partons by RPV\_ISAJET those which were produced in
the decay chain of the generated SUSY particles. We have installed
a bidirectional pointer between the reconstructed objects 
and the original particles/partons, and we have reinstalled the mother-daughter
relationships between the stored original particles/partons.
A special record containing the integrated luminosity has been also included.

As a first step in the analysis we have randomly rejected reconstructed
electrons and muons using a tabulated detection (including identification)
efficiency. The efficiencies, as a function of the transverse
momentum and $\eta$, have been extracted from
a version of ATLFAST (2.0) which was released after the bulk of
our event simulation has been carried out (see Fig.\ref{effic}). 

\vskip -0.5cm
\begin{Fighere}
\centering
  \mbox{\epsfig{file=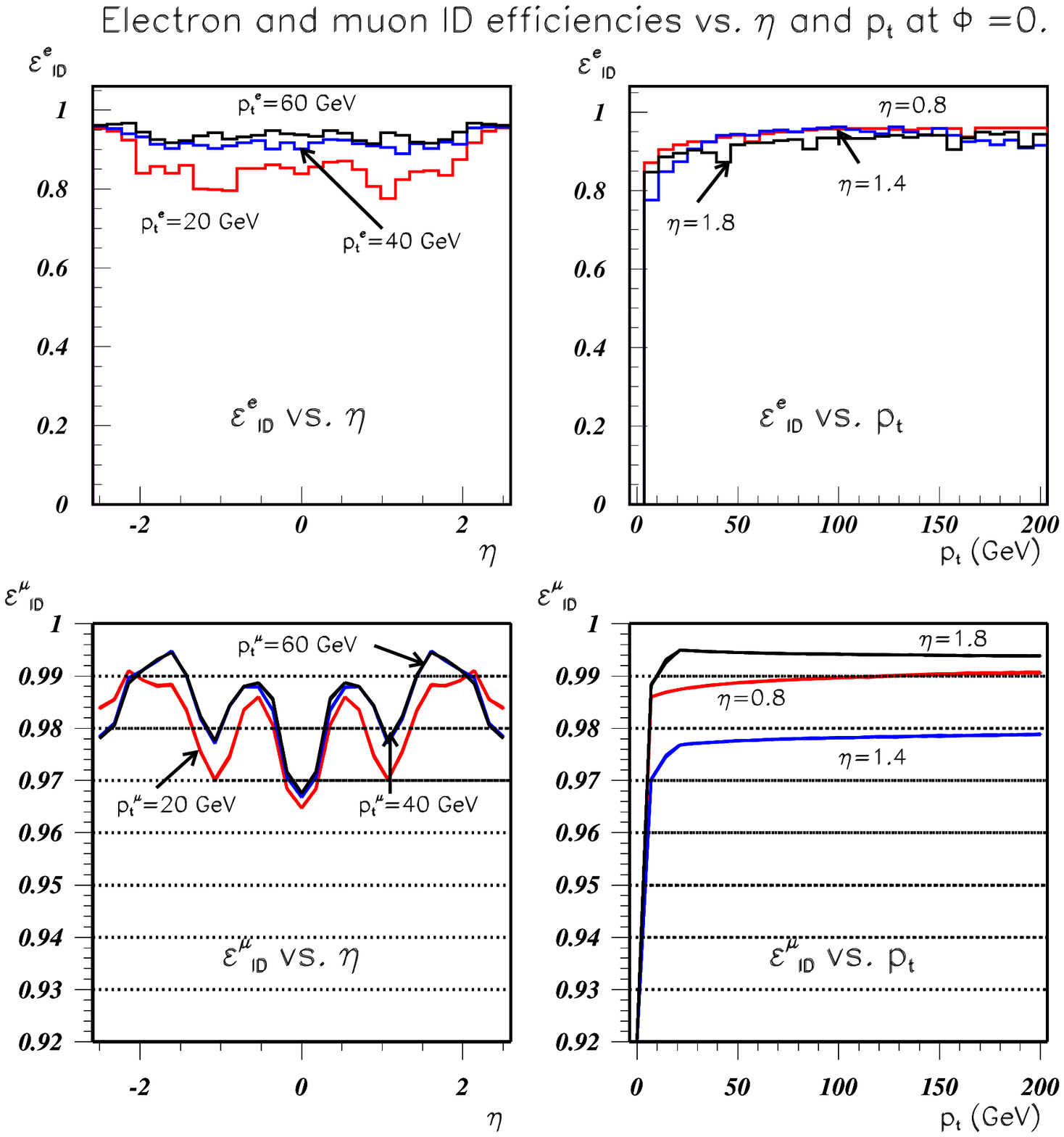,width=14cm }}
\caption{\small Detector efficiencies for muons and electrons
as a function of $p_t$ and $\eta$. The dependence 
on $\varphi$ is practically negligible.}
\label{effic}
\end{Fighere}
\vskip 0.5 cm

We have also randomly
reclassified tagged b-jets to light jets with a b-tagging efficiency of 60\%
and vice versa using the misstaging probability of $\sim$2\%. These numbers
have been obtained in a separate study by detailed Monte Carlo
simulation~\cite{btagging}.

The obtained final state reconstructed particles were submitted
to selection cuts in order to find an optimum signal of SUSY
over the background. These criteria as well as the reconstruction
algorithms are described in the forthcoming sections.


\section{Inclusive measurements}

    By measuring global variables, like e.g. the number
of leptons of a given flavour or the average $p_t$ of a lepton
in an event, or simply the number of events passing some
selection criteria, we would like to answer the following
three questions: \\
$(i)$ what is the maximum domain in the SUGRA parameter space
in which ATLAS is sensitive for $R$ parity violation; \\
$(ii)$ can we determine the approximate
energy scale of a SUGRA signal; \\
$(iii)$ can we determine the dominant type of the different couplings
which causes the signal.
For all these studies as well as for the exclusive measurements
we have fixed the values of the $\lambda_{ijk}$ couplings to $10^{-3}$.
We remind the reader that the event topology does not depend on the
particular value of $\lambda$ if it is between $10^{-2}$ and $10^{-4}$.


\vskip 0.2cm
{\it \bf Sensitivity of ATLAS in the SUGRA space}
\vskip 0.2cm

      We have generated 1000 events in each of the 16$\times$16 equally
spaced points in the $m_{1/2}$ vs $m_0$ plane the other parameters
being fixed (see Table\ref{tb:sigstat}). We have generated the
SM background events as given in Table\ref{tb:bgstat}. Since the
signature of the SUGRA event is characterized by multileptons and
missing energy, we have tried several event selection based on
the variation of the number of electrons and muons, their transverse
energy and the value of the missing energy. We have used the
quantity called significance:

\begin{equation}
S = N_{sig}/\sqrt{N_{bg}}
\label{eq:significance}
\end{equation}
to delimite the sensitive region with $S \ge 5$, where $N_{sig}$
is the number of the accepted SUGRA events and $N_{bg}$
is the number of accepted SM events. The sensitive region
extending to the highest value of $m_0$ and $m_{1/2}$ has been
obtained using the selection criteria: \\
\indent \indent 1. $N_l \ge 3$ \\
\indent \indent 2. $E_t^{miss} \ge 100$ GeV \\
\indent \indent 3. $p_t^1 \ge 70$ GeV \\
\indent \indent 4. $p_t^l \ge 20$ GeV \\
where $N_l$ is the total number of $e$ and $\mu$,
$p_t^1$ is the highest transverse momentum of the leptons,
$p_t^l$ is the transverse momentum of the other leptons in the event.
The first two cuts reject almost all SM background coming from 
the decay of heavy SM particles.
The Standard Model processes generally does not fulfill at the same time both constraints.
$Z$ pair production (by Drell-Yan or in the $t,u$ channels) 
with $Z\rightarrow l^{+} + l^{-}$ give a higher multiplicity of leptons but with a 
very low $E_{t}^{miss}$ (one of the leptons falling out the detector acceptance could mime a 
$E_{t}^{miss}$ but will decrease the lepton multiplicity).
$W^{\pm}$ pair production (by Drell-Yan or in the $t,u$ channels) and the associated production of $W$'s with
$Z$'s give a higher $E_{t}^{miss}$ through the neutrinos but proportionally less leptons.
The $t \overline{t}$ production with $W^{\pm} \rightarrow \nu_l + l^{\pm}$ and the leptons arising from b jets
reconstructed as isolated gives a higher $E_{t}^{miss}$ but is generally supressed by the isolation criteria
of leptons. The associated production $Zjj$ is similar to the above mentioned cases.

\vskip -0.0cm
\begin{Fighere}
\centering
  \mbox{\epsfig{file=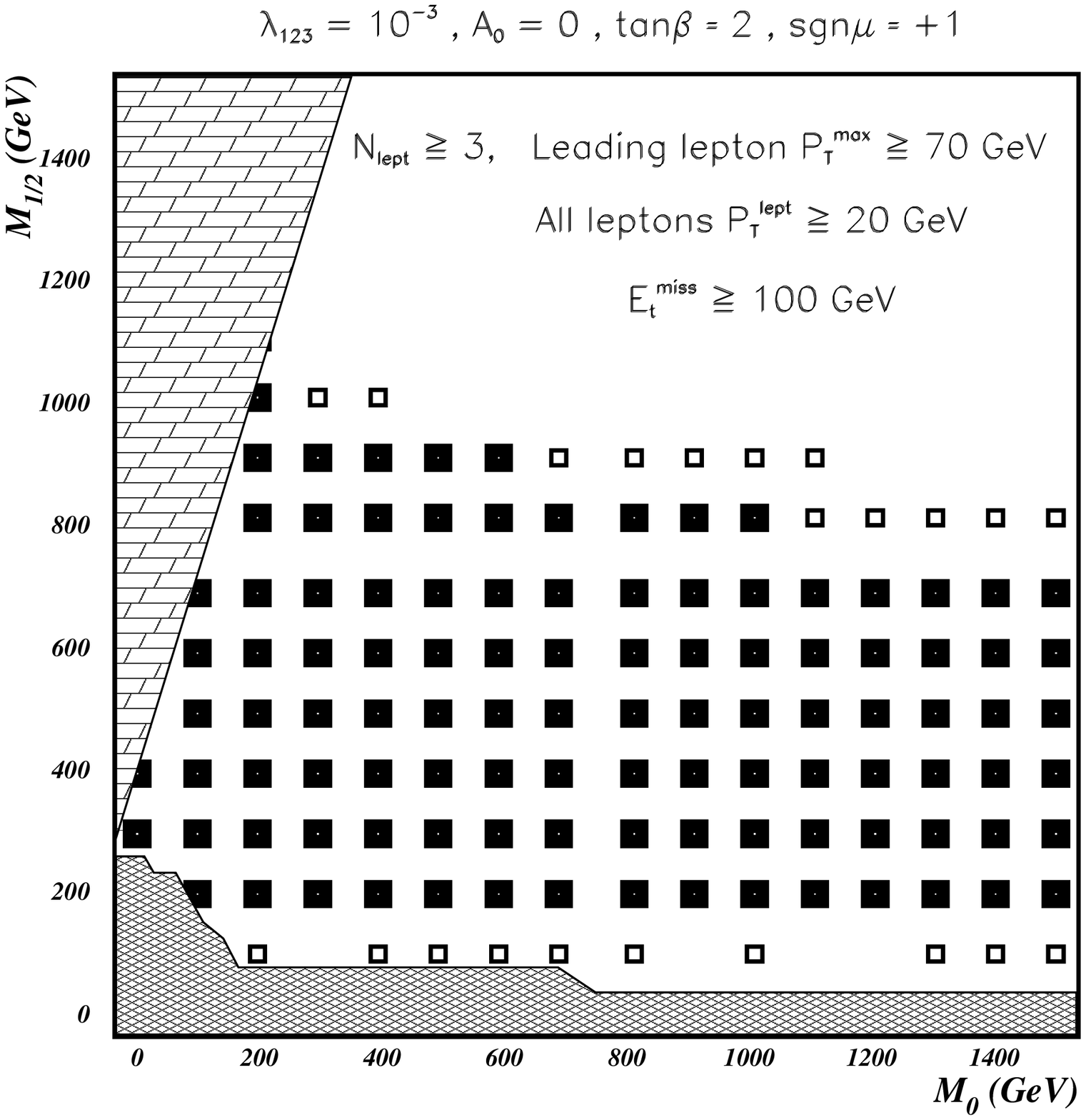,width=12.0cm  }}
\caption{\small Sensitive domain in the SUGRA parameter space.
For the explanation of the symbols see the text. The bricked domain is excluded by
theory and the cross hatched one is excluded by experimental measurements.}
\label{excl_reg}
\end{Fighere}
\vskip 0.5 cm

Fig.\ref{excl_reg} shows the domain of the sensitivity
one can obtain in 1 year of running with LHC at low luminosity.
Due to the limited number of simulated events (which is much less
than one can detect in one year) we have an uncertainty on this
region: full squares indicate the grid points where $S > 5$
at 99\% CL, no symbols at the grid points indicate
that $S < 5$ at 99\% CL, finally the open squares correspond to the
cases where one cannot make any of the two above statements.

\vskip 0.2cm
{\it \bf Energy scale of the SUGRA signal}
\vskip 0.2cm

     If a SUSY signal manifests itself, its energy scale can be 
determined in the case when $R$ parity is conserved from the
distribution of the quantity~\cite{SUGRAincl}:

\begin{equation}
M_{eff} = \sum_{i=1}^4 p_{t,i}^j + E_t^{miss}
\label{eq:meff}
\end{equation}
where $p_{t,i}^j$ are the 4 jets with the highest transverse momentum.
The missing transverse energy,
$E_t^{miss}$ originates mainly from the $\tilde\chi_1^0$ decay.
In the case of $R$ parity violation we should therefore
replace $E_t^{miss}$ by the decay products of the $\tilde\chi_1^0$, 
i.e. we have used instead of the definition (\ref{eq:meff})
the following one:

\begin{equation}
M_{eff} = \sum_{i=1}^4 p_{t,i}^j + \sum_{i=1}^4 p_{t,i}^l + E_t^{miss}
\label{eq:meffR}
\end{equation}
where $p_{t,i}^l$ are the 4 leptons with the highest transverse momentum.
The distributions of $M_{eff}$ for the LHC point No 1 and 5
can be seen in Fig.\ref{meff_p1p5}. The point No 1 (3rd line in 
Table \ref{tb:sigstat}) is associated with a high mass scale,
whereas the point No 5 (5th line in 
Table \ref{tb:sigstat}) corresponds to a medium mass scale.
This is well reflected in Fig.\ref{meff_p1p5}.

\begin{Fighere}
\centering
  \mbox{\epsfig{file=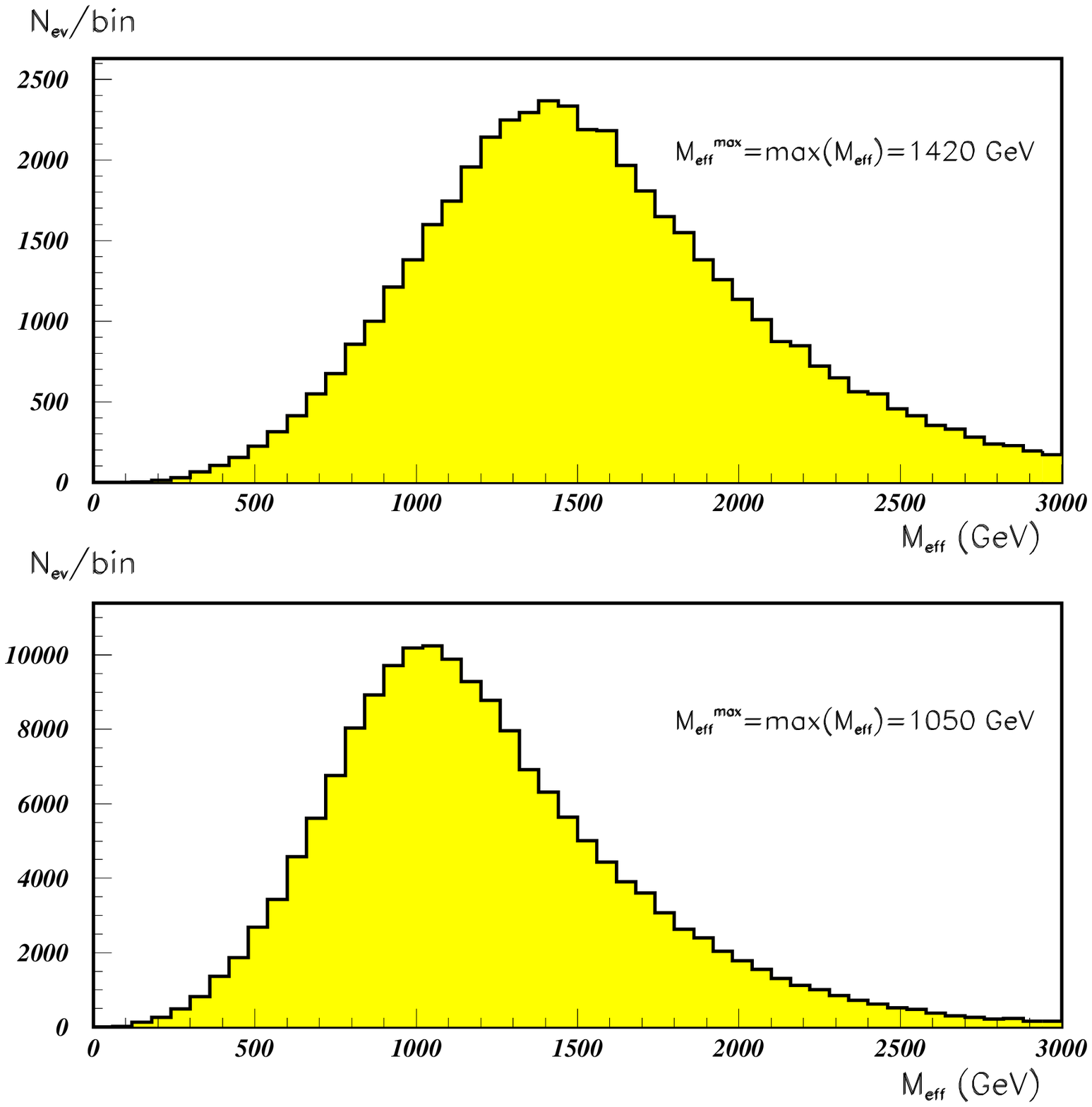,height=10cm,width=15cm  }}
\caption{\small Distribution of $M_{eff}$ for the LHC points
No 1 and 5 after 3 years of LHC run at low luminosity.
The inclusive cuts used here for both points are ($\lambda_{122}=10^{-3}$): 
$N_{leptons} \geq 4$ and $E_{t}^{miss} \geq 50 GeV$. 
The maximum in these distributions ($M_{eff}^{max}$) depends strongly on the mass
parameters of the models being a good observable for the mass scale.}
\label{meff_p1p5}
\end{Fighere}
\vskip 1cm

In order to see how the $M_{eff}$
distributions are correlated with the SUGRA mass scale,
we have chosen randomly 30 points in the region
of $0 \le m_0 \le 500$ GeV and  $0 \le m_{1/2} \le 500$ GeV. 
At each points we have generated 10 000 events (see
2nd line in Table \ref{tb:sigstat}) and determined
the maximum of the $M_{eff}$
distributions: $M_{eff}^{max}$. 
The events were selected requiring more than 3 leptons
and missing transverse energy higher than 50 GeV in an event.
In the first plot of Fig.\ref{meffmod}
we show the correlation of this quantity with the SUGRA mass scale, 
$M_{SUGRA}$ which we have defined as being the highest mass of the strongly
interacting SUSY partners :
\begin{equation}
M_{SUGRA} = \mbox{min} (m_{\tilde g}, m_{\tilde q_R}, m_{\tilde b_1}, m_{\tilde t_1})
\label{eq:mSUGRA}
\end{equation}
The observed strong correlation is even more pronounced in the 
second plot of Fig.\ref{meffmod}
where the distribution of the ratio of $M_{eff}^{max}/M_{SUGRA}$
is shown.

\vskip -1.cm
\begin{Fighere}
\centering
  \mbox{\epsfig{file=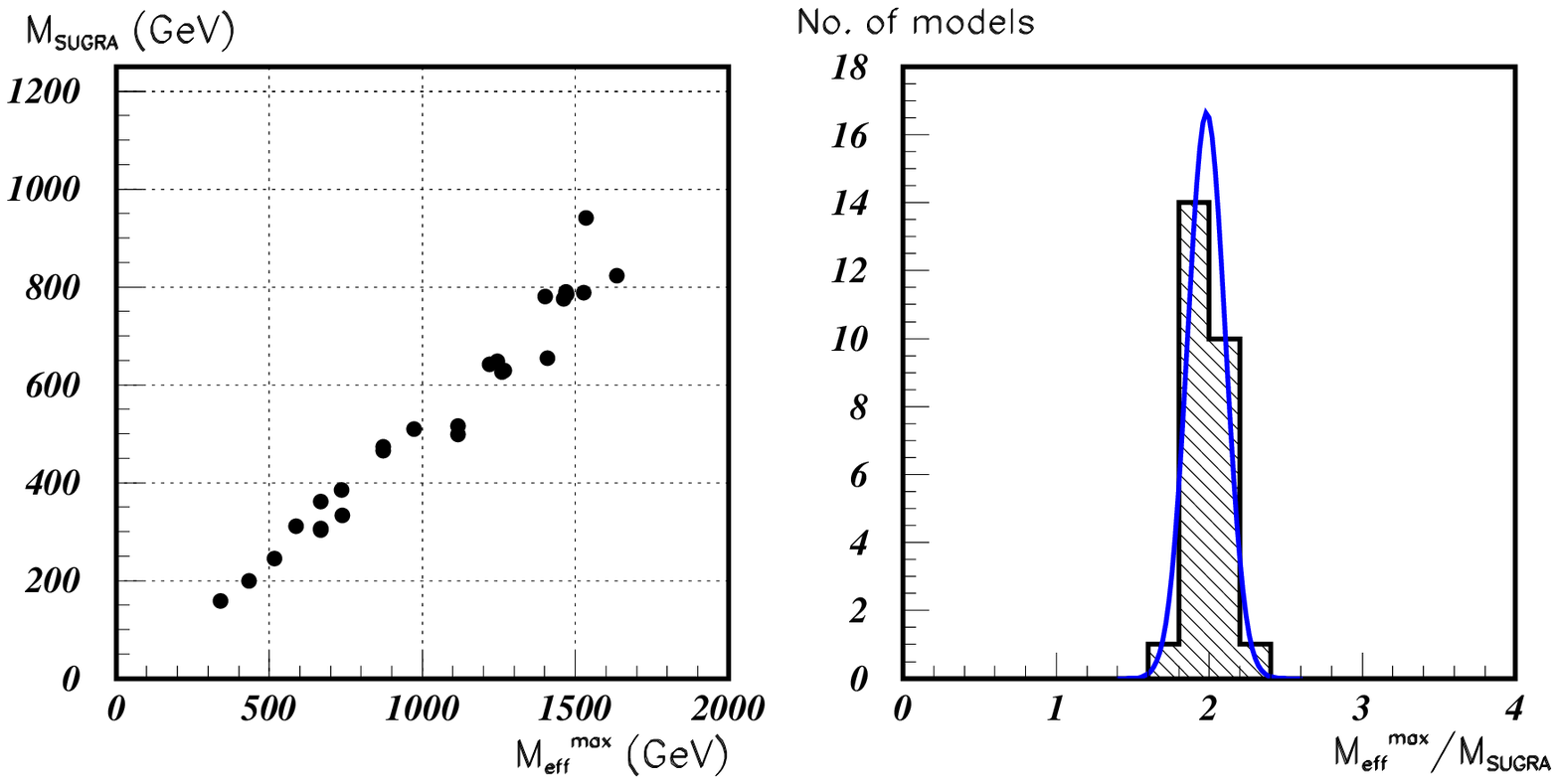,width=15cm  }}
  \vskip -7.0cm
\caption{\small Corellation between $M_{eff}^{max}$ and $M_{SUGRA}$ and
the distribution of the ratio $M_{eff}^{max}/M_{SUGRA}$ for 30 SUGRA points
(for other explanations see the text).}
\label{meffmod}
\end{Fighere}
\vskip 1cm

We have introduced another quantity to characterize the SUGRA energy scale
in the case of $R$ parity violation. Indeed, the algebraic sum of the
lepton transverse momentum divided by the number of leptons:

\begin{equation}
p_t^{l,norm} =  (\sum_{i=1}^4 p_{t,i}^l)/N^l
\label{eq:ptnorm}
\end{equation}
gives a correlation with $m_0$ and $m_{1/2}$ as shown in Fig.\ref{ptnorm},
where the symbols delimite the regions where $p_t^{l,norm}$ is higher than
a certain value. On this plot we have selected events
with $N_{lept} \geq 3$, 
$E_t^{miss} > 50$ GeV, and required a minimum value of 15 GeV for the 
momentum of any of the leptons and 50 GeV for the leading one.

\vskip 0.2cm
{\it \bf Sensitivity for the type of the coupling}
\vskip 0.2cm

As mentioned in Section 2.1 counting the number of stable
leptons not only can reveal the signal of $R$ parity violation
but also can give a hint on the type of the coupling which
is realized. In order to show this we have simulated events
in three different points of the SUGRA space (LHC No 1,3 and 5,
see lines 3-5 of Table \ref{tb:sigstat}) and for two different
couplings: $\lambda_{122}$ and $\lambda_{123}$.
In the Table \ref{tb:4lept} the number of events are given
one can observe for the two classes: $0 e + 4 \mu$ and
$4 e + 0 \mu$ in one year of LHC running at low luminosity.
The events satisfy the following criteria: \\
\indent \indent $ 20 \le p_t^l \le 250$ GeV and $E_t^{miss} \ge 250$ GeV at points 1 and 5 \\
\indent \indent$ 10 \le p_t^l \le 100$ GeV and $E_t^{miss} \ge 150$ GeV at point 3.

\begin{Fighere}
\centering
  \mbox{\epsfig{file=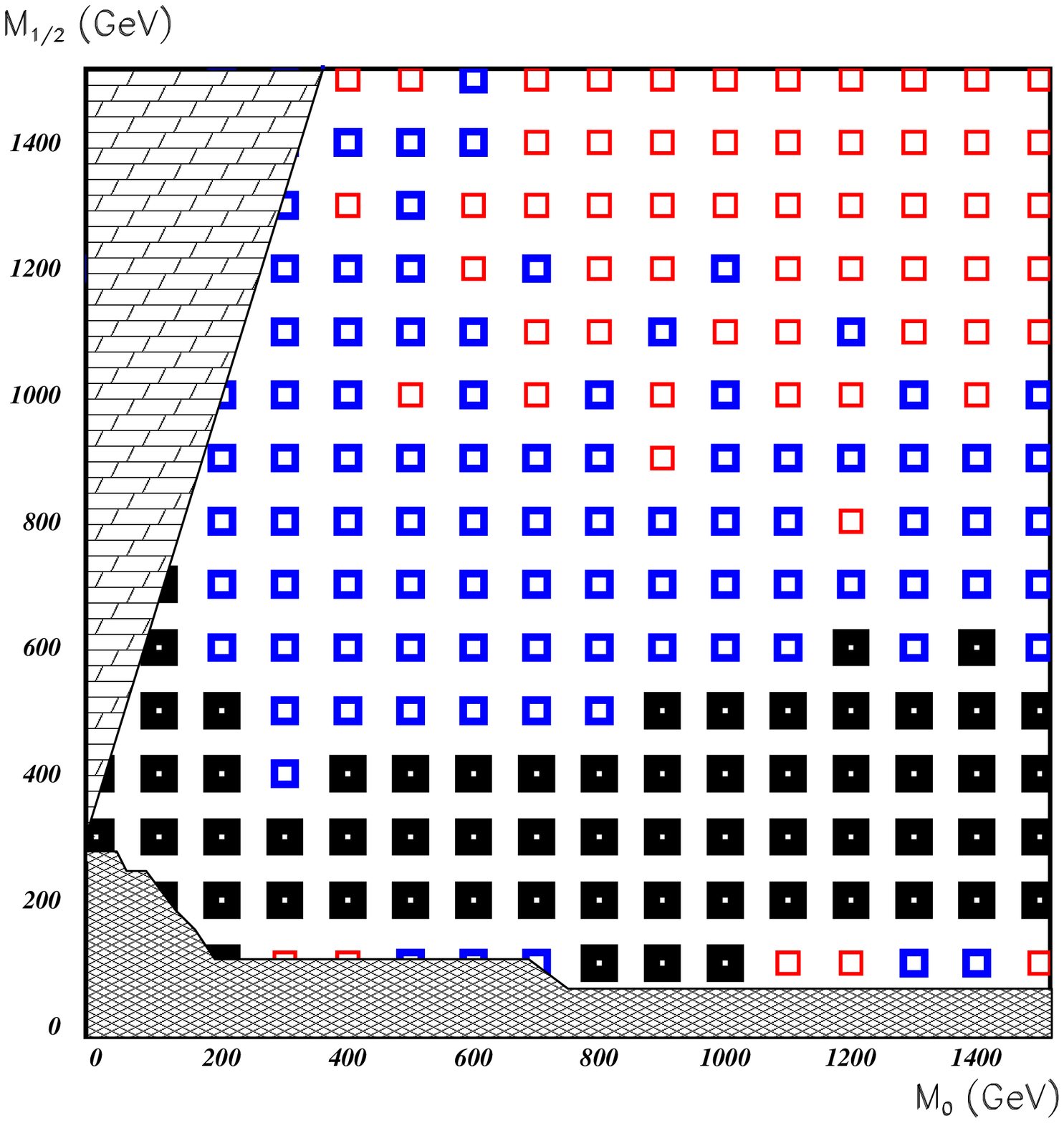,width=13.5cm  }}
\caption{\small In the case of $\lambda_{123}=10^{-3}$, $A_0 = 0$, $tan \beta = 2$,
$sign \mu = +1$ and for the global cuts :
$N_{lept} \geq 3$, $E_{t}^{miss} \geq 50$ GeV, $p_{t}^{l} \geq 15$ GeV and 
$p_{t}^{l, max} \geq 50$ GeV one can "divide" the $(m_{0}, m_{1/2})$ space in
$p_t^{l,norm}$ domains: 
-- full squares: $p_t^{l,norm} \in (30, 150)$ GeV;
-- fat squares: $p_t^{l,norm} \in (150, 250)$ GeV;
-- open squares: $p_t^{l,norm} \in (250, 500)$ GeV. Bricked domain is excluded by theory and
the hatched one is excluded by the present experimental limits.}
\label{ptnorm}
\end{Fighere}

\begin{Tabhere}
\centering
\caption{\small  Number of events with different number of electrons and muons }
\vskip 0.5cm
\begin{tabular}{|l|c|c|c|c|} \hline 
            & \multicolumn{2}{ c|}{ $\lambda_{122}$ }  & \multicolumn{2}{ c|}{ $\lambda_{123}$ }  \\ \cline{2-5} 
            & $0 e + 4 \mu$ & $4 e + 0 \mu$            & $0 e + 4 \mu$ & $4 e + 0 \mu$            \\ \hline
Point No 1  & $934 \pm 16$    & $3 \pm 1 $             & $76 \pm 4$      & $68 \pm 4$             \\ \hline
Point No 5  & $3010 \pm 52$   & $20 \pm 4 $            & $289 \pm 17$    & $304 \pm 18$           \\ \hline
Point No 3  & $52636 \pm 1760$ & $3356 \pm 444$        & $12400 \pm 600$ & $15565 \pm 672$        \\ \hline
\end{tabular}
\label{tb:4lept}
\end{Tabhere}
\vskip 0.5cm

It is obvious that one can clearly distinguish which of the two couplings
are realized in Nature. This is further illustrated in Fig.\ref{4lept},
where we have plotted the distribution of the number of muons/event for the two
different couplings where the total number of electrons and muons is
equal to four.\\
To conclude this section we have demonstrated that even if $R$ parity is violated
ATLAS can detect the signal of SUGRA in a large domain of the parameter space.
This domain is compatible with that obtained in the case when $R$ parity is
conserved. We can equally well establish the mass scale of the SUGRA compared
to the case of $R$ parity conservation. Finally, one has a possibility to
determine the type of the coupling if only one causes the violation of the
$R$ parity.

\begin{Fighere}
\centering
  \mbox{\epsfig{file=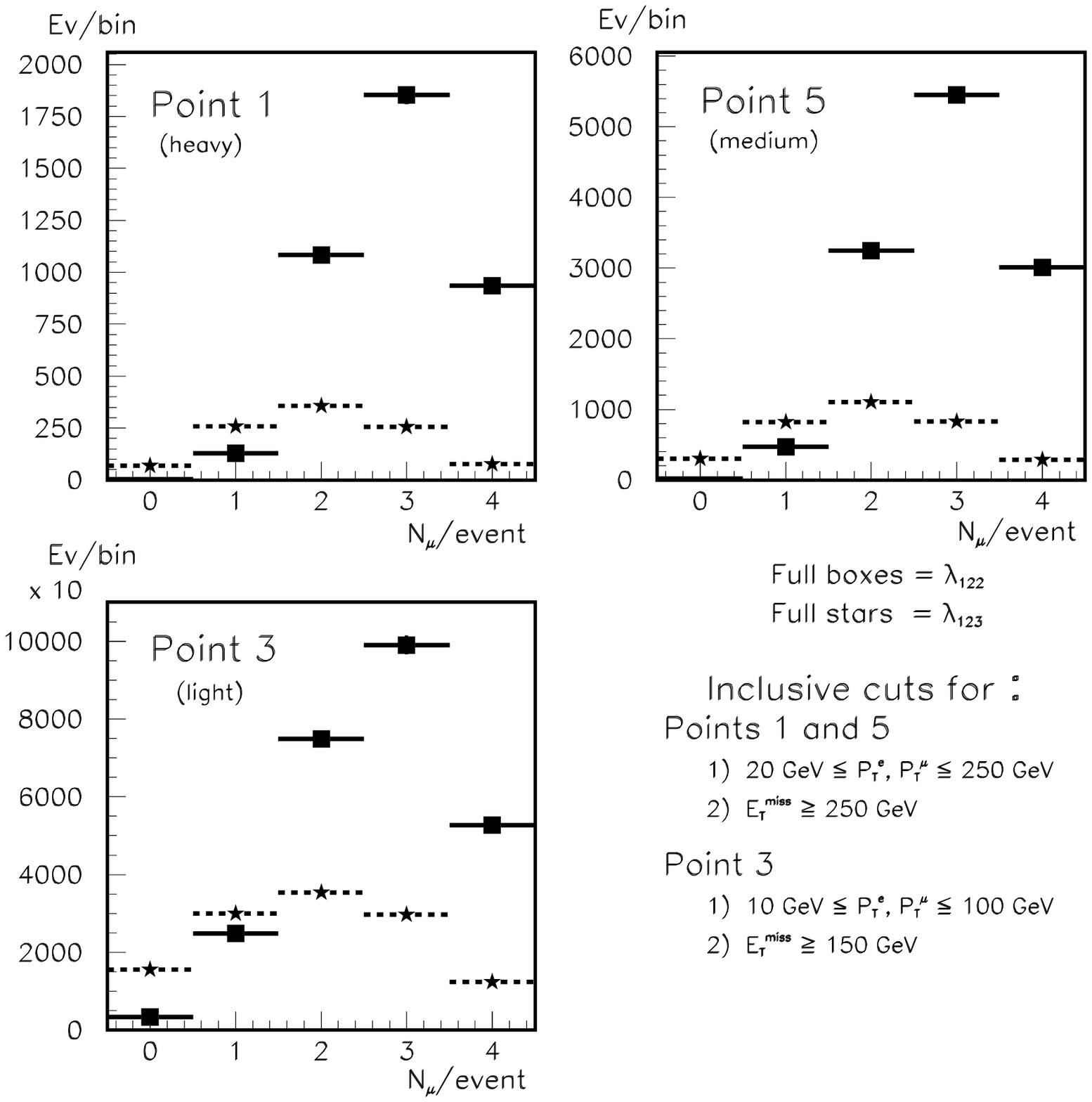,width=15cm  }}
  \vskip -0.5 cm
\caption{\small The distribution of the number of muons per event if the total number
of electrons and muons is equal to 4 in the case of two $\rlap/R$ couplings ($\lambda_{122}$,
$\lambda_{123}$) in the SUGRA points 1, 3 and 5.}
\label{4lept}
\end{Fighere}
\vspace {0.5cm}


\newpage
\section{Exclusive measurements}

\vskip 0.5 cm

In this chapter we show that one can determine the parameter
values of the SUGRA model, similarly to the case when $R$ parity is
conserved~\cite{SUGRA12}-\cite{SUGRA5}. 
The quantities to be used for this purpose, in general, 
are the masses of the reconstructed SUSY particles as well
as their observed production cross sections and branching ratios.  
We shall use only the first type of characteristics in this
analysis since determination of cross sections and branching ratios
is more sensitive to the acceptances and in many cases detailed
simulations are needed. In 3 out of the 5 LHC points, which are sufficiently different
to illustrate the methods in various conditions, we shall first
show how to reconstruct the SUSY particles, determine their masses,
and finally fit the model parameters to these mass values.

The reconstruction of the SUSY particles are difficult because there
are always at least two final state particles which cannot be detected.
These are obviousely the LSP's in the case of $R$ parity conservation,
but if $R$ parity is violated through the terms (\ref{eq:Lagl}) there is
always at least one undetectable neutrino produced 
in each LSP decay (see also Table \ref{tb:BRs}).
To overcome this difficulty we use the fact that in a 3-body decay:
\begin{equation}
A \longrightarrow a\ b\ c
\label{eq:Aabc}
\end{equation}
the invariant mass $m_{bc}$ of two of the three final state particles, e.g.
$b$ and $c$ gives a clear endpoint. The endpoint, $m_{bc}^{end}$
is related to the masses of $A$ and that of the undetected particle, $a$:
\begin{equation}
m_{bc}^{end} = m_A - m_a
\label{eq:mAa}
\end{equation}
since in the restframe of $A$ the three-momentum of $a$ (or that
of the ($bc$) system) is zero.
This equation is particularly useful if the mass of $a$ is zero: in this
case one can directly estimate the mass of the particle $A$ by measuring
the endpoint of the $m_{bc}$ distribution. Of course, in the practice
this measurement is difficult because the endpoint may be hidden
by eventual background and it can disappear if either $b$ or $c$
is unstable. This is the case when a $\tau$ is produced in the decay
of the $\tilde\chi_1^0$, thus in all couplings when a "3" appears in
the index of the $\lambda$.

Once the endpoint is established one can determine even the four-momentum 
$P^{\mu}$ of the undetected particle $a$ or that of the mother particle $A$ 
by selecting events around the endpoint:

\begin{equation}
P^{\mu}_{A,a} = \frac{m_{A,a}}{m_{bc}}(P^{\mu}_{b}+P^{\mu}_{c}) \ (\mu = 1,...,4)
\label{eq:pAa}
\end{equation}
As one can see from Equs. (\ref{eq:mAa}) and (\ref{eq:pAa}), 
if particle $a$ has zero mass (as e.g. in the case of a neutrino),
the four-momentum of the charged lepton pair is equal to that of particle $A$.
The four-momentum of particle $A$ can be used further for the reconstruction
of the parent particle, and so on, until one arrives at the beginning of the
decay chain, i.e. at the original SUSY particle. It is also obvious that at the endpoint 
the 4-momentum of particle $a$ and $A$ are parallel whatever their masses are.

Since equations (\ref{eq:mAa}) and (\ref{eq:pAa}) are strictly valid
only at the endpoint, one usually needs a large number of produced events
and in the selection of the size of the region around the endpoint
an optimum has to be found between the statistical error and
the approximate validity of the above relations. In practice one applies in 
Eq.(~\ref{eq:pAa}) $m_{bc}$ instead of $m^{end}_{bc}$ in order to "rescale" the
values of $P^{\mu}_b+P^{\mu}_c$ which are smaller than their corresponding 
values at the endpoint.

If the decay (\ref{eq:Aabc}) proceeds through a sequence of 2 two-body decays:

\begin{equation}
A \longrightarrow  B\ b \ \ \ \ \ \ \ \ B \longrightarrow a\ c
\label{eq:AaBbc}
\end{equation}
one can observe an endpoint in the $m_{bc}$ distribution, whose value is given by:\\

\begin{equation}
m^{end}_{bc}=m_A \sqrt{1-(\frac{m_B}{m_A})^2}\ \sqrt{1-(\frac{m_a}{m_B})^2}\mbox{\ \ .}
\label{eq:mend22b}
\end{equation}
\\
The quantity of $\sigma \times BR$ of the produced particle $A$ helps to
disentangle which of the two decay modes (\ref{eq:Aabc}) or (\ref{eq:AaBbc}) has
occured and to choose between (\ref{eq:mAa}) or (\ref{eq:mend22b}) to estimate the masses.

At LHC the generic production and decay chains are represented in the schemas here below: (*).
The squark or gluino undergoes the cascade decay:

\begin{tabbing}
oooooooooooooooooooooo\=oooooooooooo \kill
\> $\tilde q\longrightarrow$  \= $\tilde g  + q$ \\
\> \>
\begin{picture}(11,10)(0,0)
\put(2.0,9.0){\line(0,-2){8}}
\put(2.0,1.0){\vector(1,0){9}}
\end{picture}
 \= $\tilde q'$ +  $q'$ \\ 
\> \>\>
\begin{picture}(11,10)(0,0)
\put(2.0,9.0){\line(0,-2){8}}
\put(2.0,1.0){\vector(1,0){9}}
\end{picture}
 \= $\tilde \chi_1^{\pm}$ +  $q''$ \\ 
\> \>\>\>
\begin{picture}(11,10)(0,0)
\put(2.0,9.0){\line(0,-2){8}}
\put(2.0,1.0){\vector(1,0){9}}
\end{picture}
\=$\tilde \chi_1^{0}$ +  $W^{\pm}$ oooooooooooooooooooo\= (*) \kill  
\> \>\>\>
\begin{picture}(11,10)(0,0)
\put(2.0,9.0){\line(0,-2){8}}
\put(2.0,1.0){\vector(1,0){9}}
\end{picture}
\=$\tilde \chi_1^{0}$ +  $W^{\pm}$  \> (*) \\ 
\> \>\>\>\>
\begin{picture}(11,10)(0,0)
\put(2.0,9.0){\line(0,-2){8}}
\put(2.0,1.0){\vector(1,0){9}}
\end{picture}
$l^+$ +  $l^-$  +  $\nu$  \\ 
\> \>\>
\begin{picture}(11,10)(0,0)
\put(2.0,9.0){\line(0,-2){8}}
\put(2.0,1.0){\vector(1,0){9}}
\end{picture}
 \= $\tilde \chi_2^0$ +  $q$ \\ 
\> \>\>\>
\begin{picture}(11,10)(0,0)
\put(2.0,9.0){\line(0,-2){8}}
\put(2.0,1.0){\vector(1,0){9}}
\end{picture}
$\tilde \chi_1^{0}$ +  $(l^+l^-)/h^0$ \\ 
\> \>\>\>\>
\begin{picture}(11,10)(0,0)
\put(2.0,9.0){\line(0,-2){8}}
\put(2.0,1.0){\vector(1,0){9}}
\end{picture}
$l^+$ +  $l^-$  +  $\nu$  \\ 
\end{tabbing}
and the $R$ parity violation occurs at the end of the decay chain in the
decay of the $\tilde \chi_1^{0}$. 

For the exclusive measurements we have chosen 3 out of the 5 LHC points to
study. Point 1 has a high, point 3 has a low and point 5 has a medium mass scale.
Another crucial difference between these points, coming from
the mass parameters $m_0$, $m_{1/2}$ and sign($\mu$), is reflected
in the decay of $\tilde\chi_2^0$.
In point 3, $\tilde\chi_2^0$ decays mainly through a three body decay   
$\tilde\chi_2^{0} \rightarrow \tilde\chi_1^0 + l^{\pm} + l^{\mp}$
(with a virtual $Z$ or $\tilde l$).
In points 1 and 5 the main decay ($65 \% \div 80 \%$) of $\tilde\chi_2^0$ proceeds
through a two-body decay $\tilde\chi_2^{0} \rightarrow \tilde\chi_1^0 + h^{0}$,
but also with a non-negligible branching ratio ($20 \% \div 30 \%$) 
it decays in a sequence of 2 two-body decays
$\tilde\chi_2^{0} \rightarrow \tilde \l_{R}^{\pm} + l^{\mp} \rightarrow 
\tilde\chi_1^0 + l^{\pm} + l^{\mp}$ with a $\tilde l_R^{\pm}$ being real. 

Therefore point 3 has a completely different event topology,
namely a very large number of leptons in the final
state, in comparison with the other two points. This is
the reason that we present here an analysis in point 3 although
it is already excluded by the LEP limit on the Higgs mass~\cite{LEP_Higgs}.
In a nearby point, which is not yet excluded all our conclusions can
be considered as valid.

At each LHC points we have considered two different couplings:
$\lambda_{122}$ and $\lambda_{123}$. In the case of the 
first one the reconstruction of the SUSY particles is easier
than in the second one, where a $\tau$ particle appears
always among the decay products. This means more missing energy
in the second case and also the absence of a clear endpoint
in the invariant mass of the opposite-sign and different flavor (OSDF) lepton pair.
This in turn makes very difficult if not
impossible the reconstruction of the $\tilde\chi_1^0$ .

The values of the SUGRA parameters as well as the number
of generated events is listed in Table\ref{tb:sigstat}.
The masses of the SUSY particles in the three points are
listed in Table~\ref{tb:masses}. 

\vskip 0.5 cm
\subsection{The LHC points No1 and No5}
\vskip 1.0 cm

Concerning the reconstruction of the SUSY particles these two points
are very similar and therefore we treat them here together.
In the cascade decay chain (*) the predominant decay of the 
$\tilde\chi_2^0$ proceeds via 
$\tilde\chi_2^0\longrightarrow h^0 + \tilde\chi_1^0$
with a BR of $\sim 99 \%$ in point 1 and $\sim 63 \%$ in point 5, 
where the competitive decay through $\tilde l_{R}$ becomes also important.

The main inclusive cuts to select SUSY events are based on the requirement of a high
lepton multiplicity, $N_l \ge 4$ and a moderate missing transverse energy, 
$E_t^{miss}$.

The SM background is small and can be completely neglected
applying a cut on $E_t^{miss}$
as is shown in Fig.\ref{etm_p15}
where the  $E_t^{miss}$ distribution is shown for the points 1
and 5 and for $\lambda_{122}$ and $\lambda_{123}$.

\vspace{-0.5cm}
\begin{Tabhere}
\centering
\caption{\small  SUSY particle masses at points 1, 3 and 5 in GeV}
\vskip 0.5cm
\begin{tabular}{|l|c|c|c|} \hline
LHC point                  &  1     & 3    &    5   \\ \hline
$\tilde g$                 &  1008  & 299  &  769   \\ \hline
$\tilde q_L$               &   958  & 317  &  687   \\ \hline
$\tilde q_R$               &   925  & 312  &  664   \\ \hline
$\tilde b_2$               &   922  & 313  &  662   \\ \hline
$\tilde b_1$               &   855  & 278  &  634   \\ \hline
$\tilde t_2$               &   913  & 325  &  706   \\ \hline
$\tilde t_1$               &   649  & 260  &  494   \\ \hline
$\tilde l_L$               &   490  & 216  &  239   \\ \hline
$\tilde l_R$               &   430  & 207  &  157   \\ \hline
$\tilde \nu_L$             &   486  & 207  &  230   \\ \hline
$\tilde \tau_2$            &   490  & 216  &  239   \\ \hline
$\tilde \tau_1$            &   430  & 206  &  157   \\ \hline
$\tilde \chi^{\pm}_2$      &   775  & 274  &  526   \\ \hline
$\tilde \chi^{\pm}_1$      &   326  &  96  &  232   \\ \hline
$\tilde \chi^0_4$          &   778  & 275  &  529   \\ \hline
$\tilde \chi^0_3$          &   762  & 258  &  505   \\ \hline
$\tilde \chi^0_2$          &   326  &  97  &  233   \\ \hline
$\tilde \chi^0_1$ = LSP    &   168  &  45  &  122   \\ \hline
$h_0$                      &    98  &  69  &   95   \\ \hline
\end{tabular}
\label{tb:masses}
\end{Tabhere}
\vskip 0.5 cm
The "structures" seen in the hatched histograms of Fig.\ref{etm_p15} is due to 
statistical fluctuations.
In what follows we shall demonstrate that in the case of the
coupling $\lambda_{122}$ we can reconstruct all the SUSY particles
in the decay chain (*), starting with the $\tilde\chi_1^0$,
and in this respect we can achieve more than it was the case
with $R$ parity conserved. On the other hand, if $\lambda_{123}$
is nonzero, we can reconstruct only a fraction of the SUSY particles,
and the decay products of the  $\tilde\chi_1^0$, i.e. the additional
leptons, increase the background and thereby deteriorate the determination
of some particle masses. We shall show that in spite of these
difficulties the achieved precision is at worst comparable with
the one in the case of conserved $R$ parity.

\begin{Fighere}
\centering
  \mbox{\epsfig{file=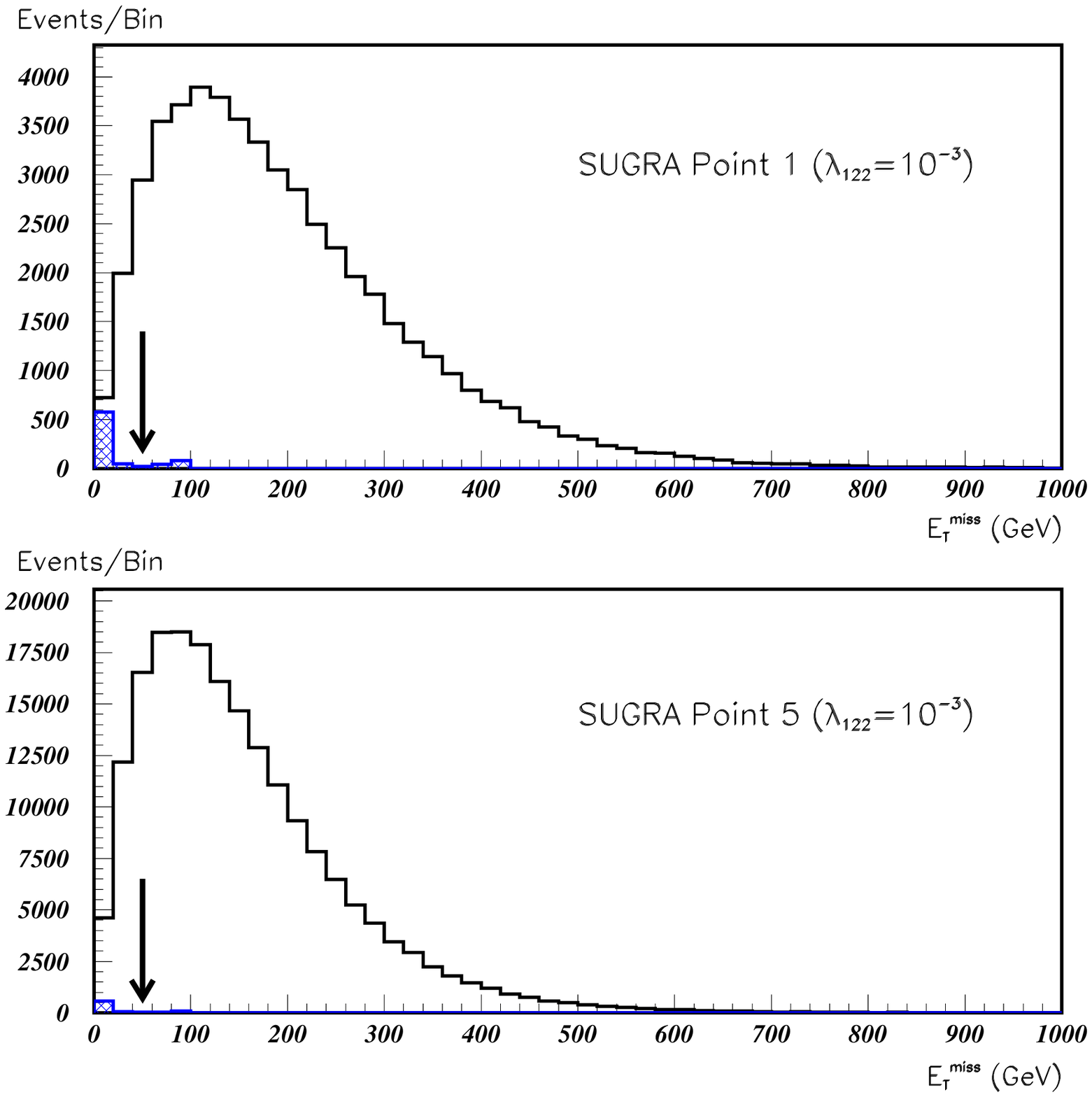,width=7.cm  } \epsfig{file=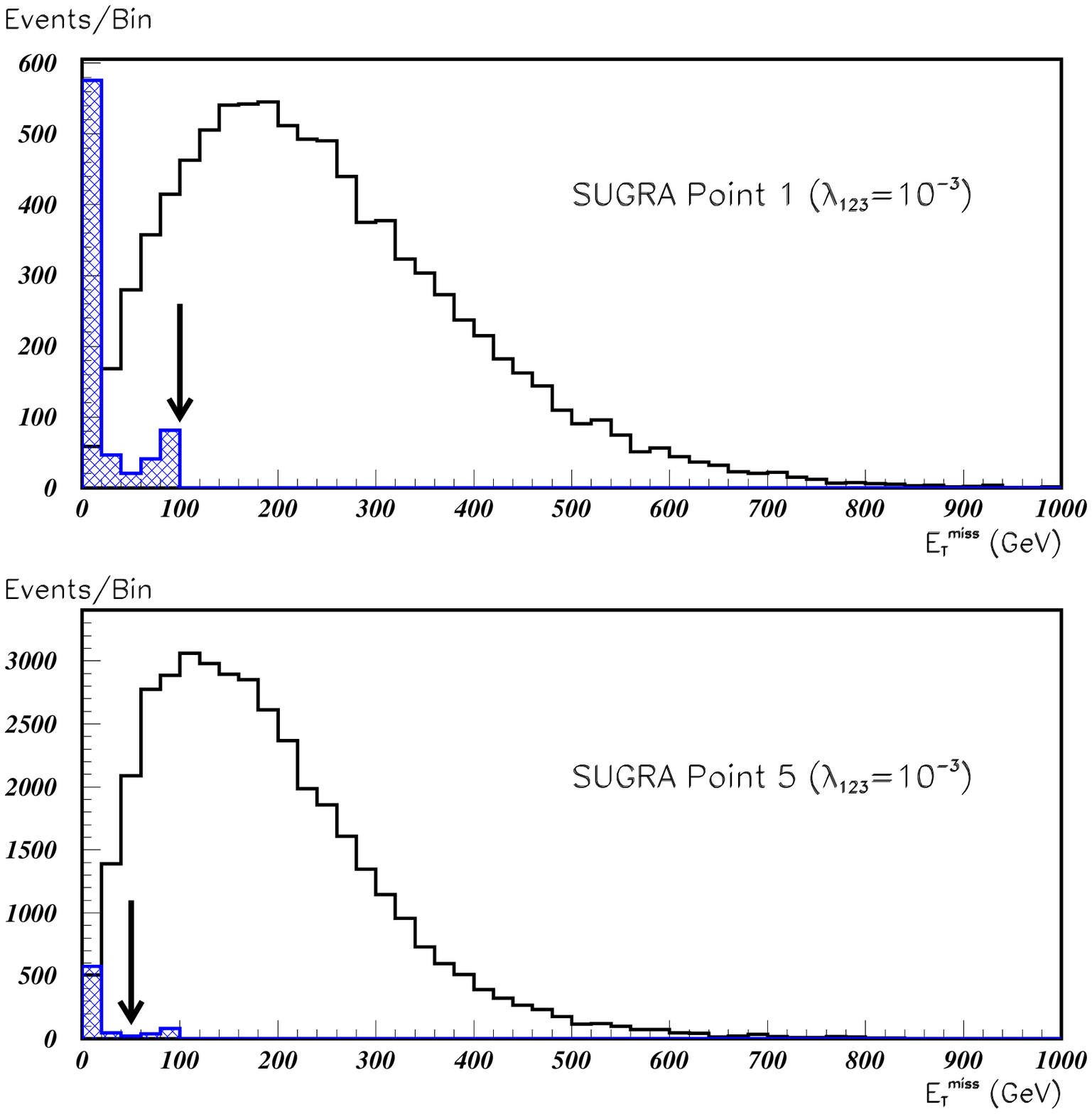,width=7.cm  }}
\caption{\small The distribution of $E_t^{miss}$ for SUSY events (open histogram) and SM events
(hatched histogram) in points 1 and 5 for $\lambda_{122}=10^{-3}$ (left plots) and
$\lambda_{123}=10^{-3}$ (right plots) for events with $N_{leptons} \geq 4$, $P_{t}^{lept} \geq 10$ GeV
and after 3 years of LHC run at low luminosity. The arrows precise the cuts to be used.}
\label{etm_p15}
\end{Fighere}

\vskip 0.8cm
{\bf \large \it \underline{The case $\lambda_{122} \ne 0$}}

\vskip 0.2cm
{\it \bf Reconstruction of $\tilde \chi^0_1 \rightarrow \nu_{e (\mu)} + e^{\pm} (\mu^{\pm}) + \mu^{\mp} $}
\vskip 0.2cm

For the reconstruction of the  $\tilde\chi_1^0$ we apply
the following selection criteria: \\
\indent \indent $(i)$ $N_l \ge 4$ (and $N_l^+ = N_l^-$ and $N_e = N_{\mu}$ for point 5), \\
\indent \indent $(ii)$ $p_t^l \ge $ 10 GeV, cos($\alpha_{l^{\pm}l^{'\mp}}$) $\geq$ 0.5, \\
\indent \indent $(iii)$ $E_t^{miss} \ge $ 50 GeV , \\
where $\alpha_{l^{\pm}l^{'\mp}}$ is the angle between any OSDF (i.e. electron -muon) leptons.\\
The invariant mass distributions of the OSDF lepton pairs
are shown in Fig.\ref{p1chi01} for point 1
and in Fig.\ref{p5chi01} for point 5, where the number of events correspond to 
3 years of LHC run at low luminosity. \\
In point 1, isolated leptons are produced practically only in the two $\rlap/R$ decays of the
$\tilde \chi^0_1$'s, therefore the background of OSDF distribution will be mainly 
$\tilde \chi^0_1$ combinatorial.
In point 5, as we mentioned before, a non-negligible production of isolated leptons (beyond 
the $\rlap/R$ decays of $\tilde \chi^0_1$'s) comes from 
$\tilde \chi^{0}_{2} \rightarrow \tilde l^{\pm}_{R} + l^{\mp} \rightarrow 
\tilde \chi^{0}_{1} + l^{\pm} + l^{\mp}$. This decay will produce two leptons of opposite sign (OS)
but of the same flavor (SF), changing the balance of leptons per flavor.
 Most of the events with an increased number of leptons, will have only one $\tilde \chi^{0}_{2}$ 
decaying leptonically (the other one, if it exists, will decay in $h^{0}$'s). To decrease the
number of bad combinations (with leptons not coming from $\tilde \chi^{0}_{1}$'s), additional cuts
like $N_l^+ = N_l^-$ and $N_e = N_{\mu}$ in each event are very efficient.

There is a clear endpoint over a moderate
background at both points 1 and 5 corresponding to the
$\tilde\chi_1^0$ mass (c.f. Table\ref{tb:masses}) in virtue
of Equ.(\ref{eq:mAa}). 
We parametrize and subtract the background with a
Maxwellian distribution and fit the resulting distribution near the
endpoint obtaining the $\tilde\chi_1^0$ mass values: 
\begin{equation}
m_{\tilde\chi_1^0}^{meas} = 169.80^{+0.2}_{-0.8}  \mbox{\ GeV at point 1}
\label{eq:mchi10p1}
\end{equation}
\begin{equation}
m_{\tilde\chi_1^0}^{meas} = 122.62^{+0.4}_{-1.0}  \mbox{\ GeV at point 5}
\label{eq:mchi10p5}
\end{equation}

\vskip -3.0 cm
\begin{Fighere}
\centering
  \mbox{\epsfig{file=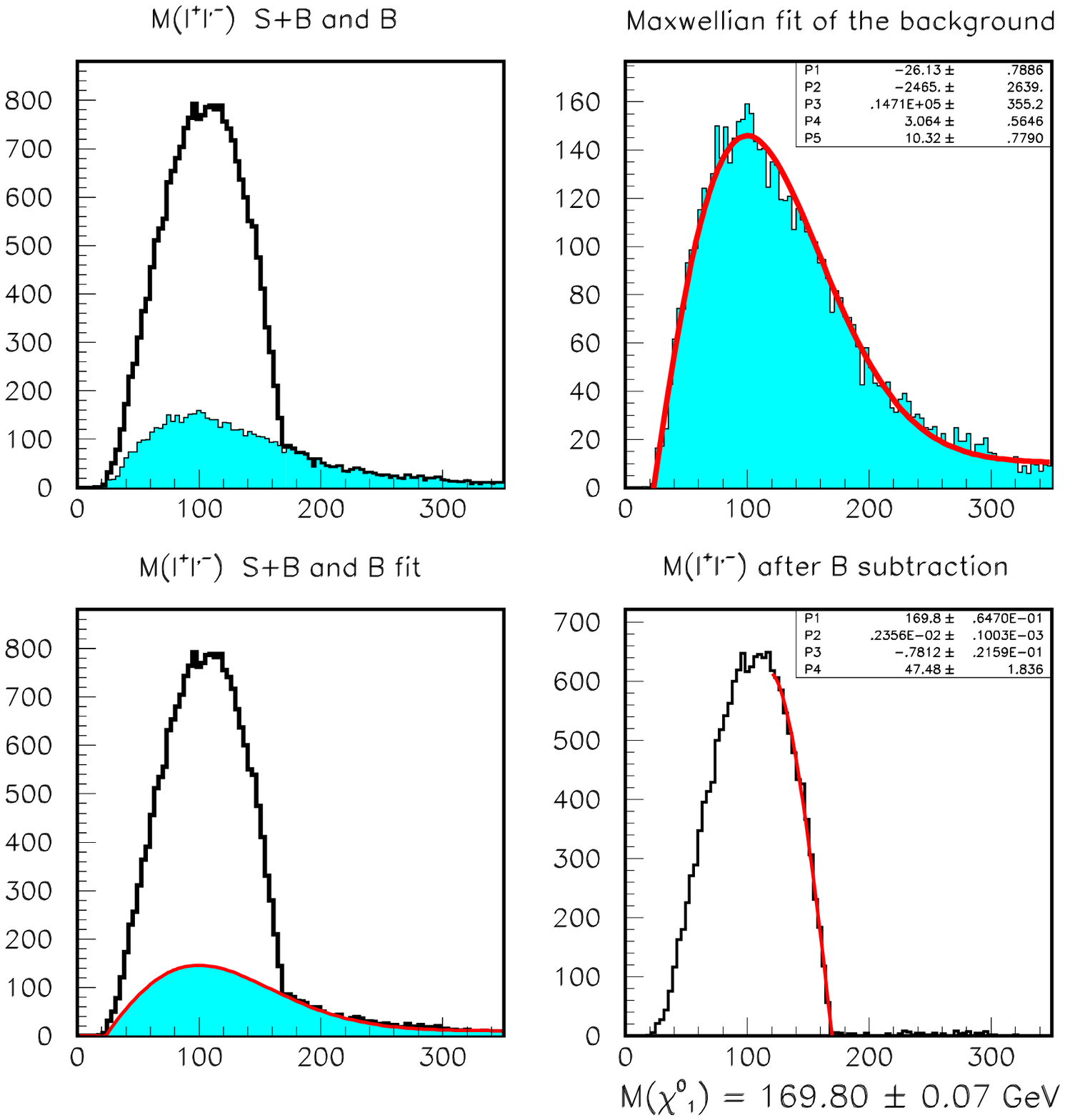,height=9.0cm,width=15cm  }}
\vskip -0.5 cm
\caption{\small The invariant mass distribution of the OSDF
lepton pairs for 3 years of LHC run at low luminosity at point 1 (for the selection
criteria see text). In the upper-left plot are represented all events (open histogram)
and the background (shadded histogram)  which is mainly SUSY combinatorial. 
The upper-right and lower-left plots show a Maxwellian-like distribution fitted to the background. 
The lower-right plot represents the result after the subtraction of the fitted background. 
The edge is fitted with a polinomial function and the error is only statistical. }
\label{p1chi01}
\end{Fighere}

\begin{Fighere}
\centering
  \mbox{\epsfig{file=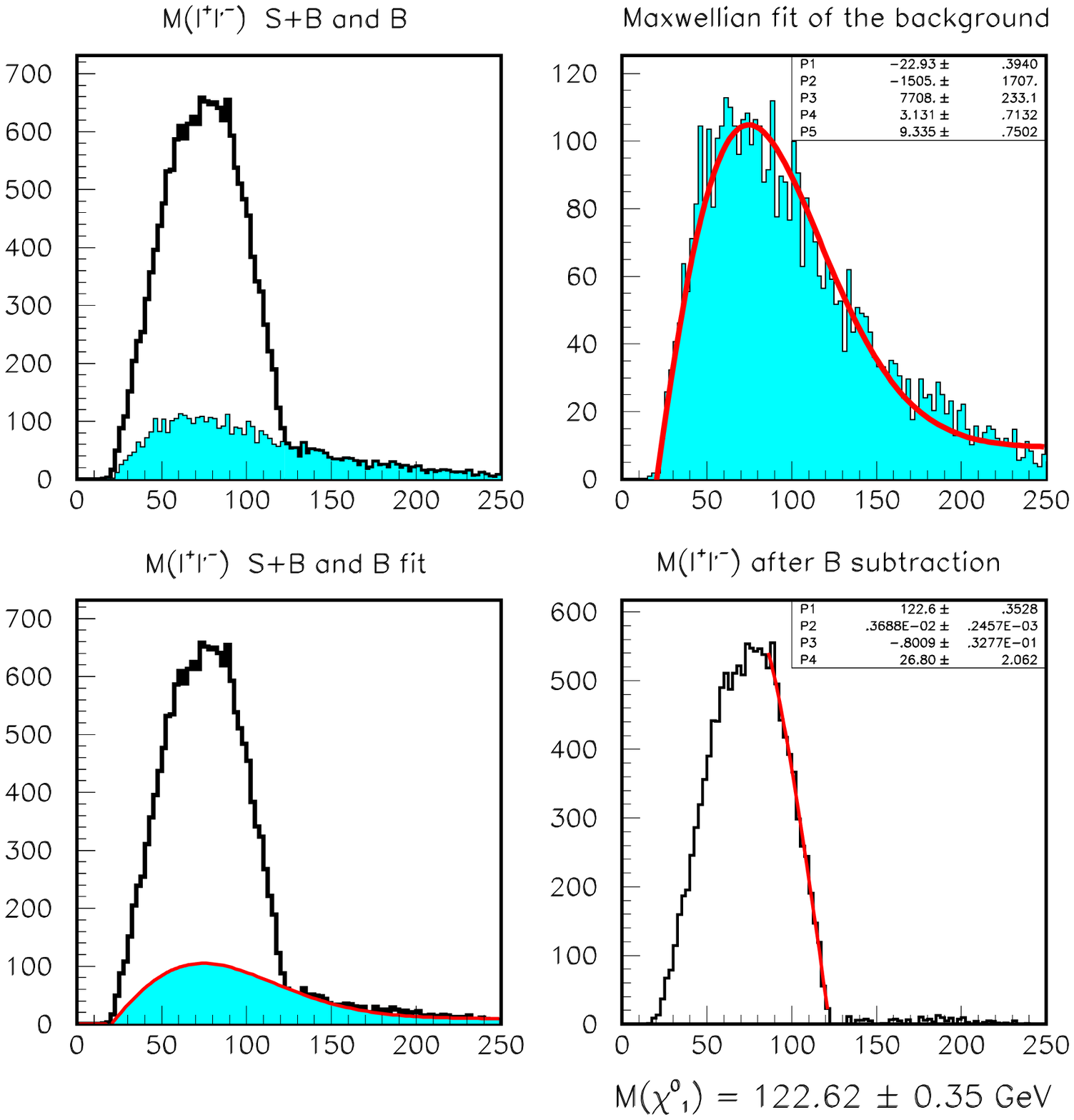,height=9.0cm,width=15cm }}
\vskip -0.5 cm
\caption{\small The invariant mass distribution of the OSDF
lepton pairs for 3 years of LHC run at low luminosity at point 5 (for the selection
criteria see text). }
\label{p5chi01}
\end{Fighere}
\vskip 1.0 cm

The estimated error contains both statistical and systematical.
This latter is due to the energy resolution of the leptons
and mainly to the finite bin-size of our histograms around the endpoint.
Since due to the bin-size one tends to overestimate the mass value the systematical
error is asymetrical.\\
For any further reconstruction using OSDF pairs (like $\tilde \chi^0_2$, $\tilde \chi^{\pm}_1$, etc) 
one will take as $\tilde \chi^0_1$ candidates the OSDF pairs with an invariant mass in 
($m_{\tilde \chi^0_1}-\Delta m_{\tilde \chi^0_1}$, $m_{\tilde \chi^0_1}$), where
$\Delta m_{\tilde \chi^0_1}=50$ GeV for point 1 and $\Delta m_{\tilde \chi^0_1}=30$ GeV for point 5.
To improve the statistics in point 5 we will not use the additional cuts (see $(i)$).

\vskip 0.2cm
{\it \bf Reconstruction of $h^0 \rightarrow b + \bar b$}
\vskip 0.2cm

To the global cuts ($N_l \geq 4$, $P_{t}^{l} \geq 10$ GeV and $E_t^{miss} \geq 50$ GeV)
we will add some specific cuts on b jets.
Since the $h^0$ decays to $b\bar b$ pairs with appr. 88\% BR its reconstruction
proceeds by the selection of these pairs, adding thus the following detection
criteria to the above ones: \\
\indent \indent $(iv)$ $p_t^b \ge 30$ GeV for point 1, $40$ GeV for point 5 respectively, 
and $p_t^b \le $ 300 GeV for both points \\
\indent \indent $(v)$ cos($\alpha_{b\bar b}$) $\ge$ 0.4 (point 1) or 0.3 (point 5) \\
where $\alpha_{b\bar b}$ is the angle between the $b$ and $\bar b$.
The obtained invariant mass distributions of the $b\bar b$ pairs are shown
in Fig.\ref{p1h0mass} and Fig.\ref{p5h0mass}. There is a clear mass peak corresponding
to the $h^0$ particle (c.f. Table\ref{tb:masses}).

\begin{Fighere}
\centering
  \mbox{\epsfig{file=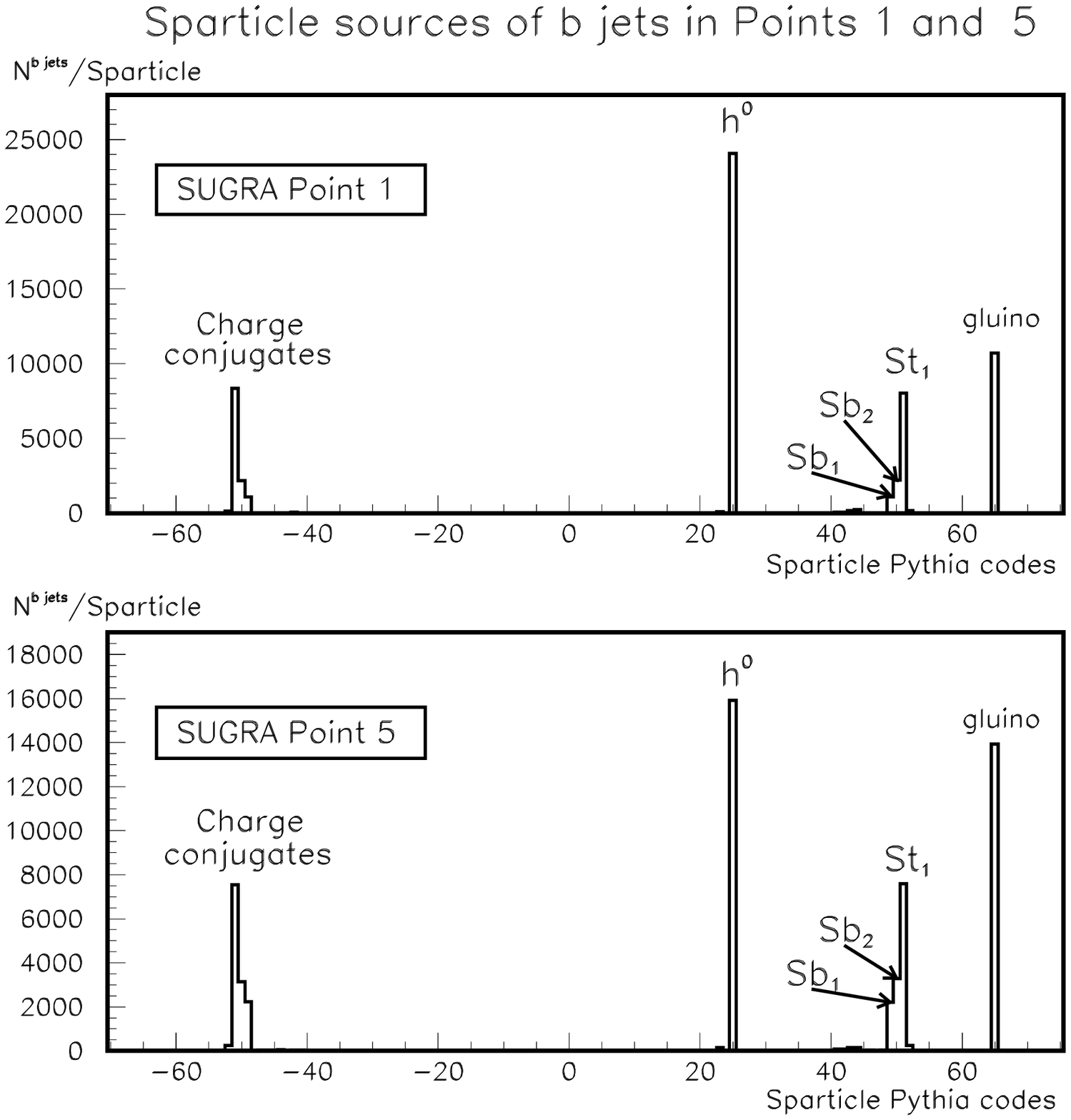,width=12cm,height=7cm }}
  \vskip -0.5 cm
\caption{\small The sparticle mothers of b jets (in Pythia codes) in the points 1 and 5
(see text).}
\label{p1p5bsource}
\end{Fighere}
\vskip 0.5 cm

As we can see from the Fig.\ref{p1p5bsource}, the b jets are mainly produced through
$h^0 \rightarrow b \bar b$ but also through
$\tilde g \rightarrow \tilde b_{1(2)} + \bar b \rightarrow \tilde \chi^0_{1(2)} + b + \bar b$ or
$\tilde g \rightarrow \tilde t_1 + \bar t \rightarrow \tilde \chi^0_{1(2)} + t + \bar t$ 
(or $\rightarrow \tilde \chi^{+}_{1} + b + \bar t$) and the subsequent decay of $t$.
In the case of conserved $R$ this SUSY background (as well as the SM $t \bar t$) is efficiently
rejected with a veto on leptons (coming from the top decay). However, in the case of
$\rlap/R$ through $\rlap/L$ couplings we cannot apply anymore such a cut (and a veto on
{\it additional} leptons doesn't increase significantly the $S/B$ ratio). 
Another source of background (albeit small) is the production
of $Z$'s (i.e. $\tilde t_{2} \rightarrow \tilde t_{1} + Z^0$ or 
$\tilde \chi^{0}_{i} \rightarrow \tilde \chi^{0}_{1} + Z^{0}$, $i=2,3,4$
or $\tilde \chi^{\pm}_{2} \rightarrow \tilde \chi^{\pm}_{1} + Z^0$, with $Z^0 \rightarrow b + \bar b$) 
resulting in a peak very close to the $h^0$ one.
This will give rise to an asymmetric $h^0$ peak with a higher width as can be seen in both 
points. A similar effect arises also from the calibration error of $b$ jets.

After havig parametrized and subtracted this
background we have fitted the mass peak with a Gaussian curve
and obtained the values: \\
\begin{equation}
m_{h^0}^{meas} = 97.08 \pm 1.5  \mbox{\ GeV at point 1}
\label{eq:mh0p1}
\end{equation}
\begin{equation}
m_{h^0}^{meas} = 94.7 \pm 1.5  \mbox{\ GeV at point 5}
\label{eq:mh0p5}
\end{equation}
The estimated systematic error is mainly due to the uncertainties in the
$b$-jet correction factor shown in Fig.\ref{p5selbjets}.

\vskip -0.5 cm
\begin{Fighere}
\centering
  \mbox{\epsfig{file=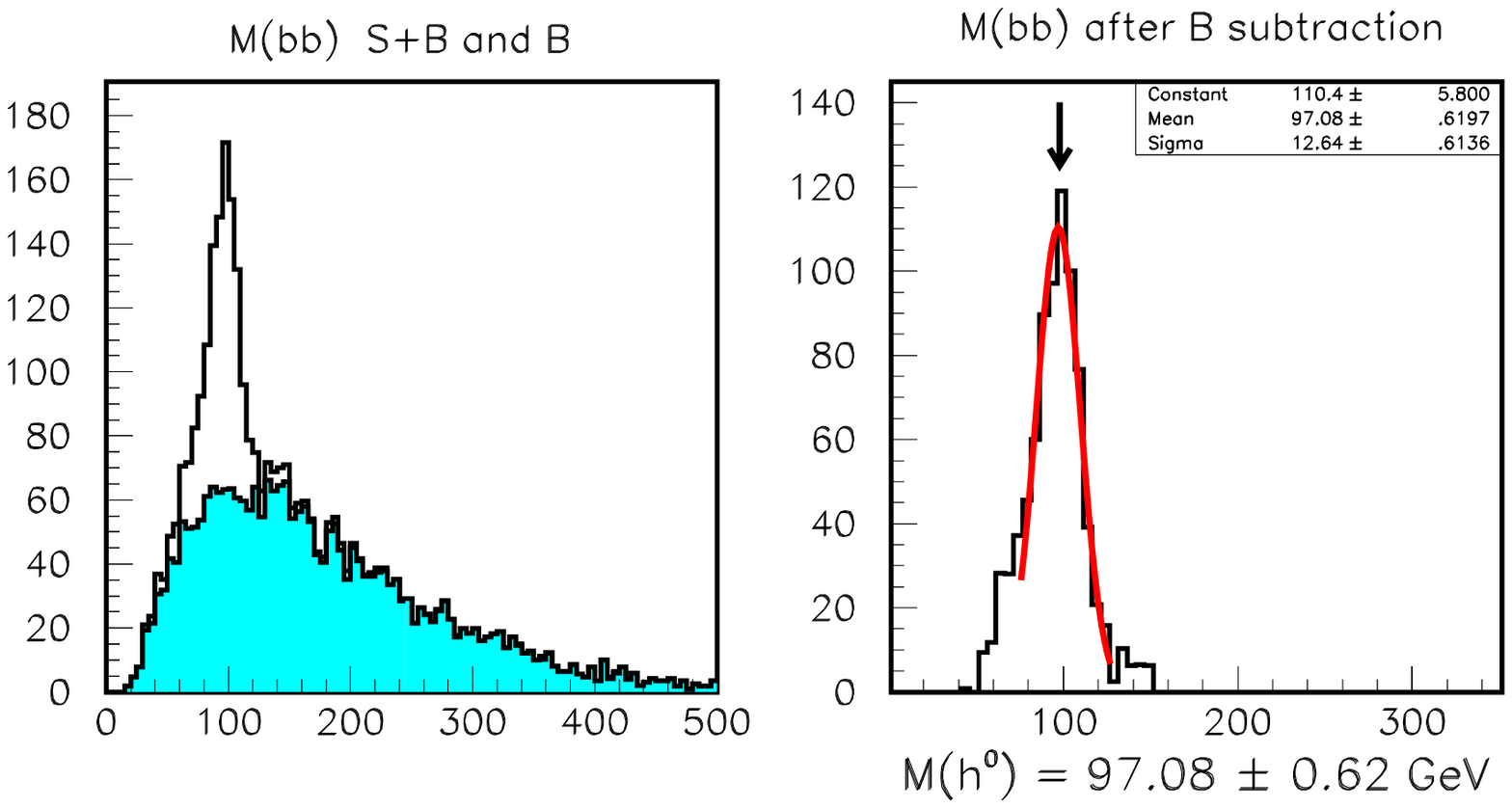,height=10.0cm,width=15cm  }}
\vskip -5. cm
\caption{\small The invariant mass distribution of the $b\bar b$
pairs for 3 years of LHC run at low luminosity at point 1.
On the left plot are represented all the events and the background (shadowed).
 The background is fitted with a Maxwellian function and subtracted. The result
is drawn in the right plot. The arrow points to  the theoretical value of $m_{h^0}$.
The peak is fitted with a gaussian. The written error contains only the statistical part.}
\label{p1h0mass}
\end{Fighere}

\begin{Fighere}
\centering
  \mbox{\epsfig{file=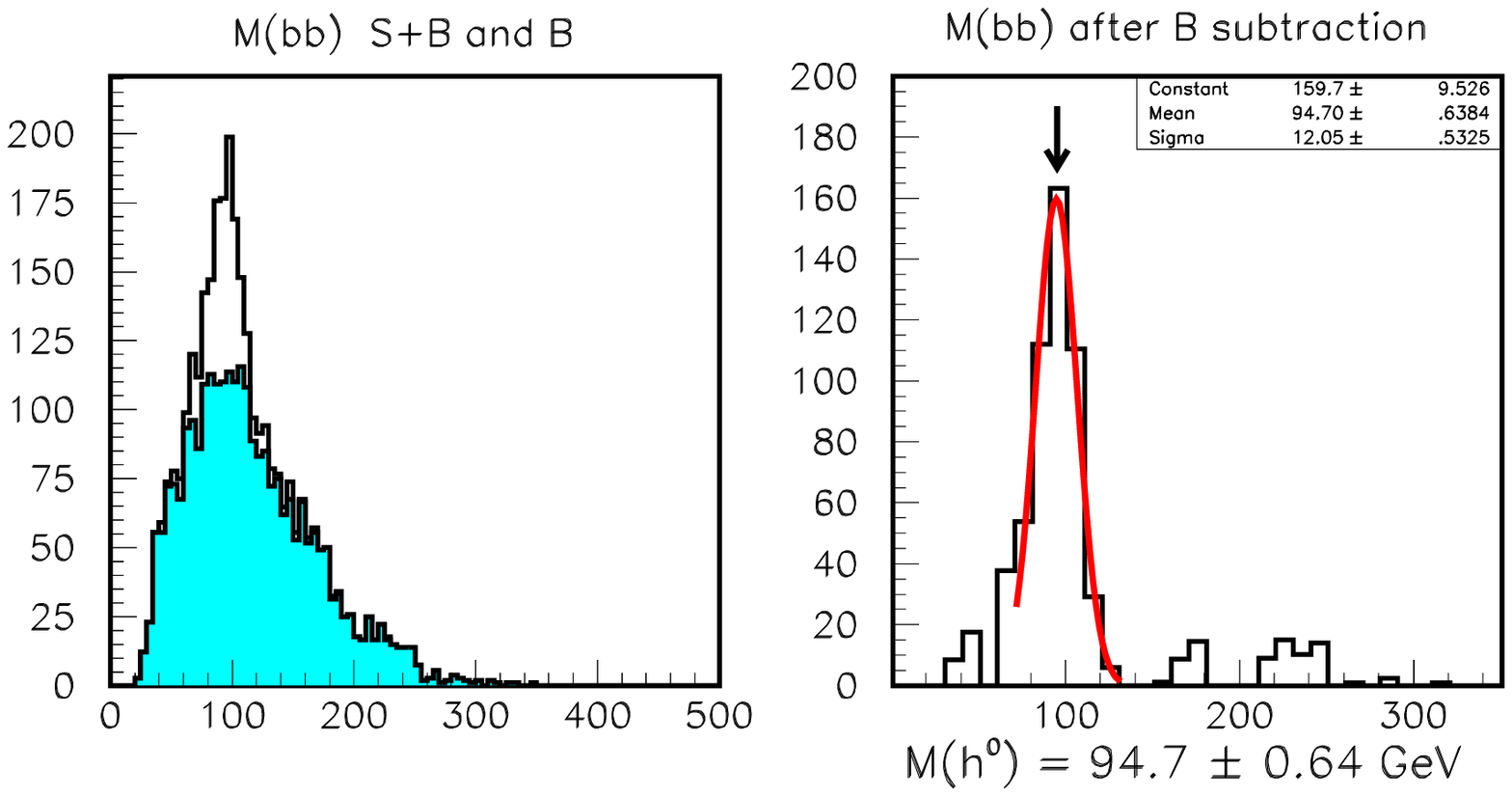,height=10.0cm,width=15cm  }}
\vskip -5. cm
\caption{\small The invariant mass distribution of the $b\bar b$
pairs for 3 years of LHC run at low luminosity at point 5.}
\label{p5h0mass}
\end{Fighere}
\vskip 1cm

\vskip 0.2cm
{\it \bf Reconstruction of $\tilde \chi^0_2 \rightarrow h^{0} + \tilde \chi^0_1$}
\vskip 0.2cm

The reconstructed $h^0$ and $\tilde\chi_1^0$ allows the reconstruction of
the mother particle, the $\tilde\chi_2^0$, since this latter decays to
the formers by 99\% BR at point 1 and 63\% BR at point 5.
We will require the $\tilde \chi^0_1$ candidates (OSDF lepton pairs) to be "close" to the endpoint :\\
\indent \indent $(vi)$ $m(OSDF) \in (m^{endp}-50,m^{endp})$ GeV - in point 1 and,\\
\indent \indent \ \ \ \ \ \ $m(OSDF) \in (m^{endp}-30,m^{endp})$ GeV - in point 5,\\
and the $h^0$ candidates ($bb$ pairs) to be around the mass peak of $h^0$ :\\
\indent \indent $(vii)$ $m(bb) \in (m^{peak}-15,m^{peak}+15)$ GeV - for both points 1 and 5.\\
Looking at the relations between $\tilde \chi^0_2$, $\tilde \chi^0_1$ and $h^0$ masses, one can expect
in point 1 a higher boost of $\tilde \chi^0_1 - h^0$ pair, comparing to point 5, therefore we demand :\\
\indent \indent $(viii)$ cos($\alpha_{h_0\tilde\chi_1^0}$) $\ge$ 0.7 (point 1) or 0.5 (point 5). \\ 
Finally one gets the invariant mass distributions of the $h_0\tilde\chi_1^0$ pairs as
shown in the Fig.\ref{mh0chi10}.
The background in this case is combinatorial on the one hand, and comes from SUSY production
of $t \bar t$ pairs with :
$t \bar t \rightarrow b + W^{+} + \bar b + W^{-} \rightarrow b + \bar b + l^{+} + l^{, -} + 
\nu_l + \bar \nu_{l^{,}} $ on the other hand.
 However, the strong correlation required separately between the leptons and between the b jets
will supress it considerably.

\begin{Fighere}
\centering
  \mbox{\epsfig{file=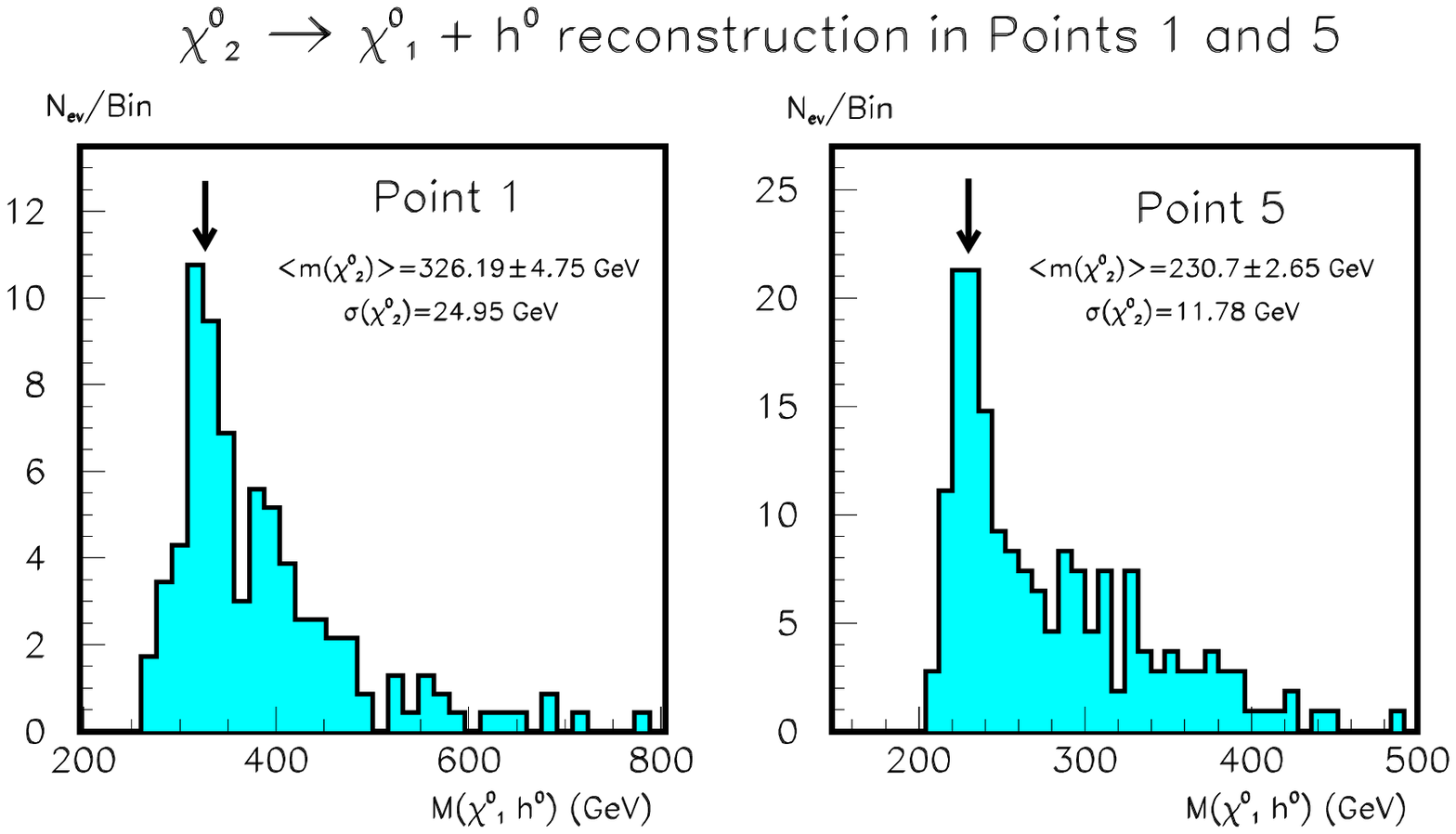,width=14cm  }}
  \vskip -7.0 cm
\caption{\small The invariant mass distribution of the $h_0\tilde\chi_1^0$
pairs for 3 years of LHC run at low luminosity in points 1 and 5. 
The tail is due to bad combinations with the wrong $\tilde \chi^0_1$.
The arrows point to the theoretical values of $m_{\tilde \chi^0_2}$.
The masses are determined by a gaussian fit around the peak. 
The error is only the statistical.}
\label{mh0chi10}
\end{Fighere}
\vskip 1.0 cm

The peak around the $\tilde\chi_2^0$
mass value (c.f. Table~\ref{tb:masses}) allows to estimate this latter
by fitting a Gaussian curve on the peak:
\begin{equation}
m_{\tilde\chi_2^0}^{meas} = 326.2 \pm 6 \mbox{\ GeV at point 1}
\label{eq:mchi20p1}
\end{equation}
\begin{equation}
m_{\tilde\chi_2^0}^{meas} = 230.7 \pm 3.9 \mbox{\ GeV at point 5}
\label{eq:mchi20p5}
\end{equation}

\vskip 1.5cm
{\it \bf Reconstruction of $\tilde \chi^{\pm}_1 \rightarrow \tilde \chi^0_1 + W^{\pm}$}
\vskip 0.2cm

The same technique allows the reconstruction of the lightest chargino ($\tilde \chi^{\pm}_1$).
As one can see in the scheme (*) one reconstructs
first the $W^{\pm}$.
The $W$ reconstruction is carried out by combining light quark jet pairs 
using the additional selection criteria (to ($i \div iii$) and ($vi$)): \\
\indent \indent $(ix)$ $p_t^j \ge $ 100 GeV and $p_t^j \le $ 600 GeV (point 1) or 350 GeV (point 5) \\
\indent \indent $(x)$ cos($\alpha_{jj}$) $\ge$ 0.9 (point 1) or 0.87 (point 5) \\
where $\alpha_{jj}$ is the angle between the $jj$ pair.
The obtained invariant mass distributions of the $jj$ pairs are shown
in Fig.\ref{mcha1}. There is a clear mass peak at the place of the $W$ mass.
Selecting the $W$ candidates around the $W$ mass peak ($\pm 15$ GeV)
and using the $\tilde\chi_1^0$ candidates close to the
endpoint (see $(vi)$) we plot
the invariant mass distribution of the $\tilde\chi_1^0 W$ pairs
in Fig.\ref{mcha1}, by requiring that the $\tilde\chi_1^0$ and the $ W$
be close in phase space: \\
\indent \indent $(xi)$ cos($\alpha_{W\tilde\chi_1^0}$) $\ge$ 0.85 in both points \\ 
where $\alpha_{W\tilde\chi_1^0}$ is the angle between the $\tilde\chi_1^0 W$ pair.

\begin{Fighere}
\centering
  \mbox{\epsfig{file=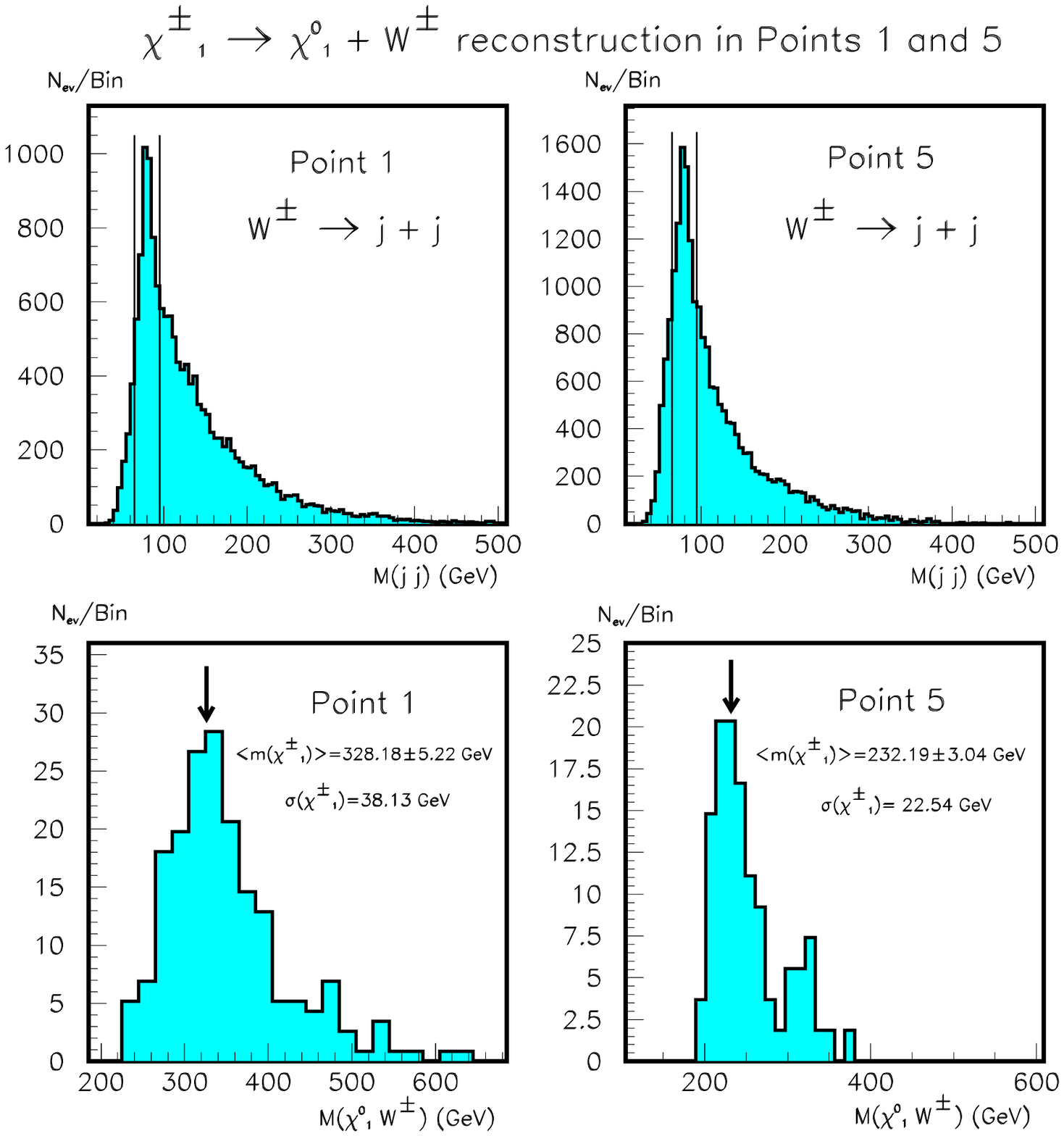,width=14.cm }}
\caption{\small The invariant mass distribution of the $jj$
pairs for 3 years of LHC run at low luminosity at point 1 
and 5 (upper plots). Selecting W candidates arround $W$ mass peak ($\pm 15$ GeV
- between the bars), we combine them with the $\tilde \chi^0_1$ candidates and obtain the
distributions represented in the lower plots for point 1 and 5 respectively.
The arrows point to the theoretical values for $m_{\tilde \chi^{\pm}_1}$.}
\label{mcha1}
\end{Fighere}
\vskip 1.0 cm

One can see a clear peak at the mass value of the lightest chargino at point 5
(c.f. Table\ref{tb:masses}). At point 1, where the statistics is much more
limited due to the heaviness of the chargino, the corresponding peak
is less clear. We have determined the mass of the chargino by fitting a
Gaussian around the mass peaks which gives:
\begin{equation}
m_{\tilde\chi_1^{\pm}}^{meas} = 328.2 \pm 6.5  \mbox{\ GeV at point 1}
\label{eq:mchap1}
\end{equation}
\begin{equation}
m_{\tilde\chi_1^{\pm}}^{meas} = 232.2 \pm 4.5  \mbox{\ GeV at point 5.}
\label{eq:mchap5}
\end{equation}

\vskip 1.5cm
 {\it \bf Reconstruction of $\tilde q_{R} \rightarrow \tilde \chi^{0}_{1} + jet$ }
\vskip 0.2cm

A competitive cascade corresponding to this decay in (*) is :\\

\begin{tabbing}
oooooooooooooooooooooo\=oooooooooooo \kill

\>( \ \ $\tilde g \longrightarrow$  \= $\tilde q_R  + q$ \ \ )  \\
\> \>
\begin{picture}(11,10)(0,0)
\put(2.0,9.0){\line(0,-2){8}}
\put(2.0,1.0){\vector(1,0){9}}
\end{picture}
\= $\tilde \chi^0_1  + q$ \indent \indent \indent \indent \indent (* *)  \\
\> \>\>
\begin{picture}(11,10)(0,0)
\put(2.0,9.0){\line(0,-2){8}}
\put(2.0,1.0){\vector(1,0){9}}
\end{picture}
$l^{\pm}$ +  $l^{\mp}$  +  $\nu$  \\ 
\end{tabbing}

Due to the fact that the squarks are very heavy, the jet produced
in the $\tilde q_R$ decay will carry an important fraction of energy. 
To avoid high combinatorial background one can ask only for one very energetic 
light jet in the event.

Applying the following additional selection criteria (to ($i \div iii$) and ($vi$))
for the light quark jets: \\
\indent \indent $(xii)$ One light jet with $p_t^j \ge 750$ GeV  (point 1) or 400 GeV (point 5) whose\\
\indent \indent \ \ \ \ \ \ \ invariant mass with any other light jet is outside the\\
\indent \indent \ \ \ \ \ \ \ $W^{\pm}$ or $Z^0$ masses ($\pm 15$ GeV);\\
\indent \indent $(xiii)$ cos($\alpha_{j\tilde\chi_1^0}$) $\ge 0.0$  in both points  \\
where $\alpha_{j\tilde\chi_1^0}$ is the angle between the $j \tilde\chi_1^0$ pair.

\begin{Fighere}
\centering
  \mbox{\epsfig{file=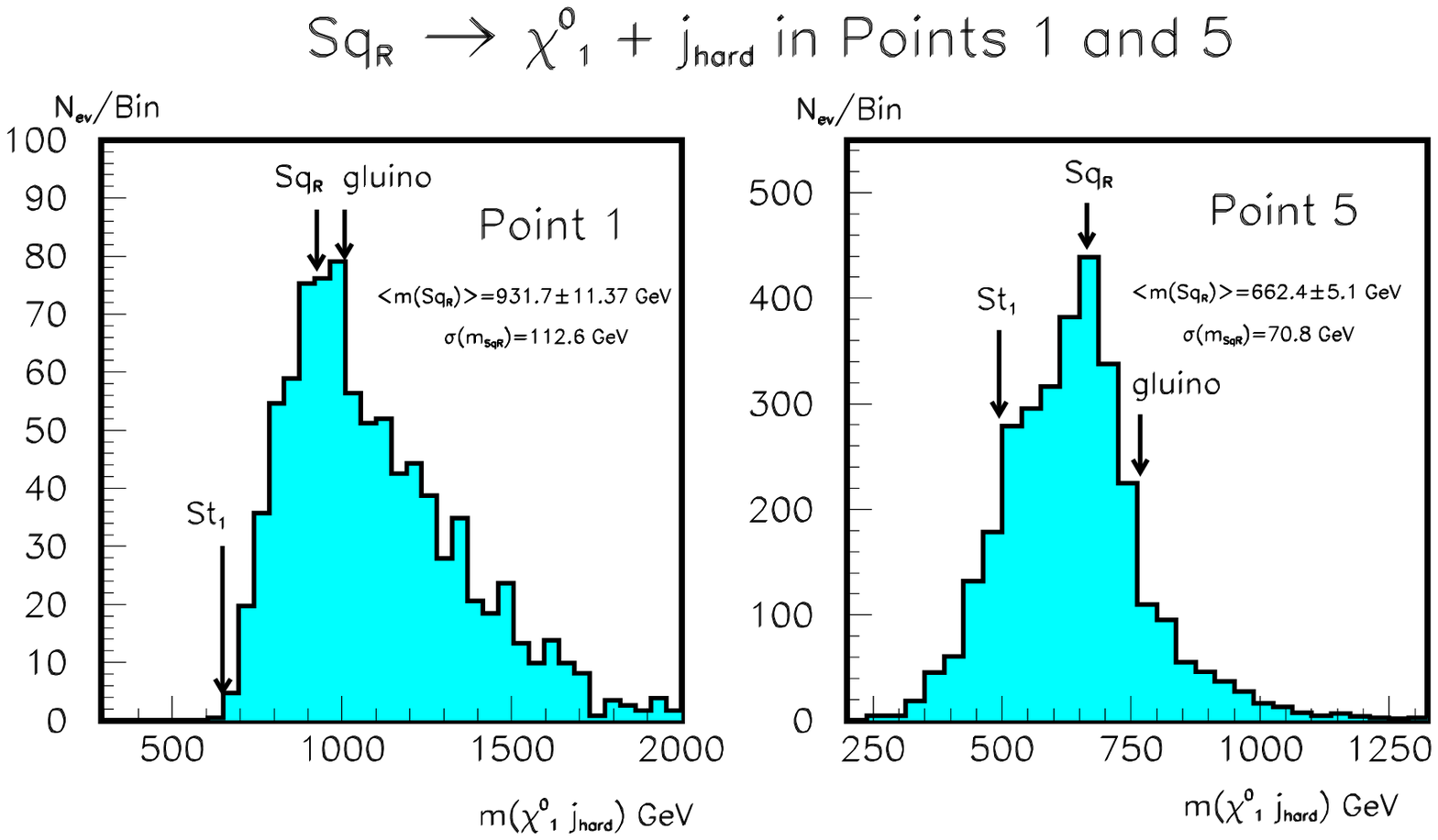,width=14cm  }}
  \vskip -0.5 cm
\caption{\small The invariant mass distribution of the $j\tilde\chi_1^0$
pairs for 3 years of LHC run at low luminosity at point 1
and at point 5. The vertical arrows point to the nominal values of masses for different sparticles.}
\label{p1p5_sqr}
\end{Fighere}
\vskip 0.5 cm

The obtained invariant mass distributions of the $j\tilde\chi_1^0$ pairs are shown
in Fig.\ref{p1p5_sqr}.
The $\tilde q_R$ can be produced directly in the hard process or in the
decay of $\tilde g$. In points 1 and 5 the jet associated with $\tilde q_R$ is very soft
comparing to the jet associated with $\tilde \chi^0_1$ and often the soft jet can fall in the
cone of the reconstructed hard jet. This is the reason of the presence of a shoulder at the
$\tilde g$ mass value in the distributions of Fig.\ref{p1p5_sqr}.
Another source of background is the decay 
$\tilde q_L \rightarrow \tilde \chi^{\pm}_1 + q$ with 
$\tilde \chi^{\pm}_1 \rightarrow \tilde \chi^0_1 + W^{\pm} \rightarrow \tilde \chi^0_1 + \nu_l + l^{\pm}$
and $l^{\pm}$ undetected. 
The left shoulder at the $\tilde t_1$ mass at point 5 corresponds to the case where the 3 jets
of the $\tilde t_1 \rightarrow \tilde \chi^{+}_1 + b \rightarrow \tilde \chi^0_1 + j + j' + b$
or $\tilde t_1 \rightarrow \tilde \chi^{0}_1 + t \rightarrow \tilde \chi^0_1 + j + j' + b$
decay chains are erroneosely combined into a single energetic one.
One can see mass peaks at the values of the right handed
squarks (c.f. Table\ref{tb:masses}). The obtained mass values will
have large errors due to the superposition of all these effects :
\begin{equation}
m_{\tilde q_R}^{meas} = 932 \pm 20 \mbox{\ GeV at point 1}
\label{eq:msqrp1}
\end{equation}
\begin{equation}
m_{\tilde q_R}^{meas} = 662 \pm 12  \mbox{\ GeV at point 5}
\label{eq:msqrp5}
\end{equation}

\vskip 0.5cm
 {\it \bf Reconstruction of $\tilde q_{L} \rightarrow \tilde \chi^{0}_{2} + jet$ in point 5}
\vskip 0.2cm

For the point 5 we can do more due to the higher statistics.
As we already mentioned in the case of $\tilde q_R$, the $\tilde q_L$ can decay with
a branching fraction of about 31 \% in $\tilde \chi^0_2$ and a light jet. This jet
is also very energetic.
In addition to the cuts $(i \div viii)$ one will select $\tilde \chi^0_2$ candidates from the 
$\tilde \chi^0_1 h^0$ distribution :\\
\indent \indent $(xiv)$ $m(\tilde \chi^0_1 h^0) \in (m_{\tilde \chi^0_2}-40,m_{\tilde \chi^0_2}+40)$ GeV \\
and only one hard light-flavoured jet with :\\
\indent \indent $(xv)$ $p_t^{jet} \geq 100$ GeV whose invariant mass with any other 
light jet is $\pm 15$ GeV \\
\indent \indent \ \ \ \ \ \ outside of the $W^{\pm}$ or $Z^0$ masses.\\

\vskip -0.5 cm
\begin{Fighere}
\centering
  \mbox{\epsfig{file=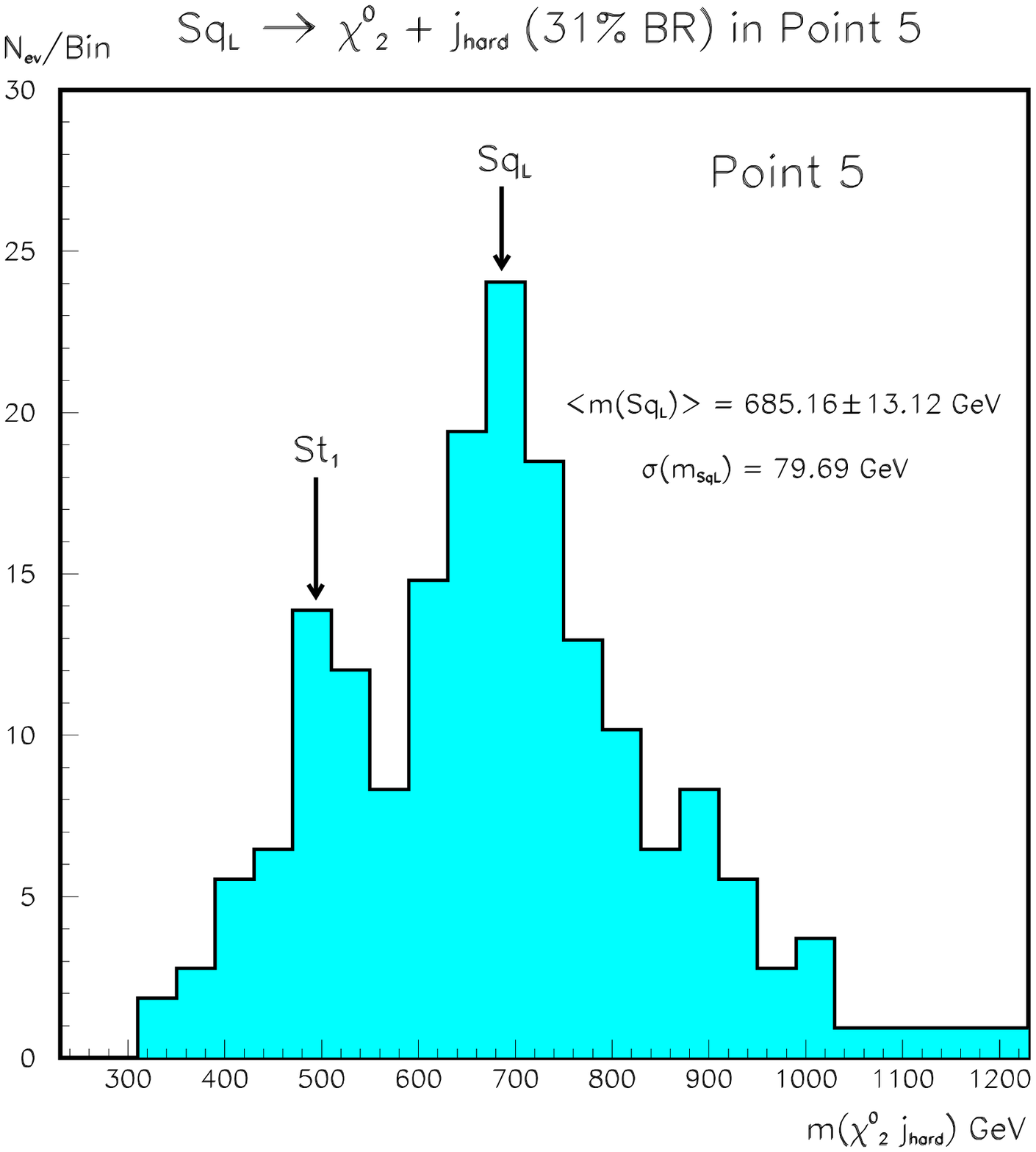,width=9.0 cm  }}
\caption{\small The invariant mass distribution of the $j\tilde\chi_2^0$
pairs for 3 years of LHC run at low luminosity at point 5. Vertical arrows point to 
the nominal values of $\tilde q_L$ and $\tilde t_1$ masses. For cuts see text.}
\label{p5_sql}
\end{Fighere}

The decay $\tilde t_1 \rightarrow \tilde \chi^0_2 + t \rightarrow \tilde \chi^0_2 + j + j + b$ 
when the 3 jets are erroneusely combined into a single hard jet, manifests itself as a peak
at the $\tilde t_1$ mass (see Fig.\ref{p5_sql}).
The gaussian fits to these peaks, after three years of LHC run at low luminosity, result in :\\

\begin{equation}
m_{\tilde q_L}^{meas} = 685 \pm 20  \mbox{\ GeV at point 5}
\label{eq:msqlp5}
\end{equation}
\begin{equation}
m_{\tilde t_1}^{meas} = 504 \pm 20  \mbox{\ GeV at point 5.}
\label{eq:mst1p5}
\end{equation}

\vskip 0.5cm
 {\it \bf Reconstruction of $\tilde l^{\pm}_{R} \rightarrow \tilde \chi^{0}_{1} + l^{\pm}$ in point 5}
\vskip 0.2cm

In the point 5, the $\tilde l^{\pm}_{R}$ decays $\sim$ 100 \% in 
$\tilde \chi^{0}_{1}$ and $l^{\pm}$.
In order to obtain the mass of the right handed slepton, we have combined
the $\tilde\chi_1^0$ candidates with a "soft" lepton not participating in the
$\tilde\chi_1^0$ reconstruction and satisfying: \\
\indent \indent $(xvi)$ $ 10  \le p_t^l \le $ 200 GeV   \\
\indent \indent $(xvii)$ cos($\alpha_{l\tilde\chi_1^0}$) $\ge$ 0.5   \\
where $\alpha_{l \tilde\chi_1^0}$ is the angle between the $l \tilde\chi_1^0$ pair.
The obtained invariant mass distribution is shown in Fig.\ref{p5_slr}.
Besides the combinatorial one, the background arises from the decay
$\tilde \chi^{\pm}_1 \rightarrow \tilde \chi^0_1 + W^{\pm} \rightarrow \tilde \chi^0_1 + l^{\pm} + \nu_l$,
which results in an {\it endpoint}
in the $l^{\pm} \tilde \chi^0_1$ distribution at the mass of $\tilde \chi^{\pm}_1$.

\begin{Fighere}
\centering
  \mbox{\epsfig{file=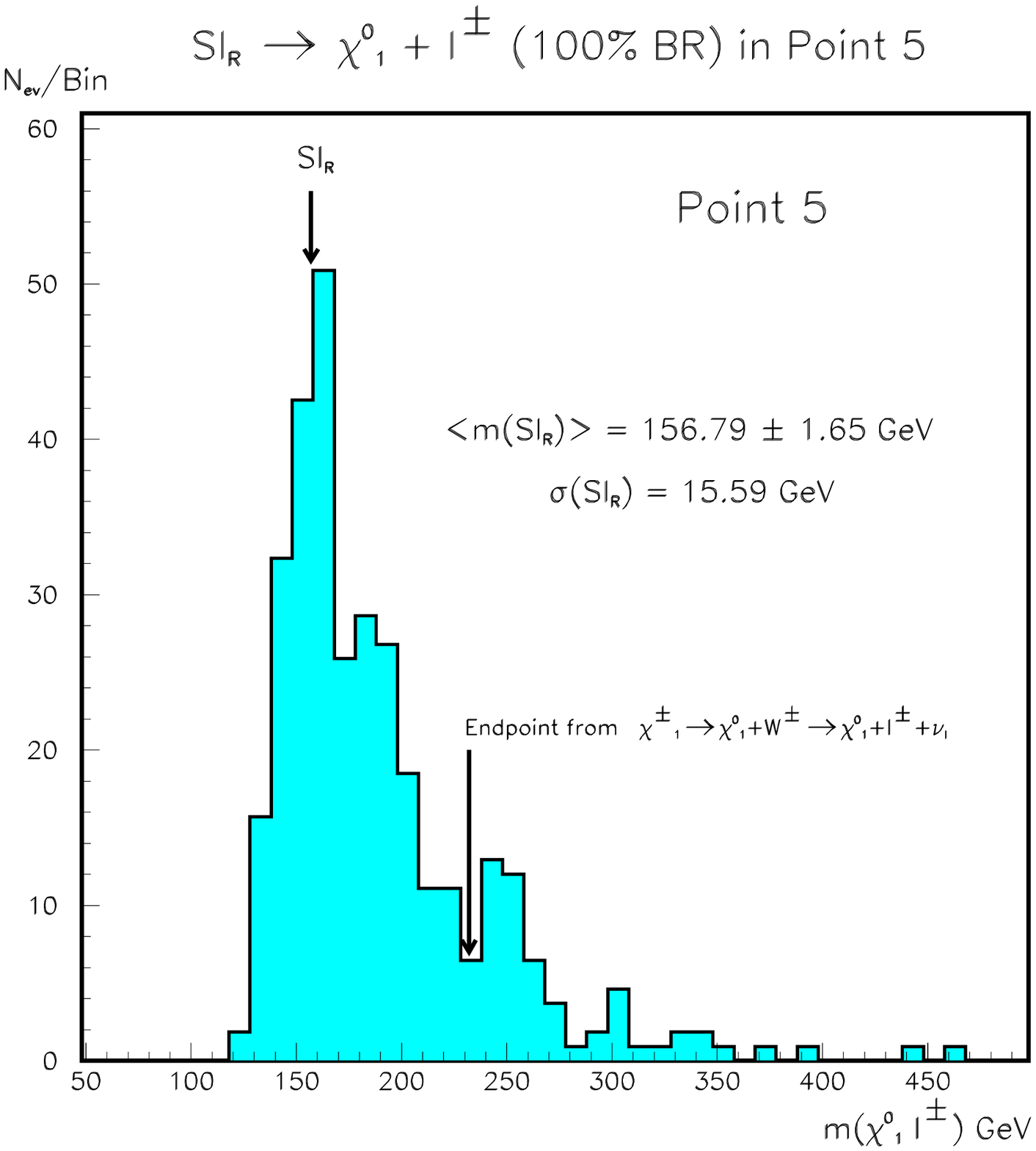,width=9.5cm  }}
\caption{\small The invariant mass distribution of the $l\tilde\chi_1^0$
pairs for 3 years of LHC run at low luminosity at point 5.
The arrows point to the nominal values of $m_{\tilde l_R}$ and $m_{\tilde \chi^{\pm}_1}$. }
\label{p5_slr}
\end{Fighere}

A gaussian fit to the mass peak gives :

\begin{equation}
m_{\tilde l_R}^{meas} = 156.8 \pm 1.8  \mbox{\ GeV at point 5}
\label{eq:mslrp5}
\end{equation}
which agrees with the value of the right handed slepton (c.f. Table\ref{tb:masses}).


\vskip 1.5cm
{\underline{\bf The case $\lambda_{123} \ne 0$}}
\vskip 0.2cm

As stated earlier, in this case there is always a $\tau$
particle among the decay products of the $\tilde\chi_1^0$,
and this spoils the endpoint in the OSDF lepton pair
mass distribution which would permit us to reconstruct $\tilde\chi_1^0$
and all the other sparticles further (see Fig.\ref{p5122on123} and
compare with the upper left plot of Fig.\ref{p5chi01}).
Therefore, the strategy in the case of $\rlap/R$ couplings implying the third
family of leptons must be changed, and consequently we shall {\it abandon the
direct reconstruction of $\tilde\chi_1^0$, and return to
the strategy developed in the case of conserved R parity}.
In other words, we shall discard the decay products of $\tilde\chi_1^0$
in the first stage of the reconstruction. \\

\begin{Fighere}
\centering
  \mbox{\epsfig{file=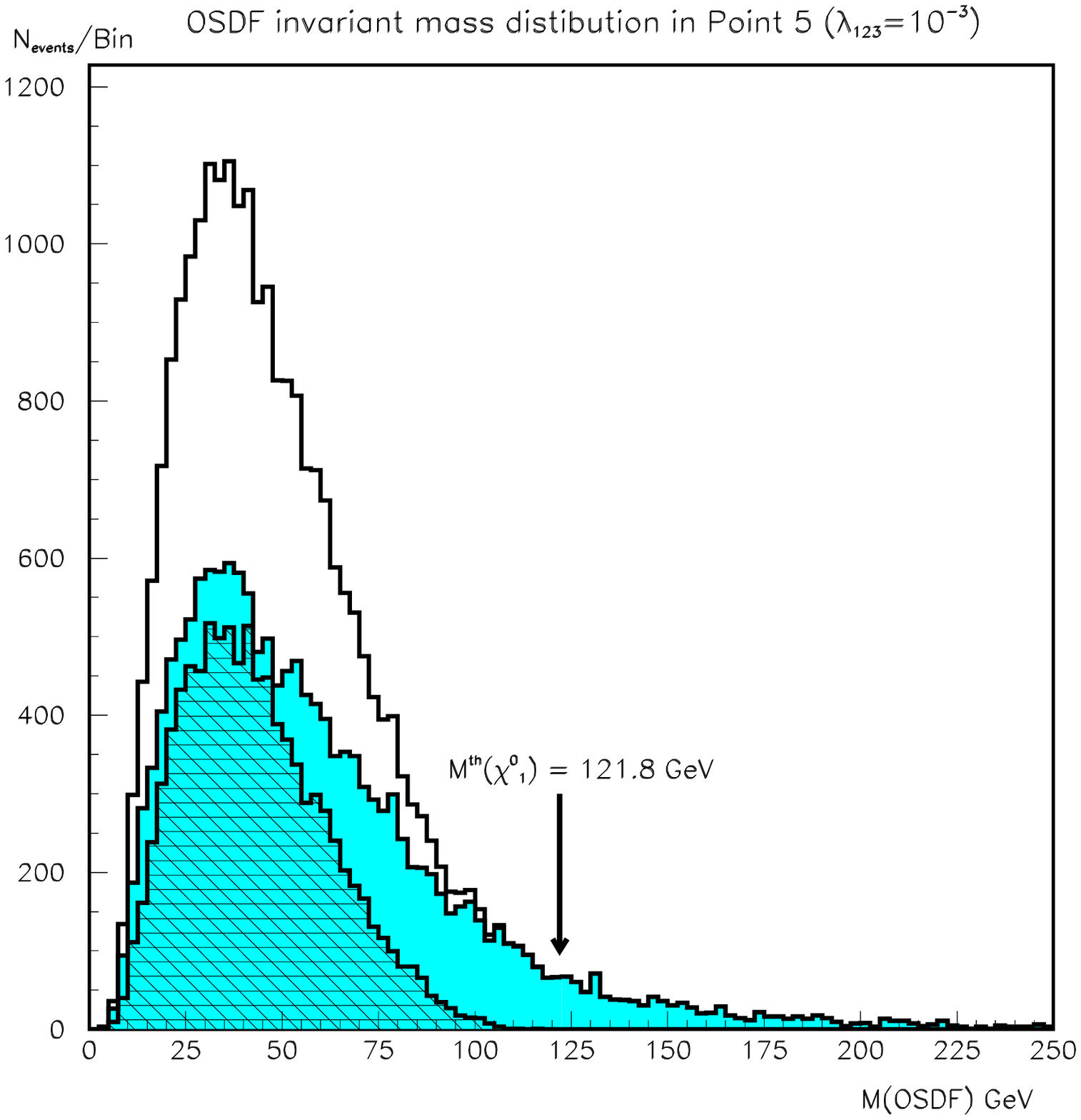,width=10cm  }}
\caption{\small The invariant mass distribution of OSDF leptons in the case of
$\rlap/R$ coupling $\lambda_{123}=10^{-3}$ for 3 years of LHC run at low luminosity
at point 5.  {\it Open histogram} - all events;
{\it shadded histogram} - background and {\it cross hatched histogram} -signal.
The arrow points to the nominal value of $\tilde \chi^0_1$ mass.
The presence of a $\tau$ lepton in the $\tilde \chi^0_1$ decay will spoil the 
endpoint at the mass of $\tilde \chi^0_1$. }
\label{p5122on123}
\end{Fighere}
 
\vskip 1.0cm
{\it \bf Reconstruction of 
$\tilde \chi^0_2 \rightarrow \tilde l^{\pm}_R + l^{\mp} \rightarrow 
\tilde \chi^0_1 + l^{\mp} + l^{\pm}$ 
($\tilde \chi^0_1 \rightarrow \nu_{e(\mu)}+\mu^{\pm}(e^{\pm})+\tau^{\pm}$)
chain}
\vskip 0.2cm

The decay of the $\tilde \chi^0_2$ produces in this case at least
three isolated leptons. 
With these leptons one can form pairs of opposite sign (OS) and different
flavors (DF) - typical for $\tilde \chi^0_1$ decay, or of the same flavor (SF)
- typical for $\tilde \chi^0_2$.
As one mentioned before, $\tilde \chi^0_2$ will produce firstly an OSSF pair
of leptons through a double 2-body decay and in a second stage, 
$\tilde \chi^0_1$ will produce again one or two leptons.
In the conserved $R$ parity case one has used the OSSF combinations.
Typically, the lepton associated with $\tilde l_R$ has
a higher $p_t$ than the lepton associated with $\tilde \chi^0_1$ 
as it can be seen from Fig.\ref{p5123ptlep}. 
In these conditions, it is useful to impose, beyond the usual cuts 
(i.e. lepton multiplicity, $E_t^{miss}$, etc), a minimal difference between 
the $p_t$ of the two leptons taken in the OSSF pairs.
To decrese the number of bad combinations with leptons coming from the subsequent $\tau$ decay,
one increases slightly the $p_t^{min}$ cut on leptons.  
The OSSF leptons, coming from different decays, are less correlated 
therefore we relax the cut in angle between them. To eliminate the
events with $\tilde \chi^0_2 \rightarrow \tilde \chi^0_1 + h^0$ we demand furthermore
that no $h^0$ has to be reconstructed in the event.

\begin{Fighere}
\centering
  \mbox{\epsfig{file=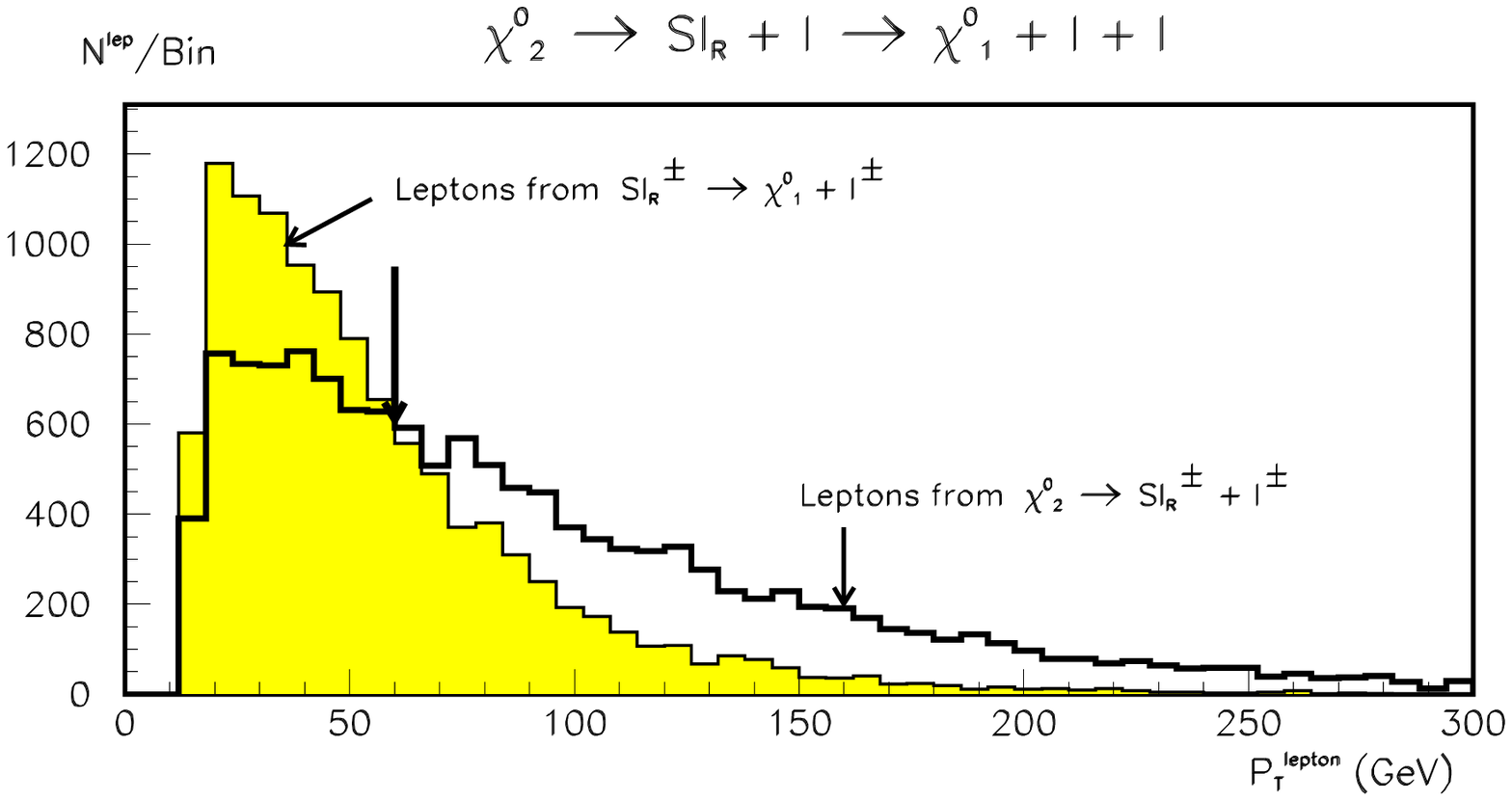,width=12cm  }}
 \vskip -6.0 cm
\caption{\small The $p_t$ distributions of the leptons 
associated with $\tilde l_R$ or $\tilde \chi^0_1$  in the 
$\tilde \chi^0_2$ decay for 3 years of LHC run at low luminosity at point 5.}
\label{p5123ptlep}
\end{Fighere}
\vskip 0.5cm

 In Fig.\ref{p5123ossf}, are presented the invariant mass distributions of 
OSSF lepton pairs, where the events were selected using the following critera: \\
\indent \indent $(i)$ $N_l \ge 4$,  \\
\indent \indent $(ii)$ $p_t^l \geq $ 15 GeV , cos($\alpha_{l^{\pm}l^{\mp}}$) $\geq$ 0,\\
\indent \indent $(iii)$ $E_t^{miss} \ge $ 50 GeV, \\
\indent \indent $(iv)$ $\Delta p_t^l = | p_t^{l_1} - p_t^{l_2} | \geq$ 60 GeV,\\
\indent \indent $(v)$ No $b b$ pair with invariant mass in $(m_{h^0}-15,m_{h^0}+15)$ GeV.\\

The huge background comes from the bad combinations with the the leptons
produced in the $\tilde \chi^0_1$ decays.
Since the $\tilde \chi^0_2$ decays into the OSSF lepton pair via two 2-body decays,
the observed endpoint in the OSSF distribution is determined by the relation of
Eq.(~\ref{eq:mend22b})
between $m_{\tilde \chi^0_2}$, $m_{\tilde l_R}$ and $m_{\tilde \chi^0_1}$.
 After the fit of the Maxwellian background and subtraction, the value of the endpoint 
is estimated as :

\begin{equation}
m_{OSSF}^{meas} = 111.9 \pm 2.5  \mbox{\ GeV}
\label{eq:mOSSFp5}
\end{equation}
where the error of about 2.5 GeV is dominated by the histogram bining (i.e. statistics).

One can isolate lepton candidates coming from the double 2-body decay of 
$\tilde \chi^0_2$ by a selection around this endpoint. 
With the remaining
leptons one can attempt to form OS lepton pairs and further combining them
with the selected OSSF pairs to reconstruct the $\tilde \chi^0_2$.
Because of the neutrinos (always present) the $\tilde \chi^0_2$
reconstruction is not complete and therefore an endpoint will appear in this 
OSSF+OS lepton invariant mass distribution depending on the $m_{\tilde \chi^0_2}$.

\vskip 0.5 cm
\begin{Fighere}
\centering
  \mbox{\epsfig{file=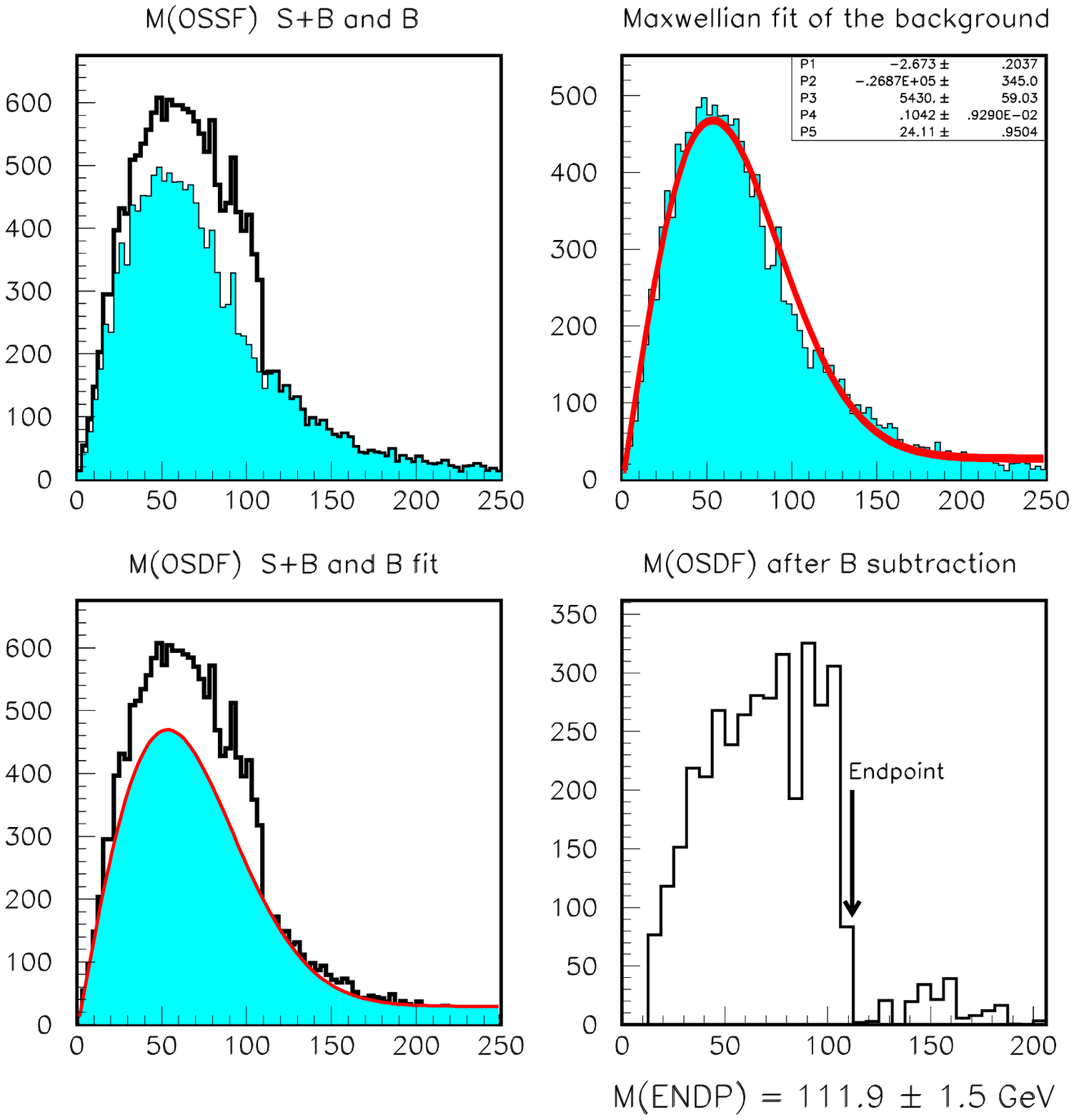,height=10.5cm,width=15cm  }}
\caption{\small The invariant mass distribution of the OSSF lepton 
pairs for 3 years of LHC run at low luminosity at point 5.
}
\label{p5123ossf}
\end{Fighere}
\vskip 0.5cm

If one applies the additional cuts to $(i \div v)$ :\\
\indent \indent $(vi)$ $m_{OSSF} \in (m^{endp}_{OSSF}-25,m^{endp}_{OSSF})$ GeV and rescaled
as explained in the introductory part of subsection 5,\\
\indent \indent $(vii)$ OS pairs constructed with leptons having $p_t^l \geq 10$ GeV,
                        $\Delta p_t^l \geq 20$ GeV and cos($\alpha_{OS}$) $\geq$ 0,\\
\indent \indent $(viii)$ cos($\alpha_{OSSF, OS}$) $\geq$ 0.85 ,\\
one obtains the distribution represented in Fig.\ref{p5123dchi0}.\\
One can observe an endpoint at:

\begin{equation}
m_{\tilde \chi^0_2} = 228.2 \pm 5  \mbox{\ GeV }
\label{eq:p5123chi02}
\end{equation}
by linear extrapolation of the distribution as is the case for a 5-body final state 
(see ~\cite{GMSB}). This endpoint corresponds to the kinematical limit of the
$\tilde \chi^0_2$ decay, namely to $m_{\tilde \chi^0_2}$.
The error of the extrapolation is reflected in the large value of the error in 
$m_{\tilde \chi^0_2}$.

\vskip 1. cm
\indent {\it \bf Reconstruction of $\tilde q_L \rightarrow \tilde \chi^0_2 + q$}
\vskip 0.2 cm

We observe that at point 5 $m_{\tilde\chi_2^0}-m_{\tilde\chi_1^0}\approx m^{endp}$
(Eq(\ref{eq:mend22b})) which means that the three-momentum of the $\tilde\chi_1^0$
is nearly zero in the $\tilde\chi_2^0$ restframe. Therefore
selecting OSSF lepton pairs near the endpoint of the distribution of Fig.\ref{p5123ossf}
one can determine $P^{\mu}_{\tilde \chi^0_2}$ using Eq.(\ref{eq:pAa}) 
with the value of $m_{\tilde \chi^0_2}$ from Eq(~\ref{eq:p5123chi02}).
These 4-momenta are then combined with a hard light jet not coming from a reconstructed
$W$ or $Z^0$. To the criteria  $(i) \div (v)$ one adds the following :\\
\indent \indent $(vi-b)$ $m_{OSSF} \in (m^{endp}_{OSSF}-30,m^{endp}_{OSSF})$ GeV and rescaled,\\
\indent \indent $(ix)$ $p_t^{jet} \geq 400$ GeV and no invariant mass with any other light jet
in $\pm$ 15 GeV \\
\indent \indent \ \ \ \ \ \ \ around $W$,\\
\indent \indent $(x)$ cos($\alpha_{j OSSF}$) $\geq$ 0. . \\
The resulting distribution is depicted in Fig.\ref{p5123sql}.
By a gaussian fit on the peak one obtains the value:

\begin{equation}
m_{\tilde q_L}^{meas} = 684 \pm 15  \mbox{\ GeV}
\label{eq:p5123sql1}
\end{equation}
\vskip -1.0 cm
\begin{Fighere}
\centering
  \mbox{\epsfig{file=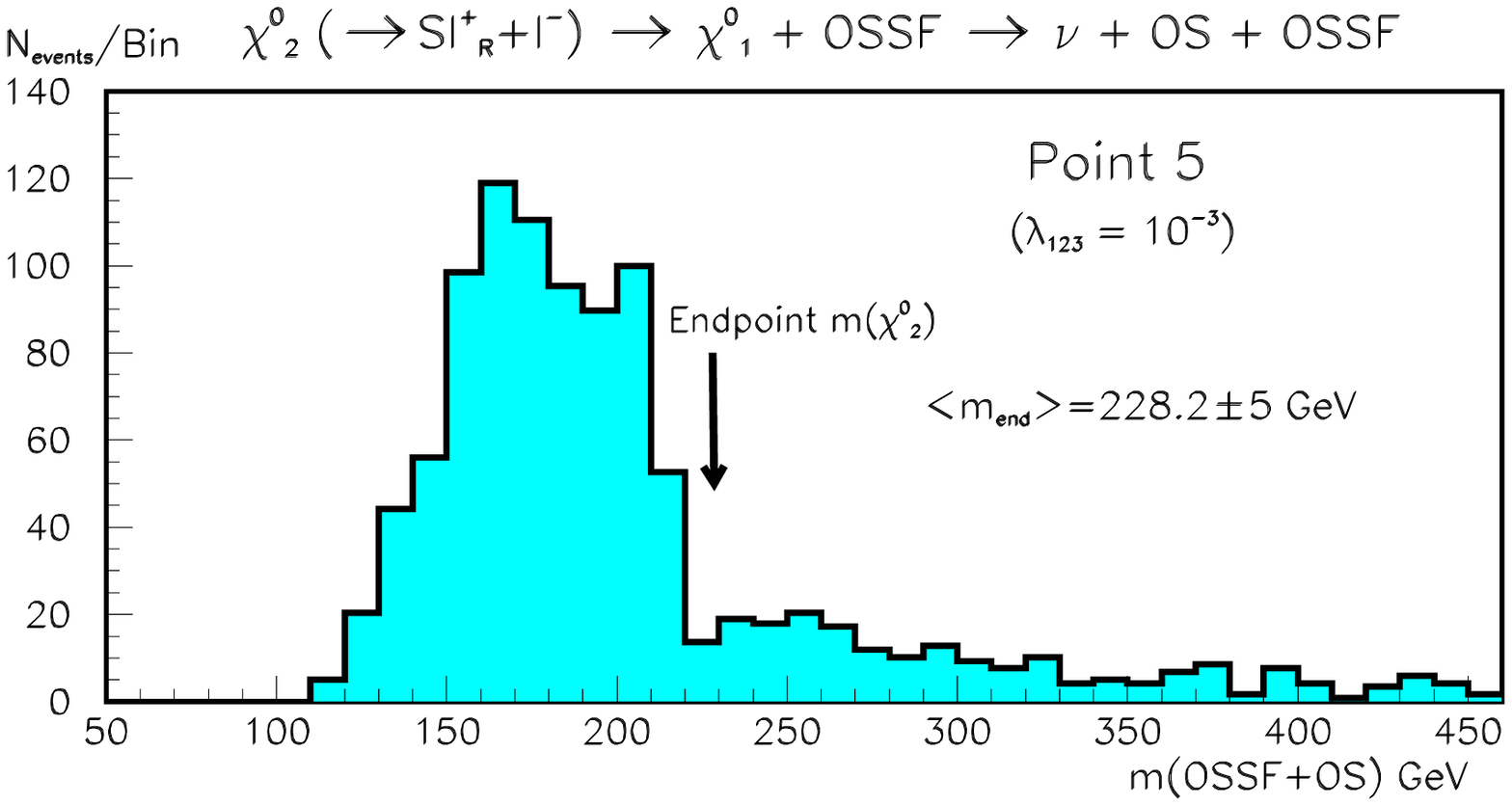,height=11cm,width=13cm  }}
  \vskip -5. cm
\caption{\small The invariant mass distribution of the OSSF+OS lepton 
pairs for 3 years of LHC run at low luminosity at point 5.
Due to the presence of neutrinos (from $\tilde \chi^0_1$ decay) an endpoint 
depending on $m_{\tilde \chi^0_2}$ and $m_{\tilde \chi^0_1}$ appears (see text).}
\label{p5123dchi0}
\end{Fighere}

\vskip -0.5 cm
\begin{Fighere}
\centering
  \mbox{\epsfig{file=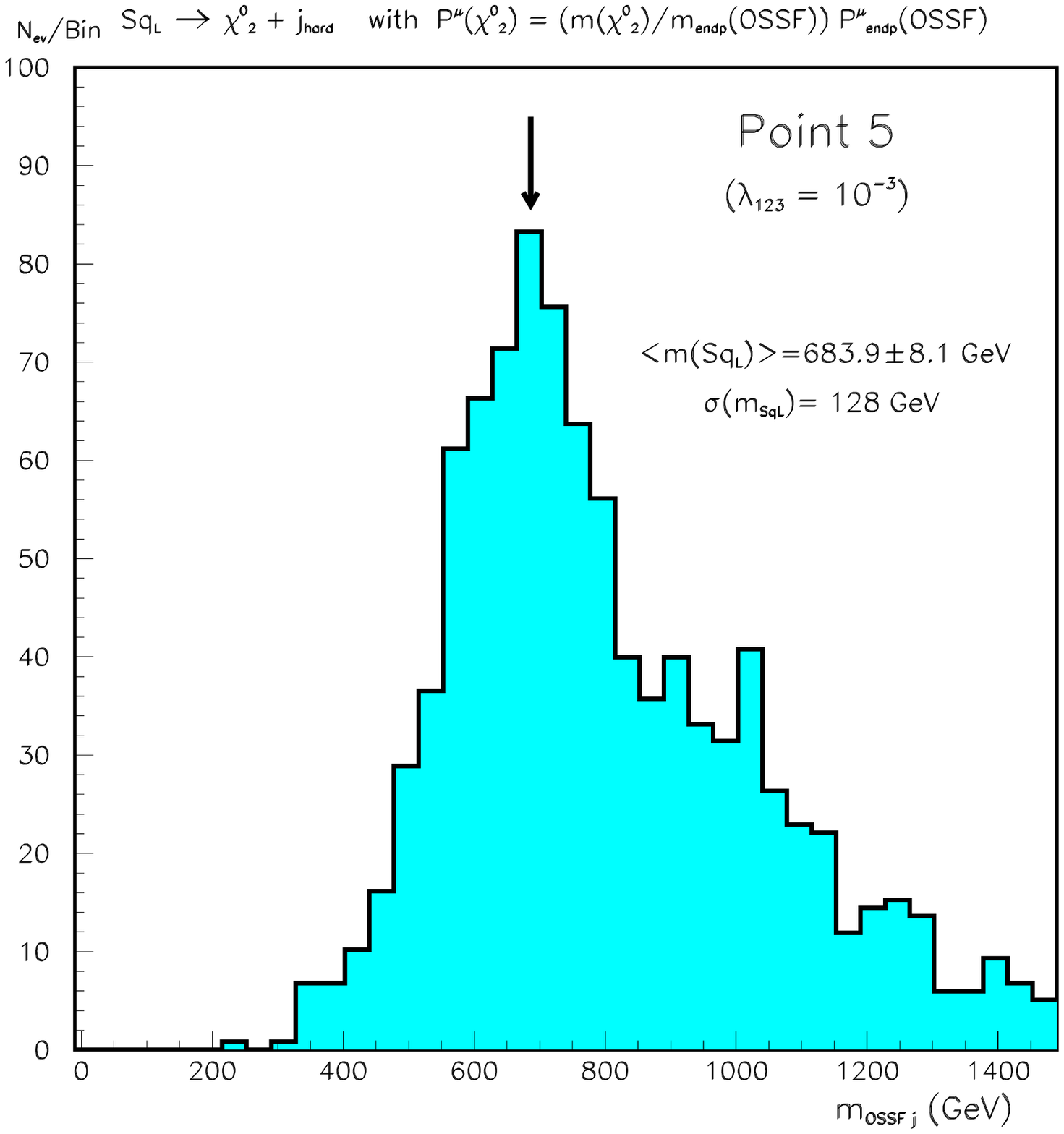,width=9cm  }}
\vskip -0.5 cm
\caption{\small The invariant mass distribution of the (jet OSSF)  
pairs for 3 years of LHC run at low luminosity at point 5.
The arrow points to the nominal value of $\tilde q_L$ mass. The mean value is obtained by 
a local gaussian fit on the main peak.}
\label{p5123sql}
\end{Fighere}
\vskip 0.5cm

$m_{\tilde q_L}$ can also be determined by combining the (OSSF+OS) pairs selected near
the $m_{\tilde \chi^0_2}$ endpoint of Fig.\ref{p5123dchi0}, whose 4-momenta are approximately equal to
$P^{\mu}_{\tilde \chi^0_2}$ according to Eq.(~\ref{eq:pAa}), with a hard light jet
not coming from a reconstructed $W$ or $Z^0$. For that one will use
the cuts $(i \div ix)$ supplemented by:\\
\indent \indent $(xi)$  $m_{\tilde \chi^0_2} \in (m^{endp}_{OSSF,OS}-50,m^{endp}_{OSSF,OS})$ GeV,\\
\indent \indent $(xii)$ cos($\alpha_{j OSSF-OS}$) $\geq$ 0 . \\
The invariant mass distribution is shown in Fig.~\ref{p5123sql2}. One obtains the value:
\begin{equation}
m_{\tilde q_L}^{meas} = 686 \pm 12  \mbox{\ GeV.}
\label{eq:p5123sql2}
\end{equation}
The agreement between the results of (~\ref{eq:p5123sql1}) and (~\ref{eq:p5123sql2}) justifies the 
procedure that we have applied in obtaining the mass of $\tilde \chi^0_2$.

\vskip 0.5 cm
\begin{Fighere}
\centering
  \mbox{\epsfig{file=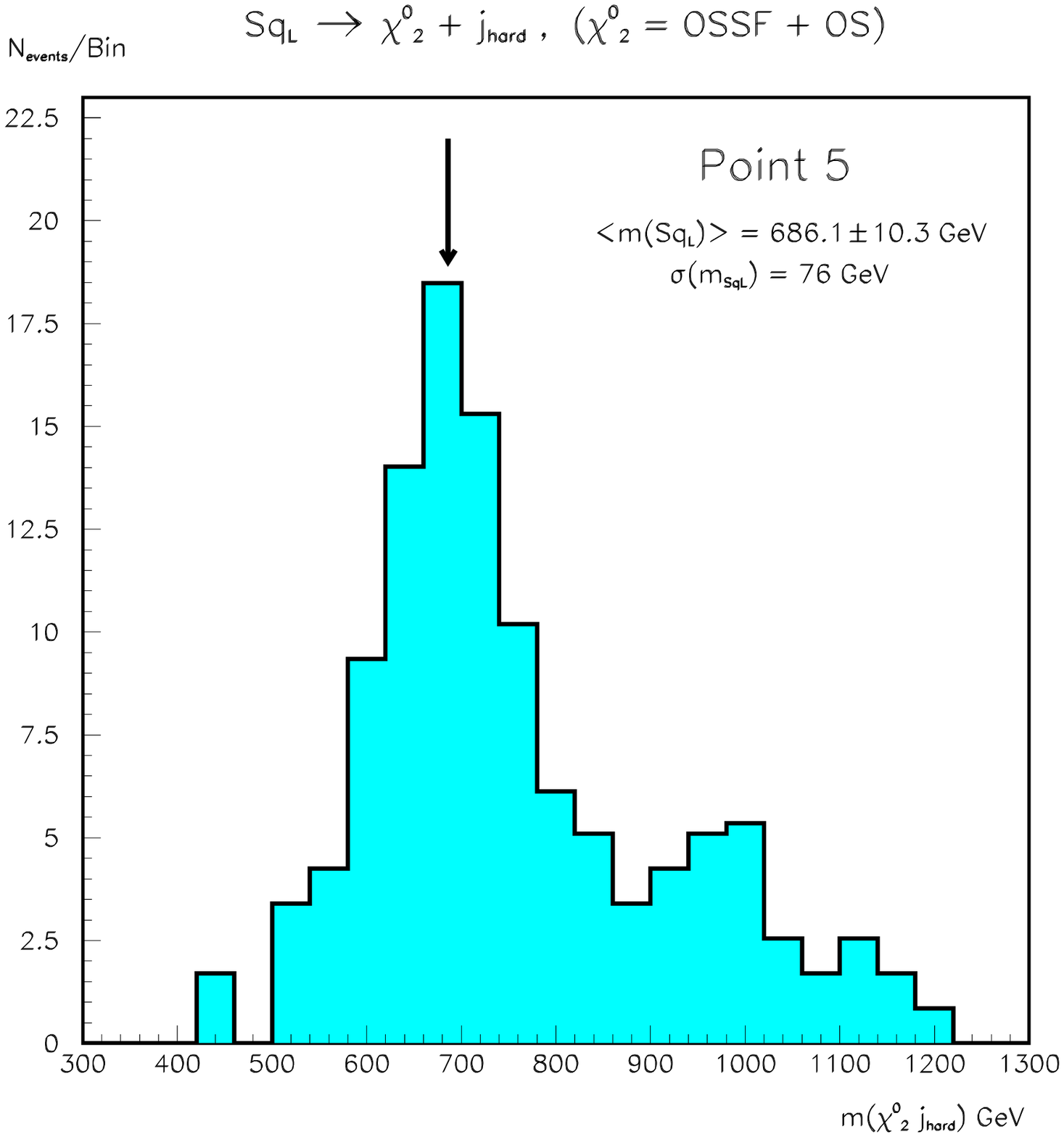,width=10cm }}
\caption{\small 
The invariant mass distribution of the (jet OSSF-OS) pairs 
for 3 years of LHC run at low luminosity at point 5. 
The arrow points to the nominal value of $\tilde q_L$ mass.
  }
\label{p5123sql2}
\end{Fighere}
\vskip 0.5cm

\vskip 1.0 cm
\indent {\it \bf Reconstruction of $h^0 \rightarrow b + \bar b$}
\vskip 0.2 cm

The reconstruction of the $h^0$ can be carried out almost with the same precision
as it was the case of the coupling $\lambda_{122}$. 
However, the applied isolation criteria reject more events due to the finite size
of the $\tau$ jets as compared to the  case of $\lambda_{122}$. 
This can be seen in the $b \bar b$ invariant mass
distributions which should be independent of the $\rlap/R$ coupling type.
Following the same procedure as in the case $\lambda_{122}$, with the selection 
criteria slightly modified ;\\
\indent \indent $(xiii)$ 15 GeV $\leq p_t^b \leq$ 300 GeV,\\
\indent \indent $(xiv)$ cos($\alpha_{bb}$) $\geq$ 0.4 GeV,\\
 one obtains the distribution shown in Fig.\ref{p5123h0}. 
After the subtraction of the Maxwellian background and the 
gaussian fit of the peak one gets:
\begin{equation}
m_{h^0}^{meas} = 94.3 \pm 1.5  \mbox{\ GeV}
\label{eq:p5123h0mass}
\end{equation}
This number includes the systematic error, mainly due to the incertitude on the energy
scale of $b$ jets.

\vskip -1.0 cm
\begin{Fighere}
\centering
  \mbox{\epsfig{file=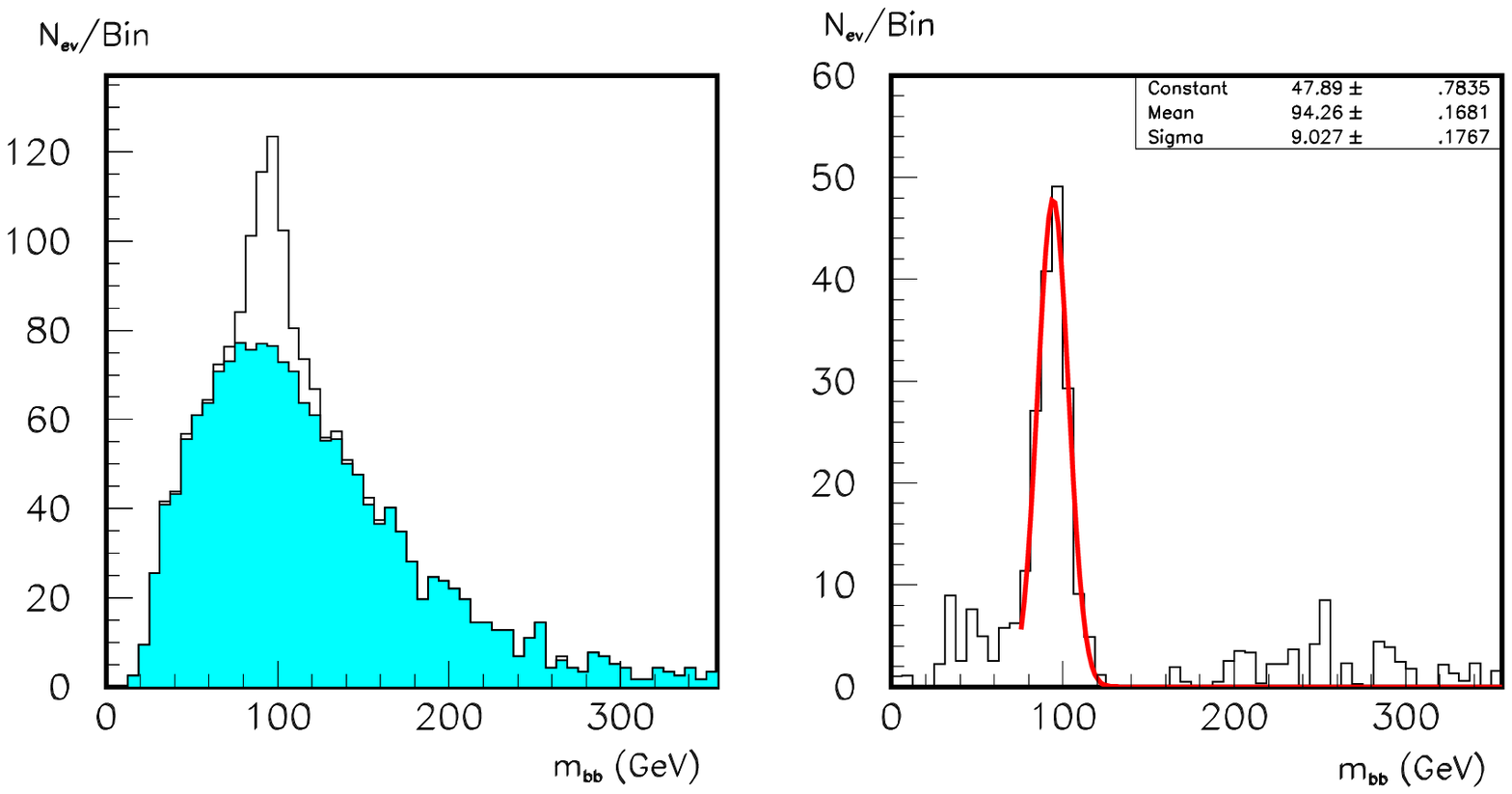,width=16cm  }}
\caption{\small The invariant mass distribution of the ($bb$)  
pairs for 3 years of LHC run at low luminosity at point 5.
}
\label{p5123h0}
\end{Fighere}
\vskip 1.0 cm


\subsection{The LHC point No3}

\vskip 1.0 cm

The salient feature of this point w.r.t. the other two
is that the SUGRA parameters $m_0$ and $m_{1/2}$ are much lower, hence 
all sparticle masses are lighter and the production
cross sections are highly increased. As a consequence we could 
generate only a fraction of the events one would be able to detect
with the ATLAS detector in one year even with low luminosity.
In addition, due to the $sign\mu=-1$ and 
$m_{\tilde\chi^0_2}-m_{\tilde\chi^0_1}\simeq 52$ GeV,
the second lightest neutralino decays into OS and SF lepton pairs and not
into a Higgs ($h^0$, with a mass $\sim$ 70 GeV). 
Therefore in this point one cannot determine the Higgs mass. 
The number of leptons in the final state being higher, the
combinatorial background is considerably larger in the
channels where the particle reconstruction involves leptons. Furthermore,
as we precised before, $\tilde \chi^0_1$ can decay 
(through $\rlap/R$ $\lambda$-type couplings) in lepton pairs of different flavors as well
as of same flavors. The consequence is a much more complicated structure
in the invariant mass distribution of the OSSF lepton pairs than in the case of conserved 
R parity.

Another feature of point 3 is that in the decay chain (*) the dominant
decay products of the gluino are the $\tilde b_1$ and the $b$. Since in the
decay of the $\tilde b_1$ the gaugino ($\tilde\chi_2^0$ or $\tilde\chi_1^{\pm}$)
is accompanied again by a $b$-quark, there is a large number of
$b$-quarks produced in each event.

In the case of conserved $R$ parity~\cite{SUGRA3} the observables which can be used
for the determination of the SUGRA parameters are {\it the functions} 
$m_{\tilde g}(m_{\tilde \chi^0_2})$, $m_{\tilde b_1}(m_{\tilde \chi^0_2})$
and the {\it difference} $\Delta m_{\tilde\chi^0} = m_{\tilde\chi_1^0}-m_{\tilde\chi_2^0}$ 
(from the invariant mass distribution of the OS and SF lepton pairs:
 $\tilde\chi_2^0\longrightarrow (l^+l^-) +\tilde\chi_1^0 $  
- see the decay chain (*) and Equ.~(\ref{eq:mAa})). 
In the $\rlap/R$ case, one can develop two strategies depending on the $\lambda$ type coupling :\\
\indent (1) \ \ For $\lambda$ couplings not producing $\tau$ jets in the LSP decays 
(i.e. $\lambda_{122}$) one will {\it directly} reconstruct the $\tilde \chi^0_1$ as in the
other LHC points. The consequence of this new information is a better fit of the SUGRA parameters
comparing with the $R$ conserved case;\\
\indent (2) \ \ For $\lambda$ couplings producing $\tau$ jets in the LSP decays
(i.e. $\lambda_{123}$) one returns to the strategy developed in the $R$ conserved case.

As in the LHC points 1 and 5, we will present distinctively the two 
representative cases: $\lambda_{122}$ and $\lambda_{123}$.


\vskip 0.8cm
{\bf \large \it \underline{The case $\lambda_{122} \ne 0$}}
\vskip 0.2cm

The analysis is focused firstly on the leptons. With these, one can form OS pairs
of the same flavor (SF) or different flavors (DF). 
Due to the facts that:\\
\indent - $\tilde \chi^0_1$ has a significantly higher production rate than the
$\tilde \chi^0_2$ and \\
\indent - only $\tilde \chi^0_1$ produces $OSDF$ lepton pairs,\\
the OSDF pairs will tag much better their origine than the OSSF ones.
Therefore, in the first stage one will reconstruct the invariant mass distribution of OSDF
leptons. This gives the $\tilde \chi^0_1$ mass through the same type of
endpoint structure as in the other LHC points. The main background in this distribution
is combinatorial. One will select OSDF lepton pairs in the
neighbourhood of this endpoint.
With the remaining leptons (originating with a higher probability from
$\tilde \chi^0_2$, but also fom $\tilde \chi^0_1$) one will form OS lepton pairs.
The endpoint in their invariant mass distribution determines 
$\Delta m_{\tilde \chi^0} = m_{\tilde \chi^0_2}-m_{\tilde \chi^0_1}$ .
Selecting OS pairs near this latter endpoint and combining them with the
$\tilde \chi^0_1$ candidates (from the OSDF distribution) one can completely reconstruct
$\tilde \chi^0_2$. Afterwards, combining $\tilde \chi^0_2$ candidates with one and
two $b$ jets one will reconstruct $\tilde b_1$ and $\tilde g$.\\

For this analysis one uses the following global cuts :\\
\indent \indent ($i$)\ \ $N_l \geq 4$, and $p_t^{e, \mu} \geq$ 10 GeV,\\
\indent \indent ($ii$) $E_t^{miss} \geq 50$ GeV.\\

\vskip 0.6 cm
{\it \bf Reconstruction of 
$\tilde \chi^0_1 \rightarrow \nu_{e(\mu)}+\mu^{\pm}(e^{\pm})+\mu^{\pm}$}
\vskip 0.2cm

Using the global cuts ($i \div ii$) and :\\
\indent \indent ($iii$) cos($\alpha_{OSDF}$) $\geq 0.85$,\\
one obtains the invariant mass distribution  of OSDF leptons as in Fig.\ref{p3122osdf}, 
where the background is mainly combinatorial.
After the fit of the Maxwellian background, subtraction and the polinomial fit of the 
resulting endpoint one obtains the $\tilde \chi^0_1$ mass :
\begin{equation}
m_{\tilde \chi^0_1}^{meas} = 44.8^{+0.1}_{-0.2}  \mbox{\ GeV}
\label{eq:p3122chi01mass}
\end{equation}

\vskip -0.5 cm
\begin{Fighere}
\centering
  \mbox{\epsfig{file=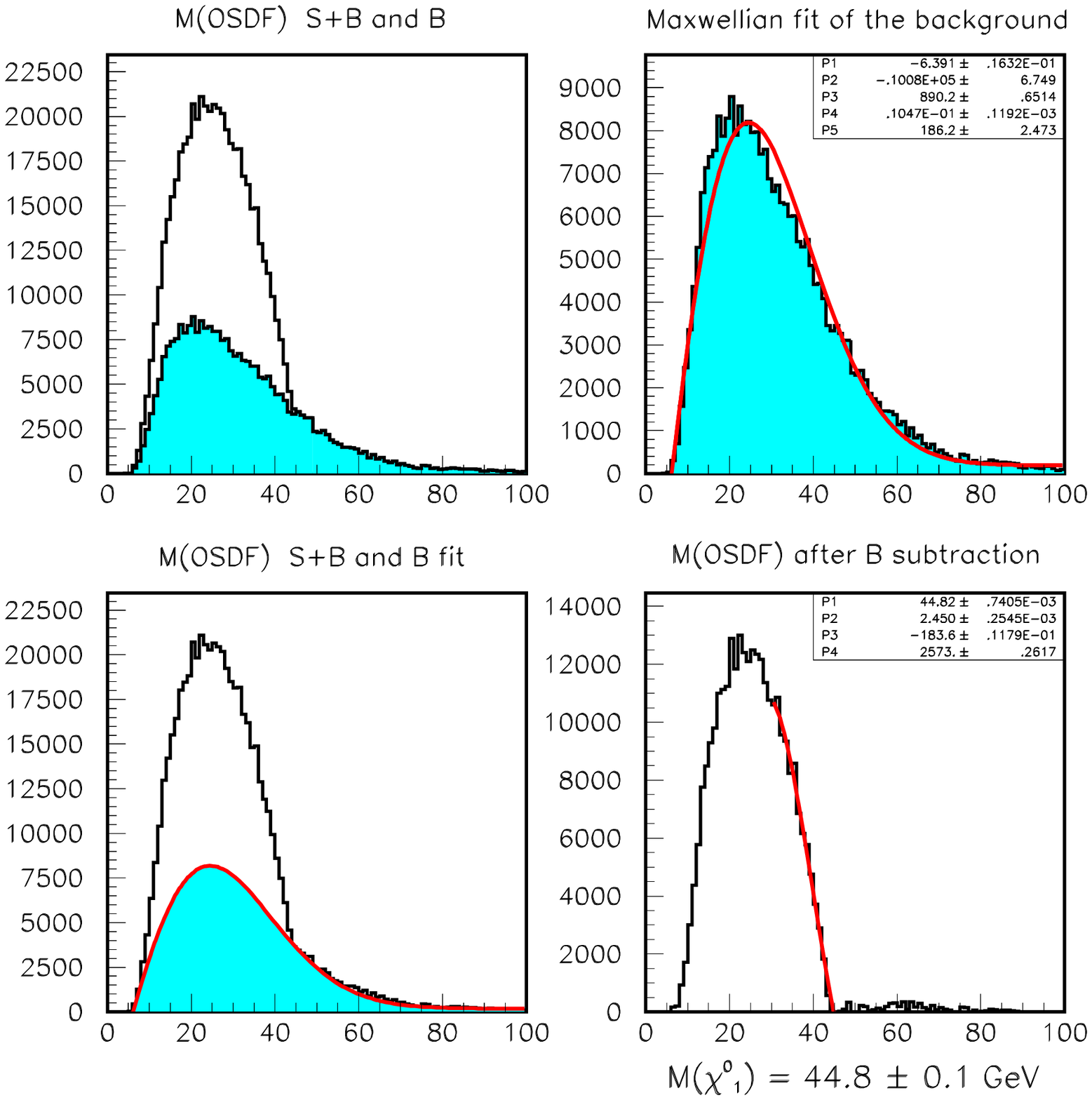,width=14cm  }}
\caption{\small The invariant mass distribution of the OSDF lepton 
pairs for 1 year of LHC run at low luminosity at point 3 in the case of 
$\rlap/R$ coupling $\lambda_{122}=10^{-3}$.
}
\label{p3122osdf}
\end{Fighere}
\vskip 0.5 cm

\vskip 0.6 cm
{\it \bf Reconstruction of 
$\tilde \chi^0_2 \rightarrow \tilde \chi^0_1 + \mu^{\pm}(e^{\pm})+\mu^{\mp}(e^{\mp})$}
\vskip 0.2cm

The $\tilde \chi^0_1$ candidates are identified by requiring OSDF lepton pairs with :\\
\indent \indent ($iv$) 
$m(OSDF) \in (m_{\tilde \chi^0_1}^{meas}-10, m_{\tilde \chi^0_1}^{meas})$ GeV,\\
with rescaled 4-momenta according to Eq.(\ref{eq:pAa}).
One then selects leptons produced directly from the 3-body decay 
of $\tilde \chi^0_2$ among the remaining OS lepton pairs demanding that the angle
between the leptons ($\alpha_{OS}$) satisfies :\\
\indent \indent ($v$) cos($\alpha_{OS}$) $\geq 0.5$ .\\
The invariant mass distribution of these OS pairs (Fig.\ref{p3122os}) has a complex structure
due to the fact that,
beside the combinatorial background (Maxwellian type with a long tail) there is also a contribution from
the $\tilde \chi^0_1$ decay itself (in fact the contribution is 50\% of its branching ratio and it has
the shape comparable with that of OSDF distribution). Therefore the OS distribution shows 
a sharp edge at about 52 GeV, correspondig to the mass difference 
$m_{\tilde \chi^0_2} - m_{\tilde \chi^0_1}$
and a second one, less pronounced around 45 GeV, corresponding to the $\tilde \chi^0_1$ 
decay products. The tail beyond 52 GeV is due to the combinatorial background 
and we have removed it. The polinomial fit of the higher endpoint gives the value:
\begin{equation}
m^{meas}_{OS} = m_{\tilde \chi^0_2}-m_{\tilde \chi^0_1} = 53.3 \pm 0.6  \mbox{\ GeV}
\label{eq:p3122osmass}
\end{equation}

\vskip -0.5 cm
\begin{Fighere}
\centering
  \mbox{\epsfig{file=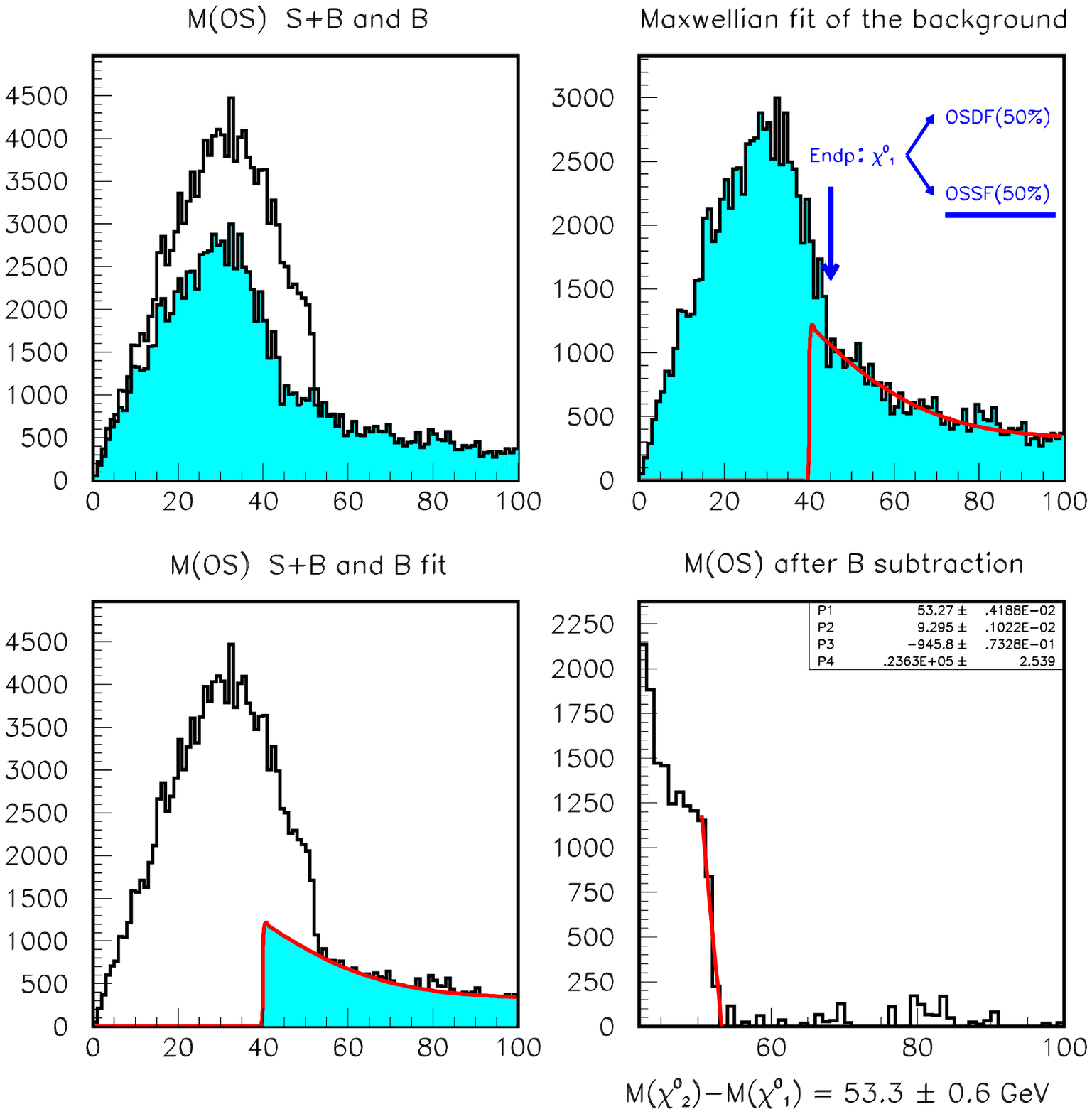,width=14cm  }}
\caption{\small The invariant mass distribution of the remaining OS lepton 
pairs, after the removal of $\tilde \chi^0_1$ candidates. The distributions corresponds to
1 year of LHC run at low luminosity at point 3 in the case of 
$\rlap/R$ coupling $\lambda_{122}=10^{-3}$.
For explanations see text.}
\label{p3122os}
\end{Fighere}
\vskip 0.5 cm
As a next step we apply the cuts ($i \div v$) and :\\
\indent \indent ($vi$) 
$m(OS) \in (m_{OS}^{meas}-10, m_{OS}^{meas})$ GeV,\\
in order to reconstruct the $\tilde \chi^0_2$ by combining the OSDF and the remainig OS lepton pairs.
We require that the angle between the two pairs should be small :\\
\indent \indent ($vii$) cos($\alpha_{OSDF, OS}$) $\geq 0.8$. \\
The obtained mass distribution is shown in the Fig.\ref{p3122chi02}. 
The gaussian fit around the peak results in the value :
\begin{equation}
m^{meas}_{\tilde \chi^0_2} = 96.7 \pm 0.2 \mbox{\ GeV}
\label{eq:p3122chi02mass}
\end{equation}
Out of the three quantities defined by the Eq.(\ref{eq:p3122chi01mass}), (\ref{eq:p3122osmass}) and 
(\ref{eq:p3122chi02mass}) we use two independent ones
(Eq.(\ref{eq:p3122chi01mass}) and (\ref{eq:p3122chi02mass})) , 
for the determination of the SUGRA parameters.

\vskip -0.5 cm
\begin{Fighere}
\centering
  \mbox{\epsfig{file=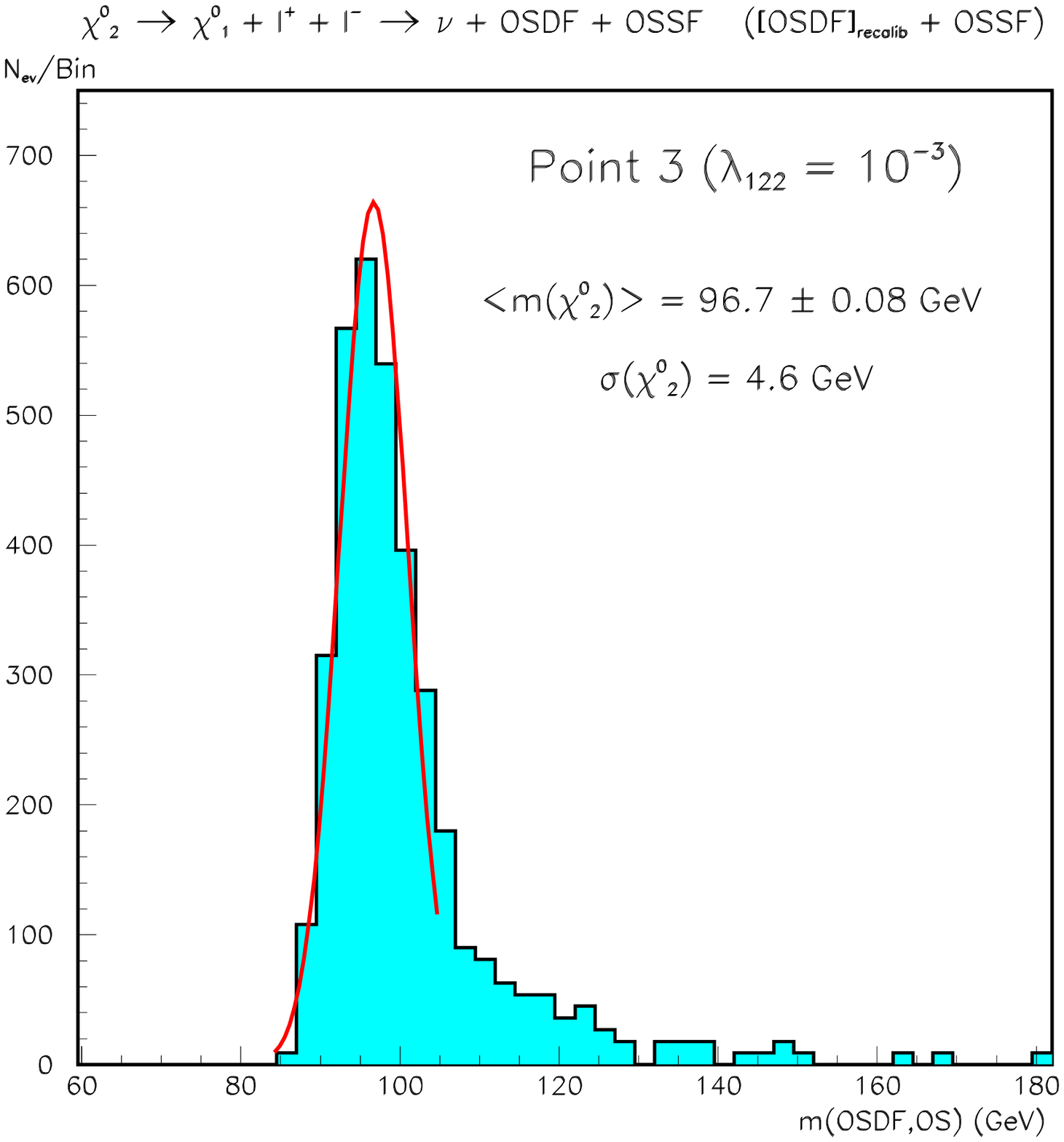,height=9.0cm,width=12.0cm  }}
\caption{\small The invariant mass distribution OSDF $+$ OS lepton pairs. 
The distributions correspond to 1 year of LHC run at low luminosity at point 3.
For explanations see text.}
\label{p3122chi02}
\end{Fighere}

\vskip 0.8cm
{\it \bf Reconstruction of the chain
$\tilde g \rightarrow \tilde b_1 + b \rightarrow \tilde \chi^0_2 + b + b$}
\vskip 0.2cm

The configuration of the sparticle mass spectrum (see Table\ \ref{tb:masses}) and the decay branching
ratios for that point, permit us to make the following remarks on the production of $b$ jets:

\indent (1) practically all $b$ jets originate from the decay chain of $\tilde g$ (with a small
fraction coming also from $\tilde t_1$ and/or $t$),\\
\indent (2) the $b$ jets from the first decay of $\tilde g$ are very soft comparing to
those produced in the second decay (and associated with $\tilde \chi^0_2$),
as it can be seen in the Fig.\ref{p3123ptbj}. Correspondingly we label the $b$ jets as follows :\\
\indent \indent ($viii$) {\it hard} $b$ jets if : $ p_t^b \geq$ 50 GeV ;\\
\indent \indent ($ix$) \ \ {\it soft} $b$ jets if : 10 GeV $\leq p_t^b \leq$ 50 GeV .\\
Selecting $\tilde \chi^0_2$ candidates around the $\tilde \chi^0_2$ peak from Fig.\ref{p3122chi02} :\\
\indent \indent ($x$) \ \ $m(OSDF,OS) \in (m^{peak}-15,m^{peak}+15)$ GeV,\\
one performs a first reconstruction of the $\tilde g$ mass as follows. First of all, one selects pairs
of $b$ jets with all $b$ jets passing the cut $p_t \geq$ 10 GeV and in each pair we identify the
{\it hard} and the {\it soft}  jet. Next we combine the $\tilde \chi^0_2$, $b_{hard}$ and
$b_{soft}$ 4-momenta if :\\
\indent \indent ($xi$) \ cos($\alpha_{\tilde \chi^0_2 b_{hard}}$) $\geq 0$ , and\\
\indent \indent ($xii$) cos($\alpha_{(\tilde \chi^0_2 b_{hard}),b_{soft}}$) $\geq 0.5$. \\

These cuts are justified by the distributions shown in Fig.\ref{p3122rawgl}a and \ref{p3122rawgl}b.

\vskip -1.5 cm
\begin{Fighere}
\centering
  \mbox{\epsfig{file=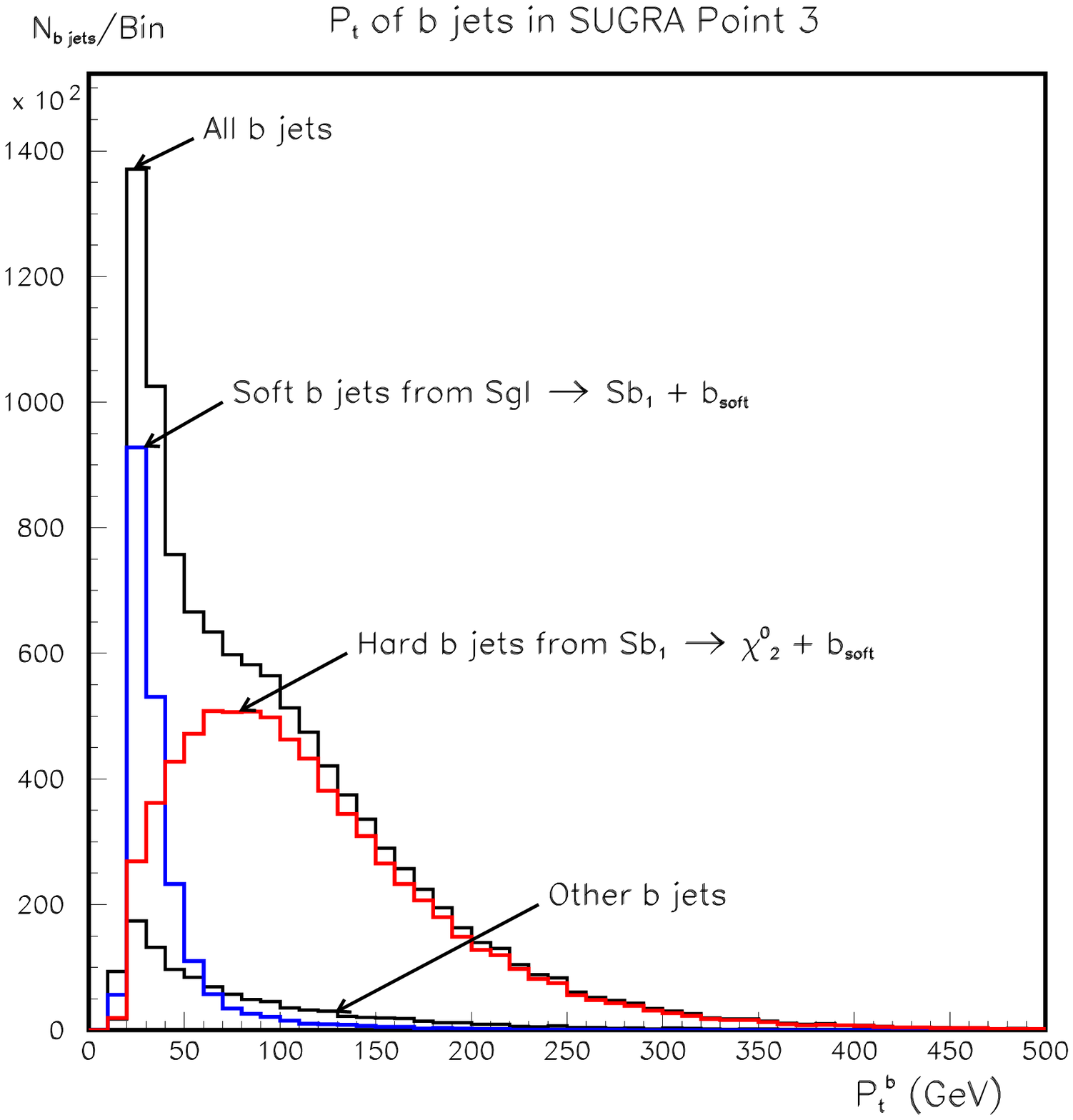,height=10.0cm,width=14cm  }}
\caption{\small The $p_t$ distribution of $b$ jets in point 3 for 1 year of LHC run at low
luminosity after applying the global cuts $N_l \geq 4$, $E_t^{miss} \geq 50$ GeV and
$p_t^b \geq 10$ GeV. One can clearely distinguish the $b$ jets from $\tilde g$
and from $\tilde b_1$.}
\label{p3123ptbj}
\end{Fighere}
\begin{Fighere}
\hspace{-1.0cm}
\centering
  \mbox{\epsfig{file=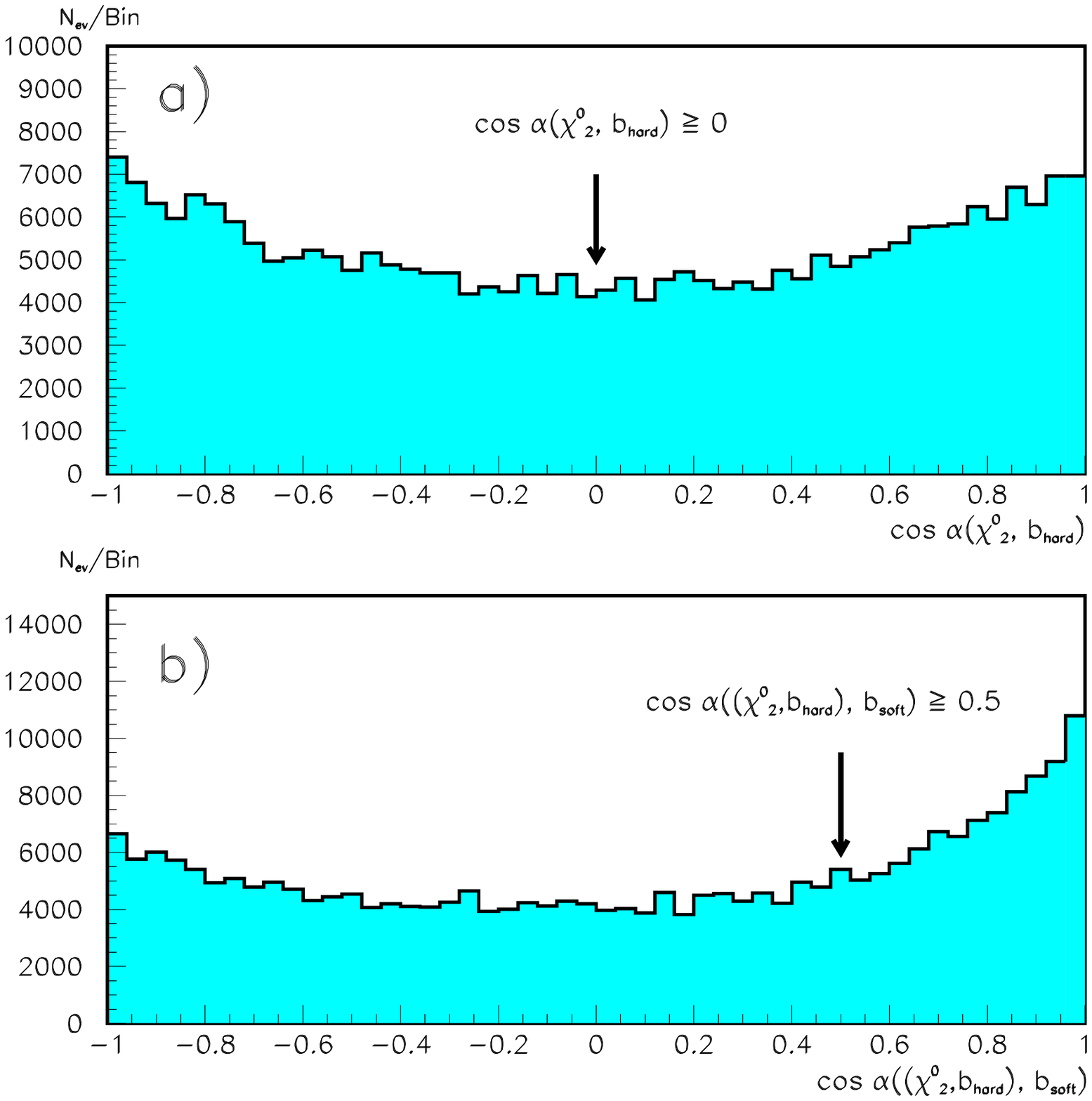,height=8.0cm }\ \
        \epsfig{file=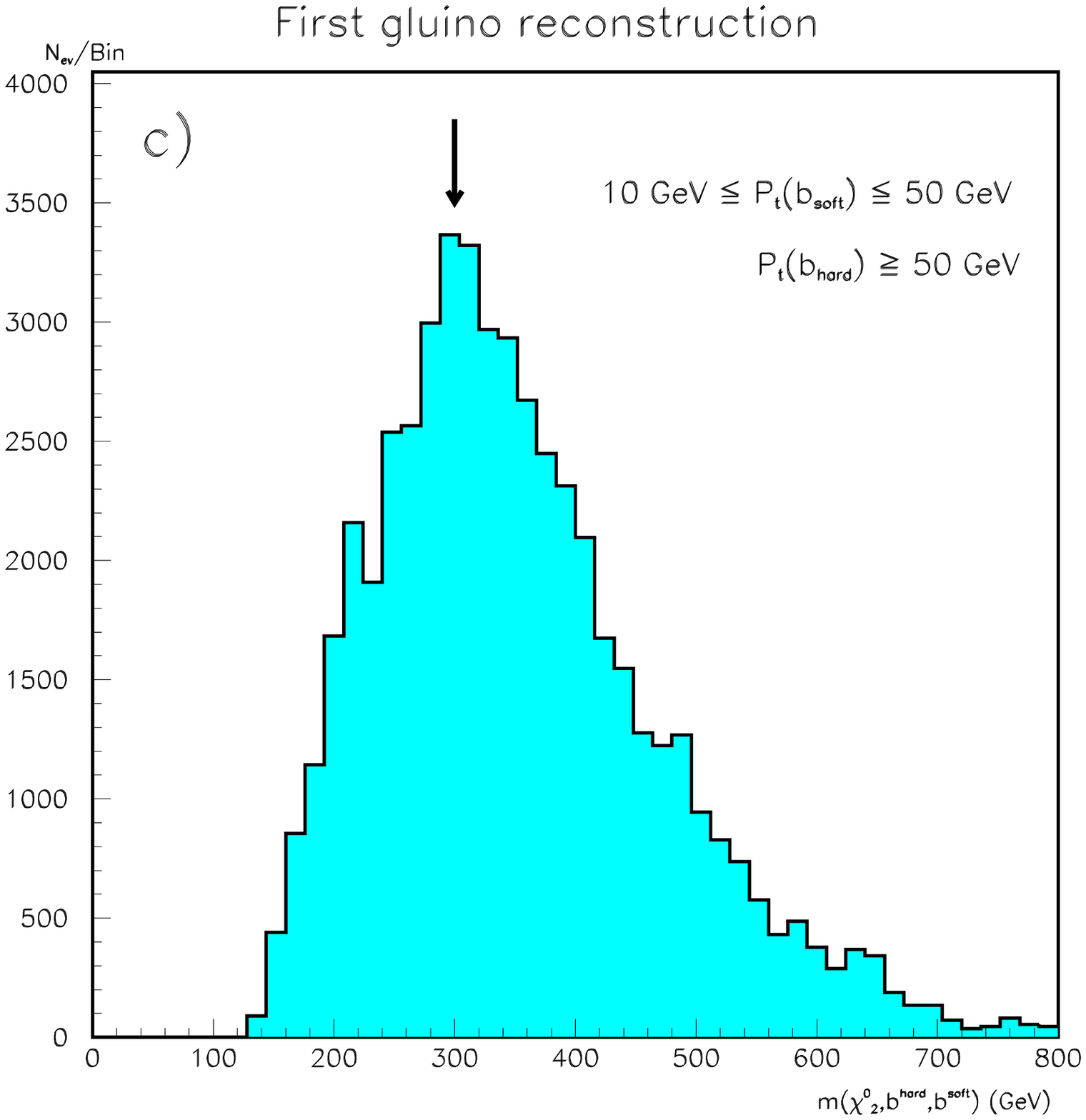,height=8.0cm }}
\vskip -0.5cm
\caption{\small The first reconstruction of the gluino mass in point 3, for
$\rlap/R$ coupling $\lambda_{122}=10^{-3}$ and after 1 year of LHC run at low
luminosity (plot c). Also shown are the angle distributions between the
$\tilde \chi^0_2$ and the hard $b$-jets (plot a) and between the ($\tilde \chi^0_2$, $b_{hard}$)
pairs and soft $b$-jets (plot b). For further comments see text.}
\label{p3122rawgl}
\end{Fighere}
\vskip 0.5 cm

The resulting mass distribution is represented in Fig.\ref{p3122rawgl}c.
For the $\tilde b_1$ reconstruction, one selects the $\tilde \chi^0_2$ candidates
according to ($x$) and the {\it hard} $b$ jets (see ($viii$)) around the 
$\tilde g$ peak of the Fig.\ref{p3122rawgl}c :\\
\indent \indent ($xiii$) $m_{\tilde g} \in (m^{peak}-10, m^{peak}+10)$ GeV, requiring this time \\
\indent \indent ($xiv$) \ cos($\alpha_{\tilde \chi^0_2 b_{hard}}$) $\geq 0.5$.\\
The result is represented in Fig.\ref{p3122sb1}. After the gaussian fit around the peak one
obtains the value :
\begin{equation}
m_{\tilde b_1}^{meas} = 276.6 \pm 3.0  \mbox{\ GeV}
\label{eq:p3122sb1mass}
\end{equation}

\vskip -0.5 cm
\begin{Fighere}
\centering
  \mbox{\epsfig{file=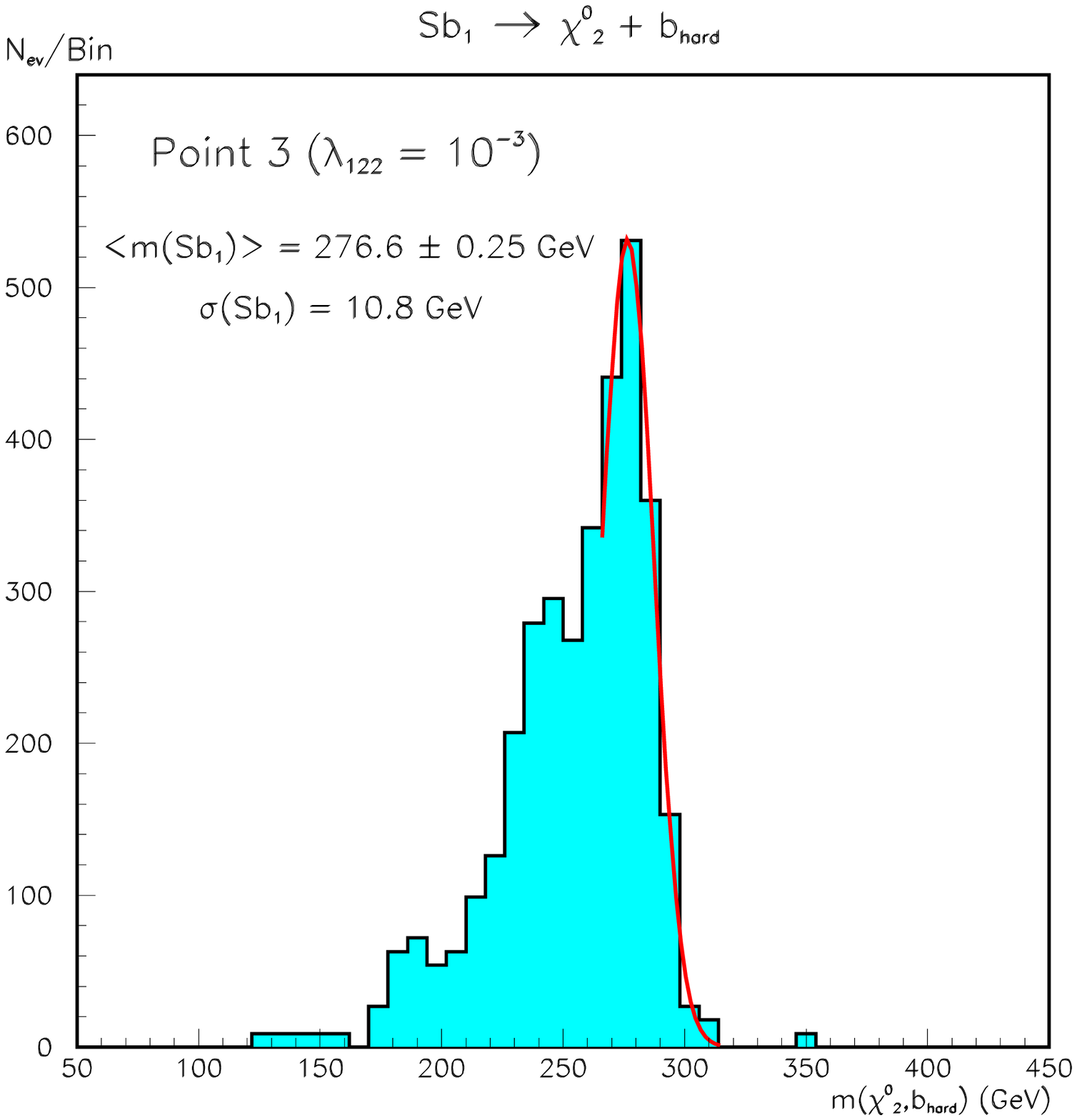,width=11cm  }}
\caption{\small The invariant mass distribution of $\tilde \chi^0_2$ and $b_{hard}$ jets.
The distribution corresponds to 1 year of LHC run at low
luminosity. For the selection criteria see text.}
\label{p3122sb1}
\end{Fighere}
\vskip 0.5 cm
This permits us to refine the reconstruction of the $\tilde g$ by tagging it with the $\tilde b_1$.
We combine the $\tilde \chi^0_2$ candidates (cut ($xi$)) with a {\it hard} $b$ (cut ($viii$))
requiring an invariant mass within the $\tilde b_1$ mass window: \\
\indent \indent ($xv$) 
$m_{\tilde \chi^0_2 b_{hard}} \in (m_{\tilde b_1}^{meas}-10, m_{\tilde b_1}^{meas}+10)$ GeV.\\
We add subsequently a {\it soft} $b$ jet with the same angular correlation as ($xii$) to obtain
the final gluino reconstruction shown in the Fig.\ref{p3122gl}. 
The gaussian fit around the peak gives the measured value :\\
\begin{equation}
m_{\tilde g}^{meas} = 301.1 \pm 3.0  \mbox{\ GeV .}
\label{eq:p3122glmass}
\end{equation}

\vskip -0.5 cm
\begin{Fighere}
\centering
  \mbox{\epsfig{file=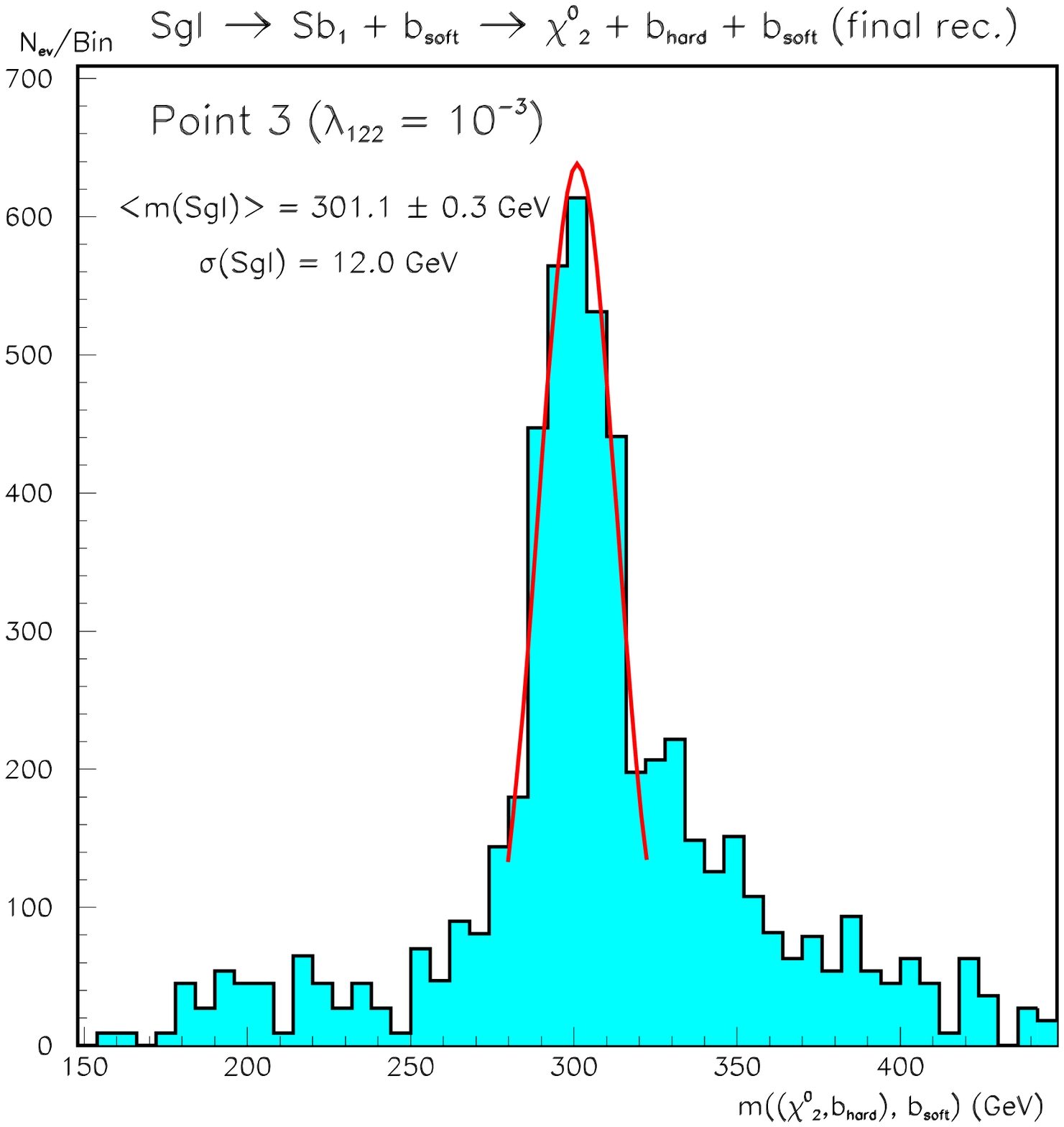,width=9cm  }}
\caption{\small The final reconstruction of the $\tilde g$ after 
having selected the ($\tilde \chi^0_2$, $b_{hard}$) pairs around the reconstructed 
$\tilde b_1$ mass peak. The distribution corresponds to 1 year of LHC run at low
luminosity.}
\label{p3122gl}
\end{Fighere}


\vskip 0.8cm
{\bf \large \it \underline{The case $\lambda_{123} \ne 0$}}
\vskip 0.5cm

{\it \bf Reconstruction of 
$\tilde \chi^0_2 \rightarrow \tilde \chi^0_1 + \mu^{\pm}(e^{\pm})+\mu^{\mp}(e^{\mp})$}
\vskip 0.2cm

Comparing to the case $\lambda_{122}$ the presence of $\tau$ jets in the decay products of 
$\tilde \chi^0_1$ spoils the endpoint structure in the OSDF lepton pair distribution and
therefore the direct reconstruction of the $\tilde \chi^0_1$.
Although the electrons and muons from the $\tau$ jets increase the combinatorial background
of the OSSF lepton pairs, due the fact that they are softer than those from the 3-body decay of
the $\tilde \chi^0_2$ they do not spoil the endpoint structure.
We select events with the usual cuts ($i \div ii$) and OSSF pairs by requiring :\\
\indent \indent ($xvi$) cos($\alpha_{OSSF}$) $\geq$ 0.5.\\
After a fit of the Maxwellian tail, subtraction and the fit of the result with a polinomial 
(see Fig.\ref{p3123dchi0}), one obtains using Eq.(\ref{eq:mAa}) :
\begin{equation}
m_{OSSF}^{meas}=m_{\tilde \chi^0_2}-m_{\tilde \chi^0_1} = 52.9^{+0.1}_{-0.3}  \mbox{\ GeV .}
\label{eq:p3123ossfmass}
\end{equation}

\vskip 0.8cm
{\it \bf Reconstruction of the chain
$\tilde g \rightarrow \tilde b_1 + b \rightarrow \tilde \chi^0_2 + b + b$}
\vskip 0.2cm

Since the $\lambda$ couplings do not affect the $b$ jets (if we neglect the difference in the
reconstruction and tagging efficiency due to the larger size of a $\tau$ jet comparing to a lepton)
the classification of $b$ jets in {\it soft} and {\it hard} types  is done by the same
criteria as in the case $\lambda_{122}$. 
The reconstruction of this chain follows  the same treatement 
as in the case $\lambda_{122}$ with the notable difference that this time one cannot 
directly reconstruct the $\tilde \chi^0_2$. Instead, we have to assume a mass value for the
$\tilde \chi^0_2$ in order to obtain its 4-momentum at the endpoint of the OSSF lepton pair 
mass distribution according to the Eq.(\ref{eq:pAa}). We have taken $m_{\tilde \chi^0_2}=97$ GeV.

\vskip -1.0 cm
\begin{Fighere}
\centering
  \mbox{\epsfig{file=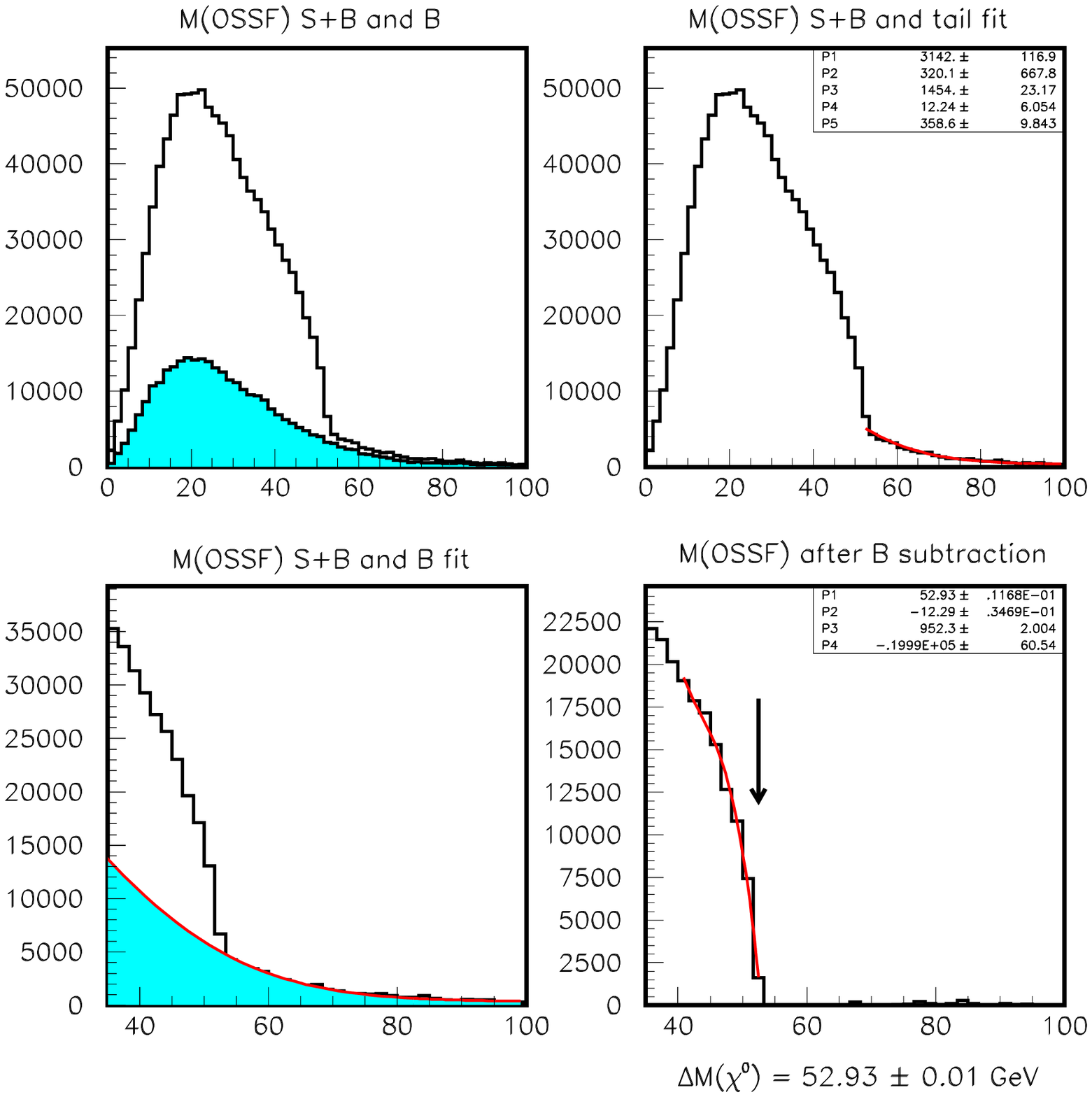,height=13.0cm,width=15.0cm  }}
\caption{\small The invariant mass distribution of the OSSF lepton pairs
for 1 year of LHC run at low luminosity at point 3 in the case of 
$\rlap/R$ coupling $\lambda_{123}=10^{-3}$ .}
\label{p3123dchi0}
\end{Fighere}
\vskip 0.5 cm

The cuts used in this case are the following :\\
\indent \indent ($x$)$^*$ \ \ $m(OSSF) \in (m^{endp}_{OSSF}-10, m^{endp}_{OSSF})$ GeV,  
for the $\tilde \chi^0_2$ candidates,\\
and for the first reconstruction of the gluino :\\
\indent \indent ($xi$)$^*$ \ cos($\alpha_{\tilde \chi^0_2 b_{hard}}$) $\geq$ 0.5, \\
\indent \indent ($xii$)$^*$ cos($\alpha_{\tilde \chi^0_2 b_{soft}}$) $\geq$ 0.5. \\

The invariant mass distribution of $\tilde \chi^0_2 b_{hard} b_{soft}$ pairs
is shown in Fig.\ref{p3123rawgl}.
Selecting the {\it hard} $b$ jets if around the gluino peak :\\
\indent \indent ($xiii$)$^*$ $m_{\tilde g} \in (m^{peak}-10, m^{peak}+10)$ GeV and,\\
\indent \indent ($xiv$)$^*$ \ cos($\alpha_{\tilde \chi^0_2 b_{hard}}$) $\geq 0.5$.\\
one obtains the $\tilde b_1$ mass peak as shown in the Fig.\ref{p3123sb1}. 
The gaussian fit around the peak gives :
\begin{equation}
m_{\tilde b_1}^{meas}(97) = 277.5 \pm 3.0  \mbox{\ GeV}
\label{eq:p3123sb1mass}
\end{equation}
The 97 inside the brackets indicates that this value was obtained assuming
$m_{\tilde \chi^0_2} = $ 97 GeV.
\vskip -0.5 cm
\begin{Fighere}
\centering
  \mbox{\epsfig{file=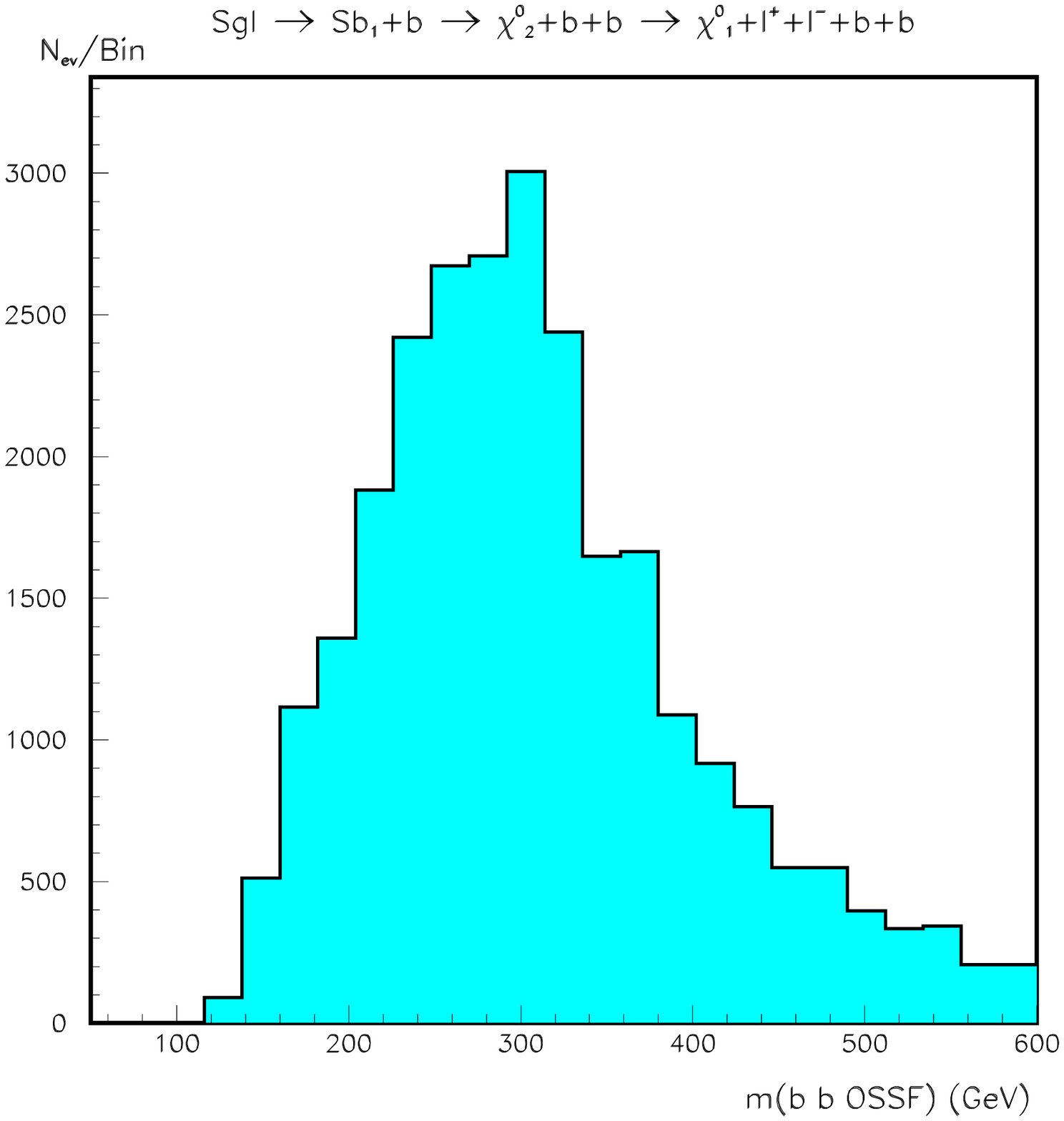,height=8.0cm,width=12cm  }}
\caption{\small The first reconstruction of the gluino in point 3, for
$\rlap/R$ coupling $\lambda_{123}=10^{-3}$ and after 1 year of LHC run at low
luminosity .}
\label{p3123rawgl}
\end{Fighere}
\vskip 0.5 cm

\begin{Fighere}
\centering
  \mbox{\epsfig{file=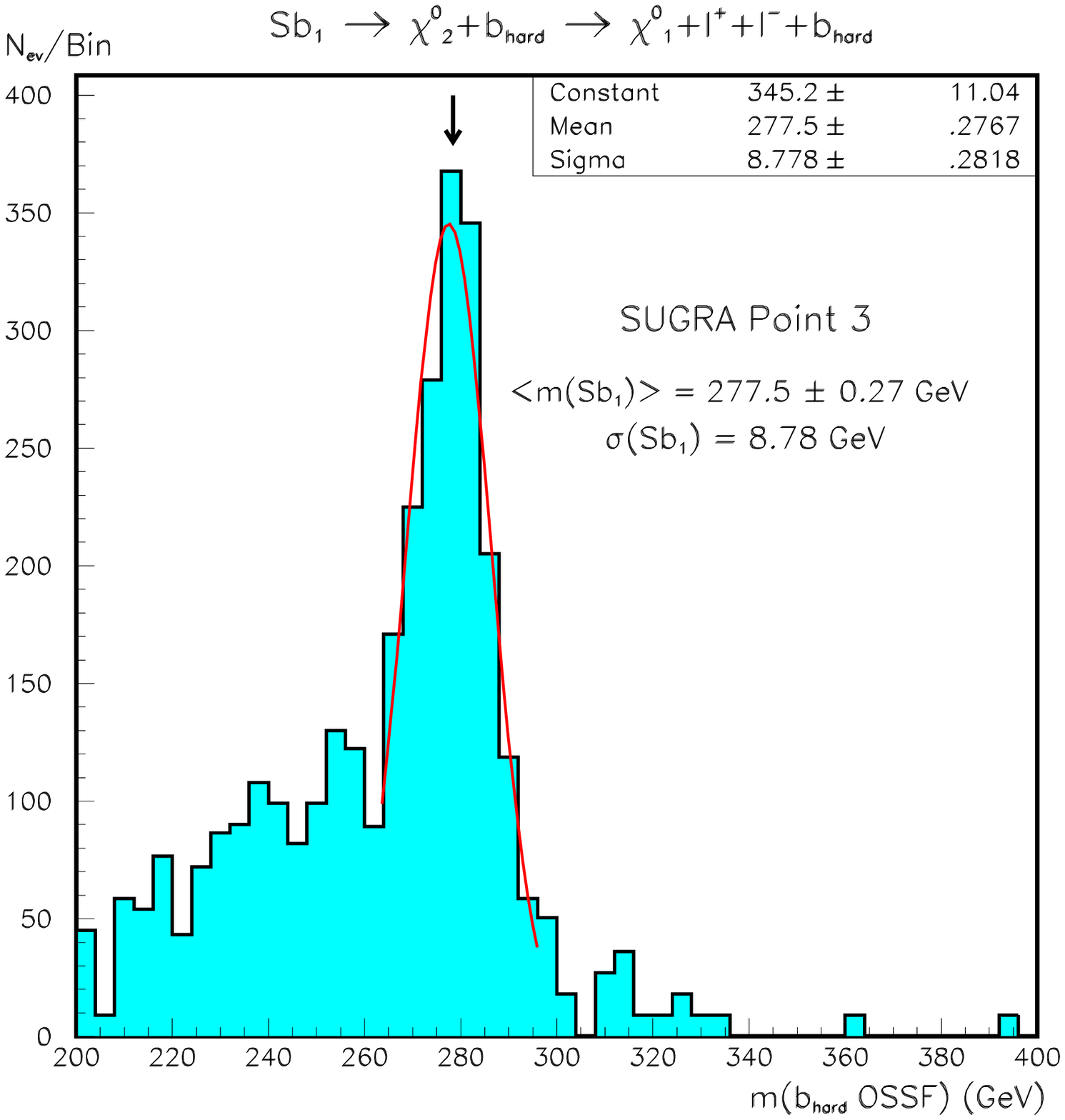,width=11cm  }}
\caption{\small The reconstruction of the $\tilde b_1$ selecting the $b$ jet around the 
$\tilde g$ peak. The distribution corresponds to 1 year of LHC run at low
luminosity.}
\label{p3123sb1}
\end{Fighere}
\vskip 0.5 cm
Finally, selecting the {\it hard} $b$ jets and the OSSF pairs around the $\tilde b_1$ peak :\\
\indent \indent ($xv$)$^*$ 
$m_{\tilde \chi^0_2 b_{hard}} \in (m_{\tilde b_1}^{meas}-10, m_{\tilde b_1}^{meas}+10)$ GeV,\\
we combine them with the {\it soft } $b$ jets with the same angle cuts as above 
(($xi$)$^*$ and $xii$)$^*$). The resulting mass distribution is shown in the Fig.\ref{p3123gl}. 
The gaussian fit around the peak gives the measured value :\\
\begin{equation}
m_{\tilde g}^{meas}(97) = 301.1 \pm 3.5  \mbox{\ GeV .}
\label{eq:p3123glmass}
\end{equation}

\vskip -0.5 cm
\begin{Fighere}
\centering
  \mbox{\epsfig{file=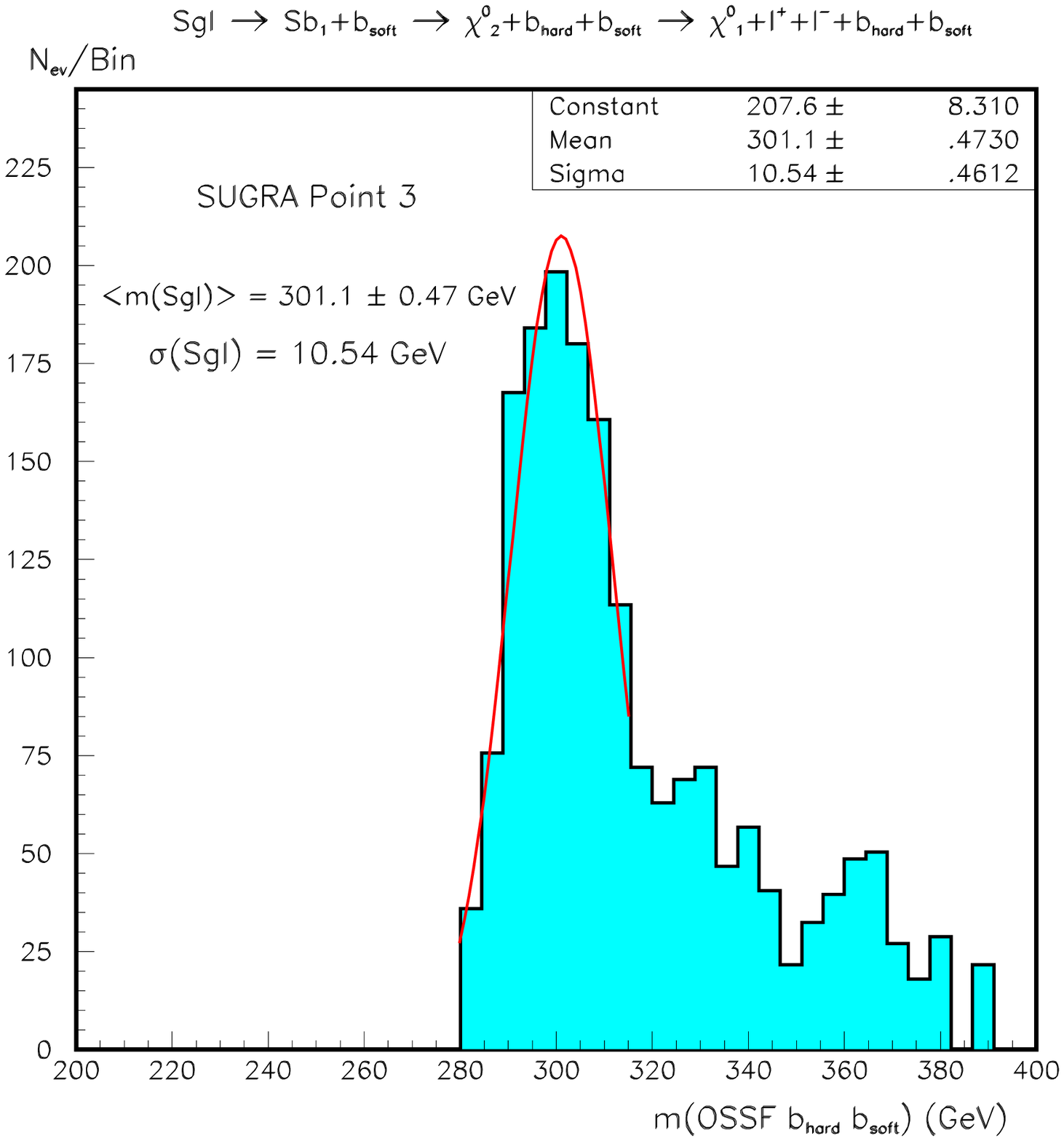,width=12.0cm  }}
\caption{\small The final reconstruction of the $\tilde g$ after having selected the OSSF lepton 
pairs and {\it hard} $b$ jets around the $\tilde b_1$ peak.
The distribution corresponds to 1 year of LHC run at low luminosity. }
\label{p3123gl}
\end{Fighere}
\vskip 0.5cm

We have repeated the above analysis for several 
different values of $m_{\tilde \chi^0_2}$. The result is
shown in Fig.~\ref{fg:p3chi02theor}. Using a linear fit one obtains the following
expressions :
\begin{equation}
m_{\tilde b_1}^{meas}(m_{\tilde \chi^0_2}) = m_{\tilde b_1}^{meas}(97) +
\theta_{\tilde b_1}(m_{\tilde \chi^0_2}-97) \mbox{ GeV}
\label{eq:p3123sb1_var_mass}
\end{equation}
\begin{equation}
m_{\tilde g}^{meas}(m_{\tilde \chi^0_2}) = m_{\tilde g}^{meas}(97) +
\theta_{\tilde g}(m_{\tilde \chi^0_2}-97) \mbox{ GeV}
\label{eq:p3123gl_var_mass}
\end{equation}
with $\theta_{\tilde b_1} = 1.48 \pm 0.52$ and $\theta_{\tilde g} = 1.55 \pm 0.55$, repectively.

\vskip -0.5cm
\begin{Fighere}
\centering
  \mbox{\epsfig{file=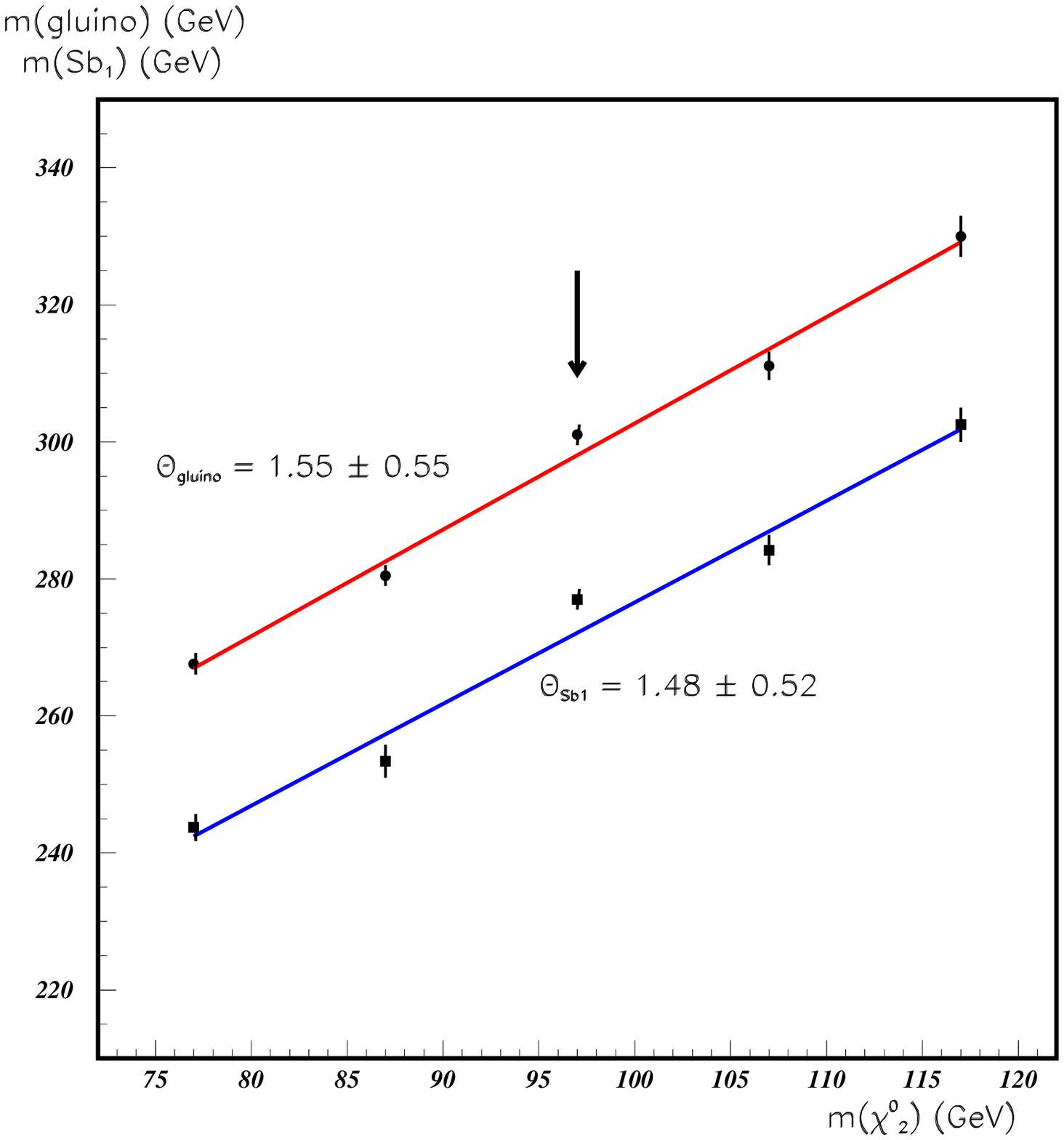,width=11.0cm }}
\caption{\small The dependence of $m_{\tilde b_1}$ (full squares) and $m_{\tilde g}$ (full circles)
as functions of $m_{\tilde \chi^0_2}$. The arrow points to the nominal value of
$m_{\tilde \chi^0_2}=97$ GeV. The factors $\theta_{\tilde b_1}$ and $\theta_{\tilde g}$ (see text) 
are extracted by a linear fit.}
\label{fg:p3chi02theor}
\end{Fighere}
\vskip 1.0 cm


\subsection{Determination of the SUGRA parameters}
\vskip 0.5 cm

In subsection 2.2 we have already reviewed the dependence of the sparticle masses
as functions of the model parameters $m_0$, $m_{1/2}$, $A_0$, tan$\beta$ and sign$\mu$.
Practically all sparticle masses are sensitive to $m_{1/2}$. $m_0$ drives mainly the sfermion masses.
sign$\mu$ affects the gaugino branching  ratios 
(i.e. $\chi^0_2 \rightarrow \chi^0_1 + l^{+} + l^{-}$ versus $\chi^0_2 \rightarrow \chi^0_1 + h^0$),
but also the sparticle mass spectrum. On the other hand, the observables are practically 
independent of $A_0$ (c.f. Fig~\ref{fg:p5alla0fit} and Table~\ref{tb:slope_p5a0}). 
An obvious way to determine the above parameters consists of finding the minimum of :
\begin{equation}
\begin{tabular}{l}
$\chi^2(m_0,m_{1/2},tan\beta,A_0) =$  \\
\\
$\sum_{\alpha,\beta=1}^{N_{obs}} [O^{th}_{\alpha}(m_0,m_{1/2},tan\beta,A_0) - O^{meas}_{\alpha}]
\sigma^{-1}_{\alpha \beta}[O^{th}_{\beta}(m_0,m_{1/2},tan\beta,A_0) - O^{meas}_{\beta}]$
\end{tabular}
\label{eq:chisquare}
\end{equation}
where $O^{th}_{\alpha}(m_0,m_{1/2},tan\beta,A_0)$ represents the theoretical evolution of the
observable $\alpha$ in function of the SUGRA parameters, $O^{meas}_{\alpha}$ is the measured value
of the same observable and $\sigma_{\alpha \beta}$ is the covariance matrix 
\footnote{In general $O^{meas}_{\alpha}$ are not independent, however at the present time we neglect the 
offdiagonal elements of $\sigma_{\alpha \beta}$.} .\\
The minimum of the $\chi^2$ can be found by scanning through the entire parameter 
space \cite{Froid_LHCC}. 
The overall minimum is practically insensitive to $A_0$. The sign$\mu$ parameter can be
determined unambiguosly \cite{Froid_LHCC}. Once the overall minimum is found one can determine the
error on the parameters at the minimum by taking the square root of the diagonal elements of the
matrix $\Delta_{ij}$ :
\begin{equation}
\Delta_{ij} = \Bigl[ \sum_{\alpha ,\beta=1}^{N_{obs}}
A^{T}_{i\alpha}(\sigma_{\alpha \beta })^{-1} A_{\beta j} \Bigr]^{-1}, \ \ \ \
A_{\alpha i} = \frac{\partial O^{th}_{\alpha}}{\partial (p_i)}, 
\ \ p_i \equiv m_0, m_{1/2}, tan \beta, A_0, \theta_{\tilde g}, \theta_{\tilde b_1}
\label{eq:paramerror}
\end{equation}

Table~\ref{tb:values} summarizes all the measured values and errors of the chosen observables 
for all studied cases. For point 3 we have added the $h^0$ mass determined by an 
independent measurement ($h^0 \rightarrow \gamma \ \gamma$) with an estimated error of 1 GeV.
As one can see from this Table in most of the cases the models are overconstrained : one disposes
of more measured quantities than the number of parameters to be determined. Comparing the measured
values of the observables with the theoretical ones (see Table~\ref{tb:masses}) one can see that
the errors are slightly overestimated.

$O^{th}(p_i)$ are the theoretical dependencies of the quantitities $O^{meas}$ collected in 
Table~\ref{tb:values}. However, there is an exception at point 3 (for $\lambda_{123} \neq 0$)
where, according to Eqs.(\ref{eq:p3123sb1_var_mass}) and (\ref{eq:p3123gl_var_mass}) one has :
\begin{equation}
\begin{tabular}{l}
$O^{th}_{\tilde g}(p_i)=m_{\tilde g}(p_i)-
\theta_{\tilde g}(m_{\tilde \chi^0_2}(p_i)-97)$, \hskip 0.5cm
$p_i=m_0, m_{1/2}, A_0, tan\beta$ \\
$O^{th}_{\tilde b_1}(p_i)=m_{\tilde b_1}(p_i)-
\theta_{\tilde b_1}(m_{\tilde \chi^0_2}(p_i)-97)$\\
\end{tabular}
\label{fitwithchi02}
\end{equation}
In this case we have to add two more terms to Eq.(\ref{eq:chisquare}), namely :
\begin{equation}
\chi^2 \longrightarrow \chi^2 + 
\Bigl(\frac{\theta_{\tilde g} - \theta^{m}_{\tilde g}}{\sigma_{\theta_{\tilde g}}}\Bigr)^2 +
\Bigl(\frac{\theta_{\tilde b_1} - \theta^{m}_{\tilde b_1}}{\sigma_{\theta_{\tilde b_1}}}\Bigr)^2
\label{newchisquare}
\end{equation}
and minimize this expression by varying also $\theta_{\tilde g}$ and $\theta_{\tilde b_1}$.
As one can see the errors on $\theta_{\tilde g}$ and $\theta_{\tilde b_1}$ do not contribute
to the errors of the other parameters in the first approximation.

 As an example, in the Figs.~\ref{fg:p5allm0fit} $\div$ \ref{fg:p5alltanbfit} we have represented 
the dependence of the observables on $m_0$, $m_{1/2}$, $A_0$ and tan$\beta$, respectively, in point 5.
Each histogram is fitted with a polinomial in a domain $\pm 50$ GeV - for $m_0$, $m_{1/2}$ and $A_0$ and
$\pm 0.5$ - for tan$\beta$, around the nominal values of point 5. The first derivatives of these
functions are used to obtain $\Delta_{ij}$ of Eq.(\ref{eq:paramerror}).

The final results are compiled in the form of relative errors in Table~\ref{tb:sugrafitp1}, taking
$A_0$ fixed at its nominal value of each point.
The obtained precision on the SUGRA parameters are in general higher than in the case of
conserved R parity \cite{Froid_LHCC}. The reason is that here we were able to determine more
observables and usually in a more direct way through the reconstruction of the LSP.

\begin{Fighere}
\centering
  \mbox{\epsfig{file=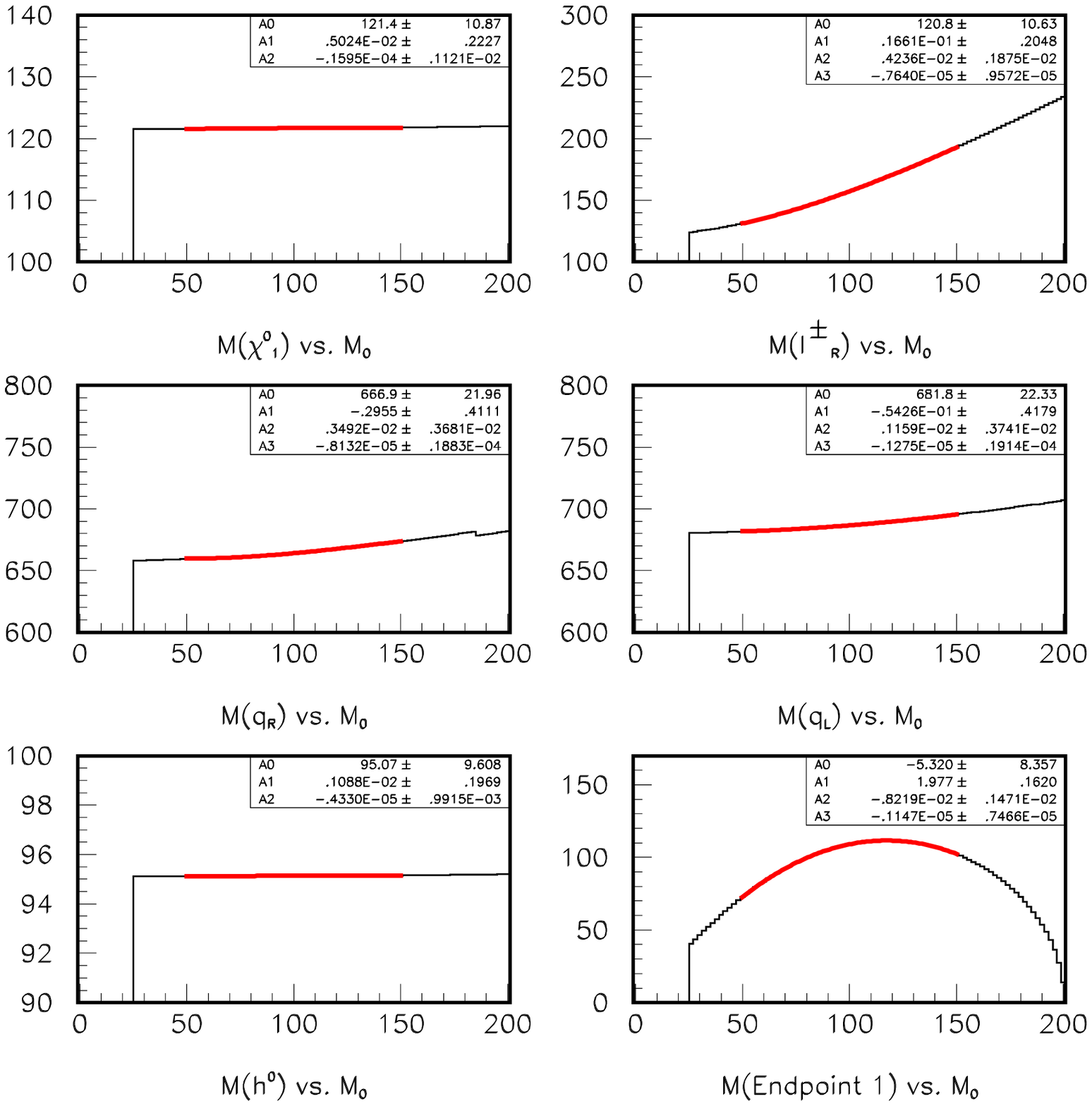,width=16.0cm }}
\caption{\small The dependence of $m_{\tilde \chi^0_1}$, $m_{\tilde l_R}$, $m_{\tilde q_R}$,
$m_{\tilde q_L}$, $m_{\tilde h^0}$ and $M^{end}_{OSSF}$ on $m_0$ smoothed by a polynomial fit.
All other parameters are fixed at those of point 5.}
\label{fg:p5allm0fit}
\end{Fighere}
\vskip 1.0 cm

\begin{Tabhere}
\centering
\caption{\small  The slopes, $\frac{\partial O^{th}_{\alpha}}{\partial m_0}$,  
 of the observables  at the
nominal value of point 5 : $m_0=100$ GeV.}
\vskip 0.5cm \hskip -0.8cm
\begin{tabular}{|c|c|c|c|c|c|} \hline\hline
$\tilde \chi^0_1$ & $\tilde l^{\pm}_R$ & $\tilde q_R$ & $\tilde q_L$ & $\tilde h^0$ & $M^{end}_{OSSF}$ \\ \hline\hline
0.001834 & 0.634610 & 0.15894 & 0.13929 & 0.001954 & 0.29879 \\ \hline\hline
\end{tabular}
\label{tb:slope_p5m0}
\end{Tabhere}

\begin{Fighere}
\centering
  \mbox{\epsfig{file=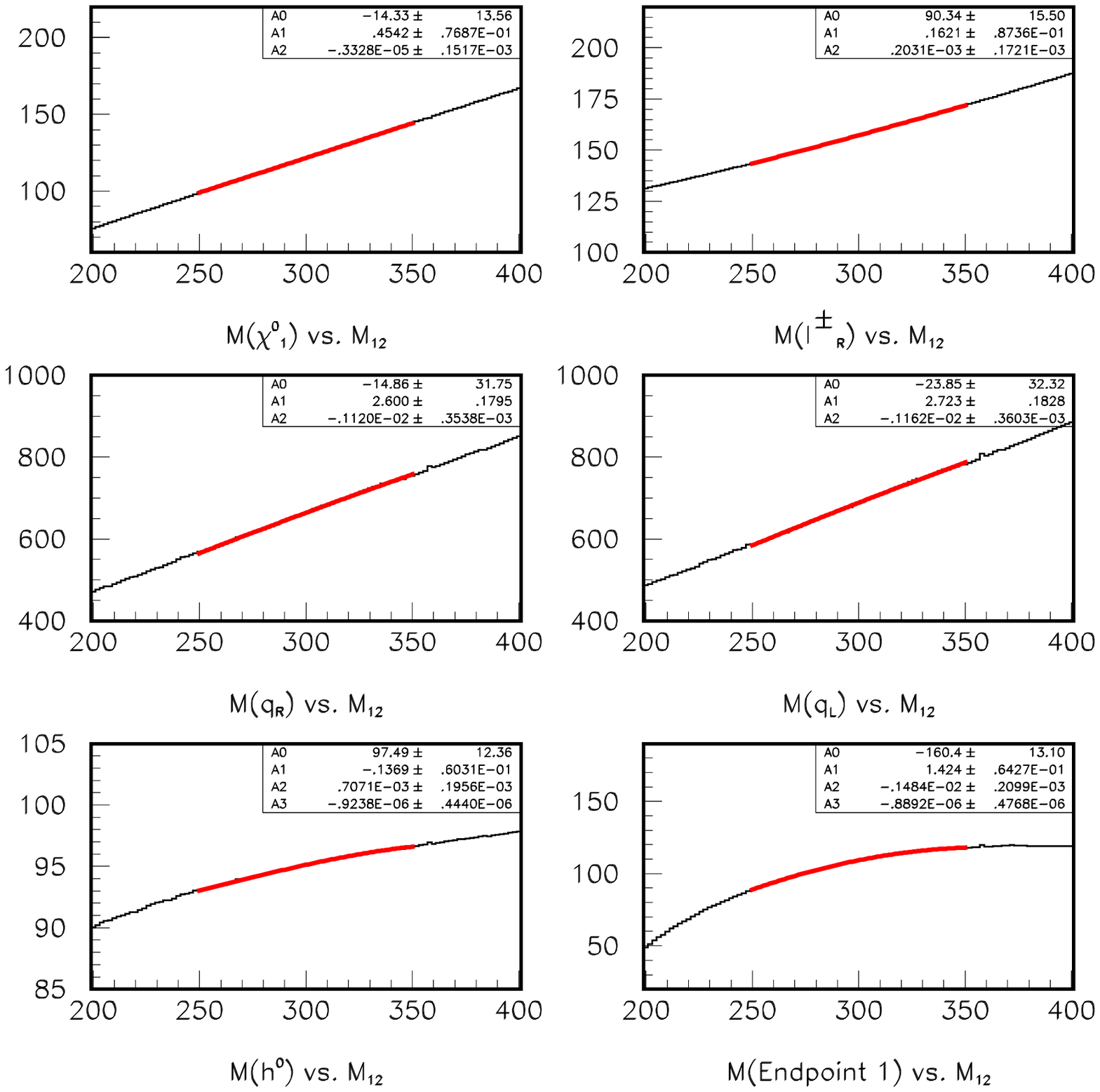,width=16.0cm }}
\caption{\small The dependence of $m_{\tilde \chi^0_1}$, $m_{\tilde l_R}$, $m_{\tilde q_R}$,
$m_{\tilde q_L}$, $m_{\tilde h^0}$ and $M^{end}_{OSSF}$ on $m_{1/2}$ smoothed by a polynomial fit.
All other parameters are fixed at those of point 5.}
\label{fg:p5allm12fit}
\end{Fighere}
\vskip 1.0 cm

\begin{Tabhere}
\centering
\caption{\small  The slopes, $\frac{\partial O^{th}_{\alpha}}{\partial m_{1/2}}$,
 of the observables  at the
nominal value of point 5 : $m_{1/2}=300$ GeV.}
\vskip 0.5cm \hskip -0.8cm
\begin{tabular}{|c|c|c|c|c|c|} \hline\hline
$\tilde \chi^0_1$ & $\tilde l^{\pm}_R$ & $\tilde q_R$ & $\tilde q_L$ & $\tilde h^0$ & $M^{end}_{OSSF}$ \\ \hline\hline
0.452232 & 0.28396 & 1.928 & 2.0258 & 0.037934 & 0.293516 \\ \hline\hline
\end{tabular}
\label{tb:slope_p5m12}
\end{Tabhere}

\begin{Fighere}
\centering
  \mbox{\epsfig{file=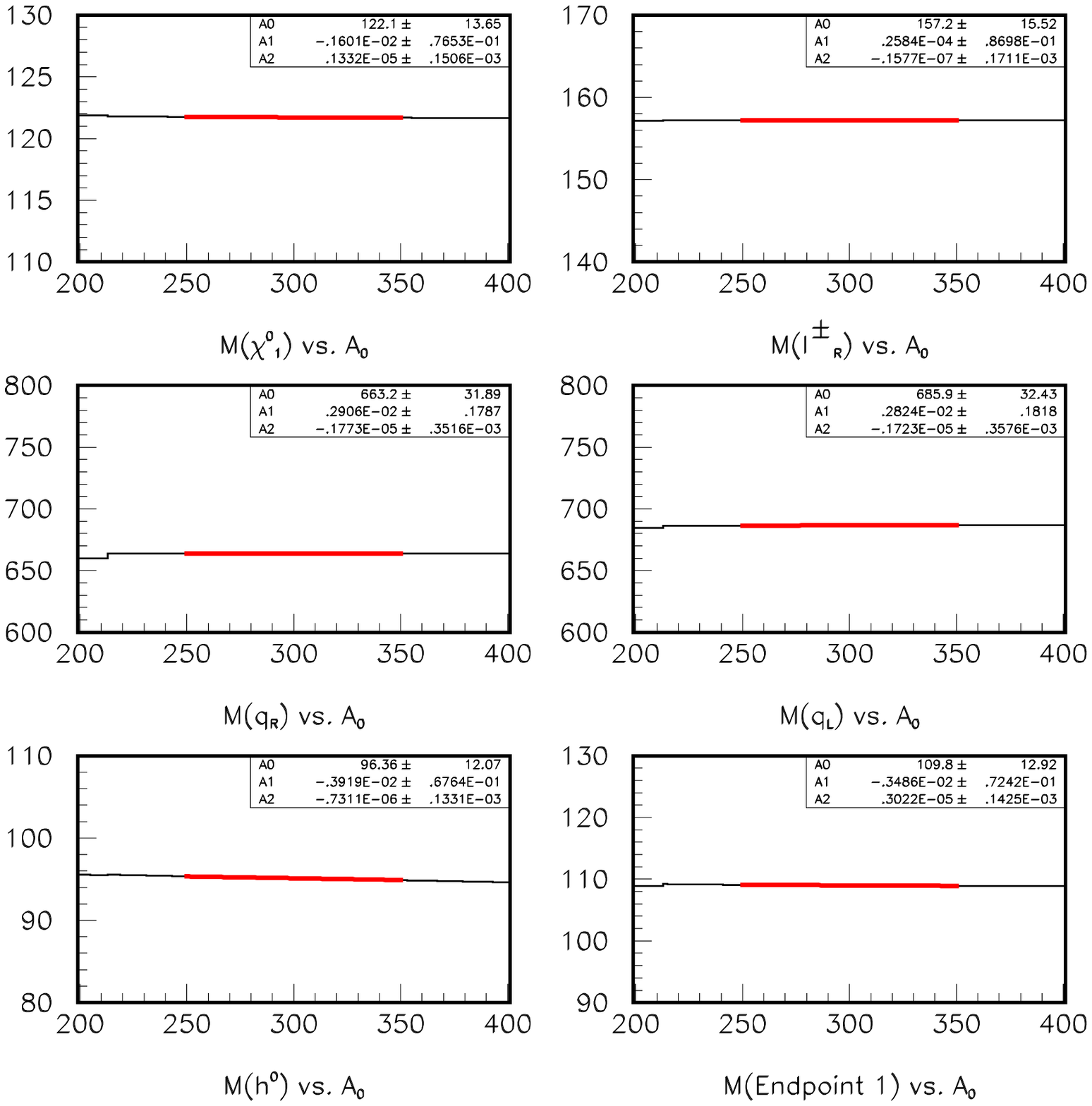,width=16.0cm }}
\caption{\small The dependence of $m_{\tilde \chi^0_1}$, $m_{\tilde l_R}$, $m_{\tilde q_R}$,
$m_{\tilde q_L}$, $m_{\tilde h^0}$ and $M^{end}_{OSSF}$ on $A_0$ smoothed by a polynomial fit.
All other parameters are fixed at those of point 5.}
\label{fg:p5alla0fit}
\end{Fighere}
\vskip 1.0 cm

\begin{Tabhere}
\centering
\caption{\small  The slopes, $\frac{\partial O^{th}_{\alpha}}{\partial A_0}$,
of the observables  at the
nominal value of point 5 : $A_0=300$ GeV.}
\vskip 0.5cm \hskip -0.8cm
\begin{tabular}{|c|c|c|c|c|c|} \hline\hline
$\tilde \chi^0_1$ & $\tilde l^{\pm}_R$ & $\tilde q_R$ & $\tilde q_L$ & $\tilde h^0$ & $M^{end}_{OSSF}$ \\ \hline\hline
-0.0008018 & 0.00002584 & 0.0018422 & 0.017902 & -0.003919 & -0.001673 \\ \hline\hline
\end{tabular}
\label{tb:slope_p5a0}
\end{Tabhere}

\begin{Fighere}
\centering
  \mbox{\epsfig{file=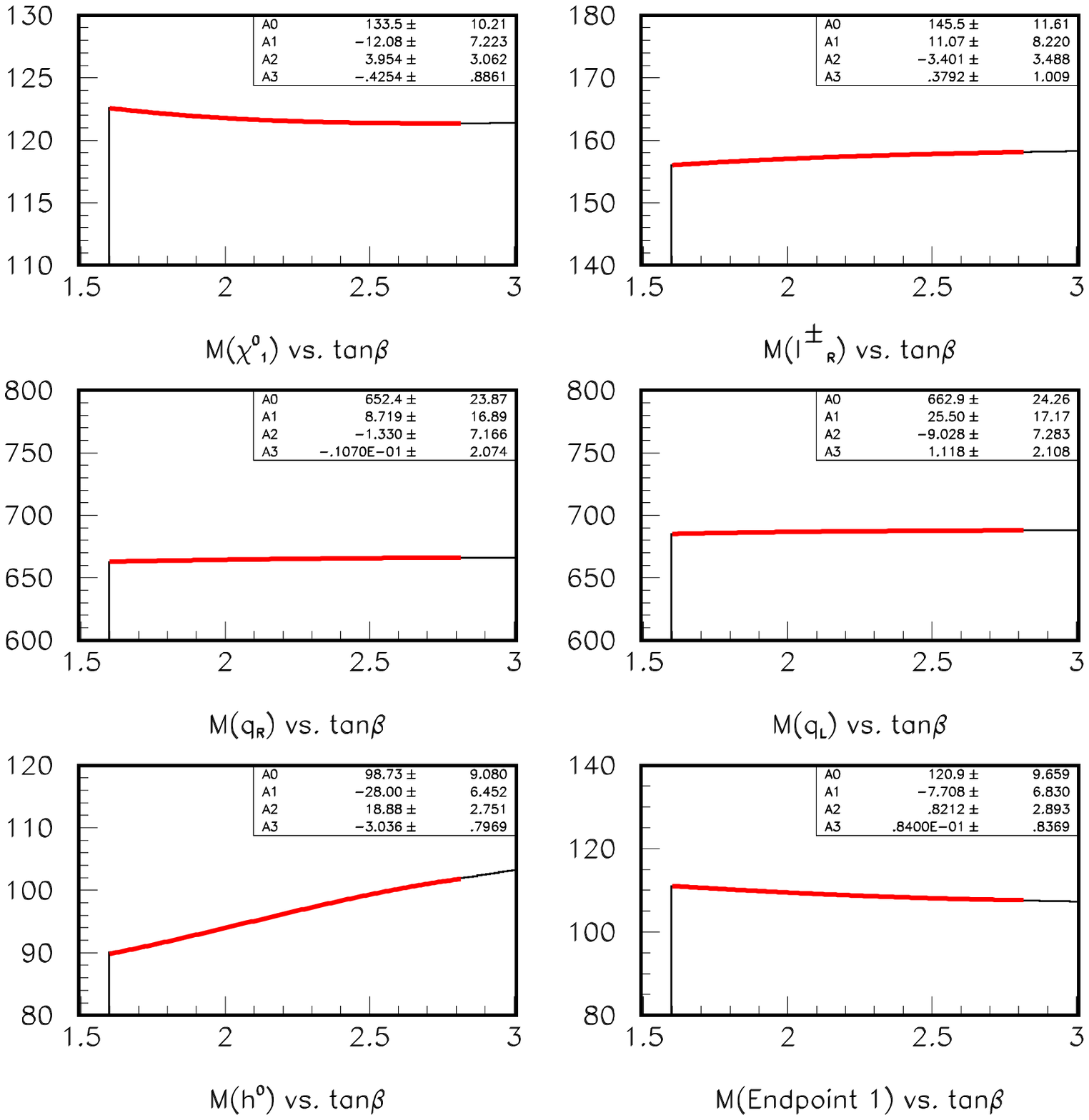,width=16.0cm }}
\caption{\small The dependence of $m_{\tilde \chi^0_1}$, $m_{\tilde l_R}$, $m_{\tilde q_R}$,
$m_{\tilde q_L}$, $m_{\tilde h^0}$ and $M^{end}_{OSSF}$ on tan$\beta$ smoothed by a polynomial fit.
All other parameters are fixed at those of point 5.}
\label{fg:p5alltanbfit}
\end{Fighere}
\vskip 1.0 cm

\begin{Tabhere}
\centering
\caption{\small  The slopes, $\frac{\partial O^{th}_{\alpha}}{\partial tan\beta}$,
of the observables  at the
nominal value of point 5 : tan$\beta=2.1$ .}
\vskip 0.5cm \hskip -0.8cm
\begin{tabular}{|c|c|c|c|c|c|} \hline\hline
$\tilde \chi^0_1$ & $\tilde l^{\pm}_R$ & $\tilde q_R$ & $\tilde q_L$ & $\tilde h^0$ & $M^{end}_{OSSF}$ \\ \hline\hline
-1.101242 & 1.802616 & 2.99144 & 2.37354 & 11.1297 & -3.14764 \\ \hline\hline
\end{tabular}
\label{tb:slope_ptanb}
\end{Tabhere}

\newpage

\begin{Tabhere}
\centering
\caption{\small  The measured values of observables in SUGRA points 1, 3 and 5 for each $\rlap/R$
coupling considered. All the values are in GeV.}
\vskip 0.5cm \hskip -0.3cm
\begin{tabular}{|c|c|c|c|c|c|} \hline\hline
$O^{meas} \pm \delta O^{meas}$ & \multicolumn{1}{c|}{Point 1} & \multicolumn{2}{c|}{Point 3} & \multicolumn{2}{c|}{Point 5} \\ \cline{2-6}

                               &$\lambda_{122}$ & $\lambda_{122}$ & $\lambda_{123}$ & $\lambda_{122}$ & $\lambda_{123}$ \\ \hline\hline
$m_{\tilde g}\pm 
\sigma (m_{\tilde g})$           &               & 301.1$\pm$3.0     & 301.1$\pm$3.5$^{3)}$ &      &               \\ \hline
$m_{\tilde q_R}\pm 
\sigma (m_{\tilde q_R})$         & 932$\pm$20    &               &               & 662$\pm$12 &               \\ \hline
$m_{\tilde q_L}\pm 
\sigma (m_{\tilde q_L})$         &               &               &               & 685$\pm$20& 686$\pm$12 \\ \hline
$m_{\tilde t_1}\pm 
\sigma (m_{\tilde t_1})$         &               &               &               & 504$\pm$20&               \\ \hline
$m_{\tilde b_1}\pm 
\sigma (m_{\tilde b_1})$         &               & 276.6$\pm$3.0   & 277.5$\pm$3.0$^{3)}$ &     &               \\ \hline
$m_{\tilde l_R}\pm 
\sigma (m_{\tilde l_R})$         &               &               &               & 156.8$\pm$1.8 &               \\ \hline
$m_{\tilde \chi^{\pm}_1}\pm 
\sigma (m_{\tilde \chi^{\pm}_1})$& 328.2$\pm$6.5 &               &               & 232.2$\pm$4.5 &               \\ \hline
$m_{\tilde \chi^0_2}\pm 
\sigma (m_{\tilde \chi^0_2})$    & 326.2$\pm$6.0     & 96.7$\pm$0.2  &               & 230.7$\pm$3.9 & 228.2$\pm$5.0 \\ \hline
$m_{\tilde \chi^0_1}\pm 
\sigma (m_{\tilde \chi^0_1})$    & 169.8$^{+0.2}_{-0.8}$ & 44.8$^{+0.1}_{-0.2}$ &  & 122.6$^{+0.4}_{-1.0}$ &               \\ \hline
$m_{\tilde h^0}\pm 
\sigma (m_{\tilde h^0})$         & 97.1$\pm$1.5  & 69$\pm$1.0$^{4)}$  & 69$\pm$1.0$^{4)}$ & 94.7$\pm$1.5 & 94.3$\pm$1.5  \\ \hline
$M^{end\ \ 1)}_{OSSF} 
\pm \sigma (M^{end}_{OSSF})$
                 &               &               &               &               & 111.9$\pm$2.5 \\ \hline
$m^{end\ \ 2)}_{OSSF} 
\pm \sigma (m^{end}_{OSSF})$
                 &               & 53.3$\pm$0.6  & 52.9$^{+0.1}_{-0.3}$ $^{3)}$ &   &         \\ \hline\hline
\multicolumn{6}{|c|}{ } \\
\multicolumn{6}{|c|}{
$^{1)}$\ $M^{end}_{OSSF} = m_{\tilde \chi^0_2}\sqrt{1-\Bigl(\frac{m_{\tilde l_R}}{m_{\tilde \chi^0_2}}\Bigr)^2}
\sqrt{1-\Bigl(\frac{m_{\tilde \chi^0_1}}{m_{\tilde l_R}}\Bigr)^2}$  } \\
\multicolumn{6}{|c|}{
$^{2)}$\ $m^{end}_{OSSF} = m_{\chi^0_2} - m_{\chi^0_1}$ } \\
\multicolumn{6}{|c|}{ 
$^{3)}$ Assuming $m_{\tilde \chi^0_2}= 97$ GeV } \\
\multicolumn{6}{|c|}{ 
$^{4)}$ From other measurements ($h^0 \rightarrow \gamma \ \gamma$) } \\
\multicolumn{6}{|c|}{ } \\ \hline\hline
\end{tabular}
\label{tb:values}
\end{Tabhere}
\vskip 0.5cm

\begin{Tabhere}
\centering
\caption{\small  The relative errors on $m_0$, $m_{1/2}$ and tan$\beta$ in SUGRA points 1, 3 and 5 for each $\rlap/R$
coupling considered.}
\vskip 0.5cm
\begin{tabular}{|c|c|c|c|c|c|c|c|} \hline\hline
  & \multicolumn{1}{c|}{ } & \multicolumn{4}{c|}{ } & \multicolumn{2}{c|}{ } \\
Relative errors on the &  \multicolumn{1}{c|}{Point 1} & \multicolumn{4}{c|}{Point 3} &
                            \multicolumn{2}{c|}{Point 5} \\
SUGRA parameters  &
                     \multicolumn{1}{c|}{ } & \multicolumn{4}{c|}{ } & \multicolumn{2}{c|}{ } \\ \cline{2-8}

                    &$\lambda_{122}$ & $\lambda_{122}$ & $\lambda_{123}$ &
		     $\lambda_{122}^{*)}$ & $\lambda_{123}^{*)}$ &
                     $\lambda_{122}$ & $\lambda_{123}$ \\ \hline\hline
$\delta m_{0} / m_0$ (\%)           &   12  &  4.4 &  7.3 &  4   &  5.8 &  2.9 & 9.7   \\ \hline
$\delta m_{1/2} / m_{1/2}$ (\%)     &   0.3 &  0.3 &  0.6 &  0.2 &  0.4 &  0.5 & 1.4   \\ \hline
$\delta tan \beta / tan \beta$ (\%) &   5   &  3.3 &  3.3 &  1.8 &  1.8 &  6   & 6.2  \\ \hline\hline
\multicolumn{8}{|c|}{ } \\
\multicolumn{8}{|c|}{ $^{*)}$ If the measurement of $m_{h^0}$ is taken into account. } \\
\multicolumn{8}{|c|}{ } \\ \hline\hline
\end{tabular}
\label{tb:sugrafitp1}
\end{Tabhere}

\newpage
\section{Conclusions}

We have studied the feasibility to detect a SUSY signal 
by ATLAS in the framework of the SUGRA model 
and to determine its parameters in the
case when $R$ parity is broken in conjunction with
lepton number violation: $\lambda \ne 0$. For this purpose we
have chosen three representativ points in the SUGRA
parameter space and two different type of couplings,
both having a value $10^{-3}$, small enough to
concentrate the effect in the LSP decay but large enough
not to see displaced vertex in this decay.

Our conclusions are the following:

1. The SUGRA signal is visible in a very large domain
of the parameter space, even beyond $m_0 \sim m_{1/2} \sim 1$ TeV.

2. The energy scale of the SUGRA signal can be determined by
inclusive measurements like the effective mass ($m_{eff}$) or
the normalised transverse momentum per lepton ($p_t^{l, norm}$).

3. In the case of couplings with absence of a $\tau$ among the
decay products of the $\tilde \chi^0_1$ (e.g. $\lambda_{122}$) one can reconstruct the
SUSY particles  and this reconstruction
can be used for a precision determination of the model parameters.
The achieved precision turns out to be
better than it was the case with conserved $R$ parity. This
is because one can reconstruct the LSP from its decay products.
At the low energy point where
the chargino or second lightest neutralino produces additional
leptons this determination is slightly handicapped by the combinatorial
background and the most complex structure.

4. In the case of LSP decay with a $\tau$ particle in the final
state the full reconstruction of the LSP, i.e. the determination of
its four momentum, is not always possible, however,
one can still estimate its mass (except at point 3 - $\lambda_{123}$).
It allows ones to determine the parameters of the SUGRA model in spite of
the large combinatorial background due to the leptonic
decay of the LSP. This determination in most of the
cases is better or at least comparable in precision with that when $R$ parity
is conserved.

\vskip 1.0cm
{\Large \bf Aknowledgements}\\
\\
We would like to thank Daniel Froidevaux for suggesting us to
carry out this study, his constant help and encouragement.
We had many useful discussions within the ATLAS SUSY working
group and would like to express our gratitude to its members, especially
to Ian Hinchliffe, Frank Paige and Giacomo Polesello.


\end{document}